\documentclass[
  reprint,
  amsmath,
  amssymb,
  aps,
  pra,
  twocolumn,
  longbibliography
]{revtex4-2}

\usepackage{macros}
\usepackage{tikz,xcolor}
\usepackage{bm}
\usepackage{braket}
\usepackage{comment}
\usepackage{algpseudocode}
\usepackage{booktabs}
\usepackage{multirow}
\usepackage{subfiles}
\usepackage{bibunits}
\usepackage[normalem]{ulem} 

\newcommand{\starttocentries}{\let\addcontentsline\oldaddcontentsline}

\newcommand{\needsref}[1]{{\textsf{{\color{orange}[Insert Ref]}}}}

\begin{document}
\title{Regularized Warm-Started Quantum Approximate Optimization\\and Conditions for Surpassing Classical Solvers on the Max-Cut Problem}

\author{Zichang He}
\email{zichang.he@jpmchase.com}
\affiliation{Global Technology Applied Research, JPMorganChase, New York, NY 10017, USA}
\author{Anuj Apte}
\affiliation{Global Technology Applied Research, JPMorganChase, New York, NY 10017, USA}
\author{Brandon Augustino}
\affiliation{Global Technology Applied Research, JPMorganChase, New York, NY 10017, USA}
\author{Arman Babakhani}
\affiliation{Global Technology Applied Research, JPMorganChase, New York, NY 10017, USA}
\author{Abid Khan}
\affiliation{Global Technology Applied Research, JPMorganChase, New York, NY 10017, USA}
\author{Sivaprasad Omanakuttan}
\affiliation{Global Technology Applied Research, JPMorganChase, New York, NY 10017, USA}
\author{Ruslan~Shaydulin}
\email{ruslan.shaydulin@jpmchase.com}
\affiliation{Global Technology Applied Research, JPMorganChase, New York, NY 10017, USA}

\date{\today}%
             
\begin{abstract}
    Demonstrating quantum heuristics that outperform strong classical solvers on large-scale optimization remains an open challenge. Here we introduce Regularized Warm-Started QAOA (RWS-QAOA), which initializes qubits by minimizing expected energy with a regularizer that penalizes near-bitstring states, preventing QAOA from stalling. We further propose a protocol that yields fixed, instance-independent parameters, enabling RWS-QAOA to operate as a non-variational algorithm in which the quantum circuit parameters are fixed and only a classical warm starting step is instance-dependent. 
    We evaluate RWS-QAOA on the Max-Cut problem for random regular graphs, where this protocol yields a constant-depth quantum circuit, across three complementary settings. 
    First, on Quantinuum's trapped-ion processor, RWS-QAOA outperforms the classical algorithms with the best provable guarantees for Max-Cut on $3$-regular graphs, namely Goemans--Williamson and Halperin--Livnat--Zwick, on $96$-node instances.
    Second, tensor-network simulations on graphs with up to $N{=}10{,}000$ nodes show that depth-$6$ RWS-QAOA, achieving an average cut fraction of $0.9167$, surpasses the best classical heuristics under matched restrictions (no local-search post-processing and no iterative refinement). Third, we remove these restrictions and benchmark against the strongest unrestricted classical heuristics, including an optimized parallel Burer--Monteiro solver that improves upon the \texttt{MQLib} implementation. Even against this stronger baseline, we project that surface-code RWS-QAOA reaches a quantum--classical runtime crossover below $0.2$ seconds on $3{,}000$-node graphs with fewer than $1.3$ million physical qubits.
    Our results show that constant-depth quantum circuits combined with a classical warm start have a credible potential to surpass classical solvers on the Max-Cut problem when executed on future quantum computers.
    
\end{abstract}
\maketitle
\begin{bibunit}[apsrev4-2-author-truncate]
\section{Introduction}

Quantum computing has demonstrated experimental advantages across diverse domains, from sampling and simulation to cryptography~\cite{quantumecho2025,liu2025certified,niroula2026digital,Liu2025,King2025,haghshenas2025digital}. While great theoretical progress has been made in quantum algorithms for optimization~\cite{Jordan2025,chakrabartiShortPath,Dalzell2023,2602.13494,2510.03385,2311.03977,2303.01471,Abbas2024}, these quantum algorithms with provable guarantees have deep quantum circuits, leading to a high crossover time against classical solvers~\cite{omanakuttan2025threshold}. The high runtime needed to outperform classical solvers limits the near-term practical usefulness of these algorithms.

Among candidate quantum algorithms for optimization, the quantum approximate optimization algorithm (QAOA)~\cite{farhi2014quantumapproximateoptimizationalgorithm} stands out due to its simplicity, enabling practical experiments on current devices~\cite{hogg2000quantumoptimization,Hogg2000,Shaydulin2023npgeq,Pelofske2023,Pelofske2024,he2024performance}. Studies indicate that QAOA can outperform classical solvers in targeted regimes, with provable polynomial~\cite{boulebnane2022solving,shaydulin2023evidence,Vaishnav2024,apte2025iterative} and even exponential speedups under specific conditions~\cite{Zhou2024}. In addition, there is empirical evidence for scaling advantages on hard instances~\cite{shaydulin2024evidence,boulebnane2024solving}. However, standard QAOA on its own may have limited practical utility as resource estimates suggest crossover time of hours when taking into account the overhead of fault tolerance~\cite{omanakuttan2025threshold}.

Standard QAOA evolves a uniform initial state through $p$ alternating layers of cost and mixer unitaries. 
To improve QAOA performance at small depth, numerous strategies for \textit{warm-starting} QAOA have been proposed~\cite{bhattacharyya2025solving,tate2023warm,augustino2024strategies,egger2021warm,yu2025warm,yuan2025iterative}, where the basic idea is to provide the QAOA with an initial state obtained through classical pre-processing. Similar ideas have also been applied to other quantum algorithms beyond QAOA~\cite{nguyen2025theoretical,PRXQuantum.6.010317}. However, existing approaches face key limitations: (i) simple warm-start schemes can produce initial states that are too close to classical bitstrings, and cause QAOA evolution to stall~\cite{He2023,cain2022qaoa}, (ii) warm starts based on semidefinite programming (SDP) incur cubic pre-processing cost in the problem size, (iii) QAOA performance depends on parameter schedules which require heavy instance-specific optimization, and (iv) no approach has demonstrated clear advantages over state-of-the-art classical solvers at large scale.  These gaps motivate the present work.

\begin{figure*}[th]
    \centering
    \includegraphics[width=1\linewidth]{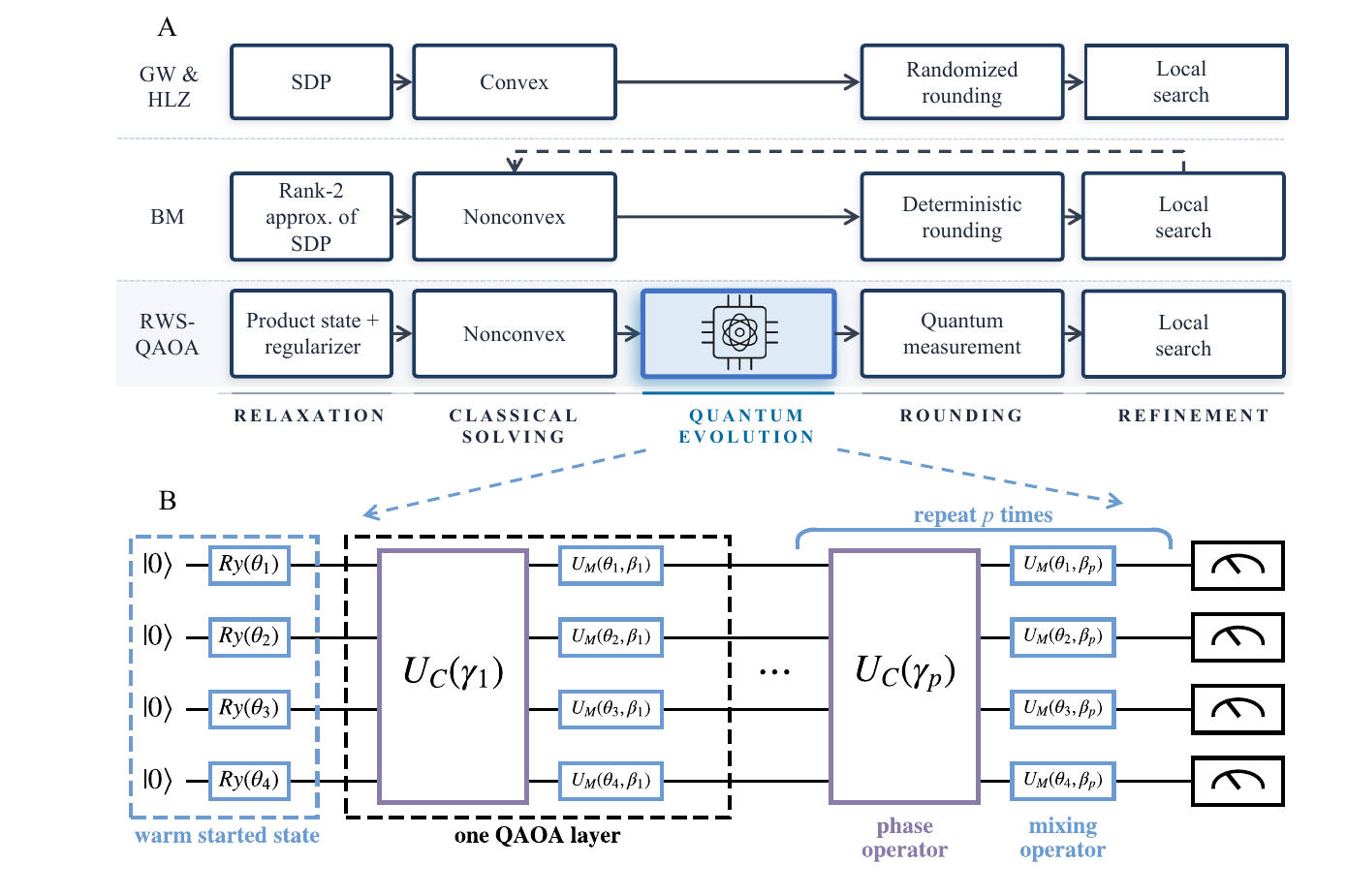}
    \caption{ 
    \textbf{Overview of the proposed RWS-QAOA algorithm.}
    (A) Algorithmic pipeline of Goemans--Williamson (GW), Halperin--Livnat--Zwick (HLZ), Burer--Monteiro (BM) and RWS-QAOA. All four methods solve a continuous relaxation and then round to a bitstring. RWS-QAOA inserts a constant-depth QAOA circuit with a fixed, instance-independent parameter schedule between the relaxation and measurement steps. 
    (B) RWS-QAOA circuit. A regularized warm-start optimization sets per-qubit $R_y(\theta)$ rotations, placing each qubit in a biased superposition. The state then undergoes $p$ alternating layers of cost and mixer evolution. 
    }
    \label{fig:algorithmic_pipline}
\end{figure*}

Our primary contribution is a \emph{regularized warm start} (RWS) protocol for QAOA. Each qubit is initialized by an $R_y$ rotation with angle $\theta_i$, and a regularization term $\lambda \sum_i \sin^2 \theta_i$, which penalizes bitstring-like extremes ($\theta_i \in \{0,\pi\}$). 
This balances two competing objectives: the energy term biases qubit orientations toward cuts favored by the graph topology, while the regularizer promotes superposition, preventing the initial state from collapsing to a classical bitstring where QAOA evolution stalls.
The resulting nonconvex optimization problem over $\boldsymbol{\theta}$ is solved via gradient-based methods, with per-iteration cost that scales linearly with the number of nonzeros in the objective. The overall circuit diagram is illustrated in \cref{fig:algorithmic_pipline}B. 

To complement these techniques, we propose a parameter-setting protocol that yields a fixed, instance-independent schedule for the QAOA layer parameters $\boldsymbol{\gamma}$ and $\boldsymbol{\beta}$, which control the evolution under the cost and mixing Hamiltonians at each depth.
By optimizing over sampled local subgraphs drawn from a representative set of instances, we obtain concentrated parameters that transfer to unseen graphs within the same problem family. Consequently, RWS-QAOA operates as a fully non-variational algorithm: once the warm-start state is prepared, the only cost associated with running the algorithm is the execution of a quantum circuit. 

We provide a comprehensive evaluation of RWS-QAOA against state-of-the-art classical solvers on the Max-Cut problem for random regular graphs across three complementary settings. 

First, we compare against classical algorithms with provable guarantees. On Quantinuum's trapped-ion processor, \texttt{Helios}~\cite{ransford2025}, RWS-QAOA achieves higher approximation ratios and success probabilities for Max-Cut on $96$-node 3-regular graphs than the state-of-the-art SDP-based approaches due to Goemans and Williamson~\cite{goemans1995} and Halperin, Livnat and Zwick~\cite{Halperin2004}. 

Second, we compare against classical heuristics. Since existing simulation techniques can only compute expected QAOA performance and cannot sample from $p=6$ QAOA state beyond a few hundred qubits, we cannot evaluate the impact of local search improvements to the bitstrings produced by QAOA. To make the comparison with classical heuristics fair, we similarly restrict their local search step. We use tensor-network simulation up to $N=10{,}000$ and $p=6$ to show that the RWS-QAOA surpasses the best classical heuristic, the Burer--Monteiro (BM) algorithm, with no local search improvement.

Finally, we remove the requirement of ``fair'' comparison and compare with best classical heuristics executed with no restrictions. We introduce an optimized parallel version of BM, improving upon the implementation in \texttt{MQLib}. We predict that RWS-QAOA for a planar surface-code architecture \cite{kitaev2006anyons,google2025quantum,gidney2025factor} will achieve a quantum--classical runtime crossover against this improved \texttt{MQLib+} implementation below $0.2$ second for $3{,}000$-node graphs, using fewer than $1.3$ million physical qubits.

The regularized warm-start approach we introduce extends naturally to any quadratic unconstrained binary optimization problem; extending it to a broader class of problems is an interesting direction for future work. 
Note that the optimal logical two-qubit depth for a degree-$D$ regular graph only depends on the degree according to Vizing's theorem~\cite{west2001introduction}. Thus, given a fixed QAOA depth $p$, the logical circuit has constant depth $\mathcal{O}(Dp)$. 
Our large-scale numerical experiments show that constant-depth quantum algorithms combined with a classical warm start have a credible potential to outperform the best classical heuristics on future quantum computers.

\section{Results}

In this work, we study quantum and classical algorithms for the Max-Cut problem, which has applications in fields such as circuit design (minimizing wire crossings)~\cite{Barahona1988} and information aggregation in machine learning~\cite{chatziafratis2021}. It is a problem that has been widely studied using quantum optimization methods~\cite{ farhi2014quantumapproximateoptimizationalgorithm, wurtz2021maxcut,  wurtz2021fixed, He2023, dupont2025benchmarking}.

The goal of Max-Cut is to partition the vertices of a graph into two disjoint sets such that the number of edges crossing the partition is maximized. The Max-Cut problem can be naturally formulated as a quadratic unconstrained binary optimization (QUBO) problem, by associating each vertex with a binary variable, and expressing the cut value as a quadratic function of these variables. For a graph $G = (V, E)$ with $N := |V|$ vertices, the Max-Cut problem can be written in the form
\begin{equation} \label{e:max_cut}
\max_{\mathbf{x} \in \{0,1\}^N} f(\mathbf{x}) = \sum_{(i,j) \in E} (x_i + x_j - 2  x_i x_j),
\end{equation} 
where each binary variable $x_i$ indicates the side of the cut to which vertex $i$ is assigned. The term $x_i + x_j - 2 x_i x_j$ equals $1$ if $x_i \ne x_j$ and 0 otherwise, so the objective counts the number of edges crossing the cut.

\Cref{e:max_cut} is an instance of the general QUBO problem 
\begin{align}\label{eq:qubo_prob}
    \min_{\mathbf{x} \in \{0,1\}^N} \mathbf{x}^\top \mathbf{Q} \mathbf{x},
\end{align}
with $\mathbf{Q} = -\mathbf{L}$, where $\mathbf{L}$ is the graph Laplacian matrix:
\begin{equation}
\mathbf{L}_{ij} := \begin{cases}
    \sum_{j : (i,j) \in E} 1 &\text{if}~i = j \\
    - 1 &\text{otherwise}.
\end{cases}
\end{equation} 

To evaluate the quality of a solution $\mathbf{x}$ to the Max-Cut problem, it is common to define performance metrics in terms of the \emph{cut fraction} 
\begin{equation}
    \mathrm{cut~fraction}(\mathbf{x}) := \frac{f(\mathbf{x})}{\lvert E \rvert}, %
\end{equation}
where $\lvert E \rvert$ is the total number of edges. Letting $f_{\rm max}$ denote the maximum cut value, the \emph{approximation ratio} of an algorithm on graph $G$ is given by
\begin{equation}
    \alpha := \frac{f(\mathbf{x}_{\mathrm{alg}})}{f_{\rm max}},
\end{equation}
where $\mathbf{x}_{\mathrm{alg}}$ is the solution returned by the algorithm. 
The \emph{success probability} of an algorithm is defined as the probability of finding the optimal cut value.
\begin{equation}
    p(\text{success}) := p(f(\mathbf{x}_{\mathrm{alg}}) = f_{\rm max}).
\end{equation}

For Max-Cut, it is NP-hard to achieve a worst-case approximation ratio exceeding $16/17$ for general graphs~\cite{Hstad2001} and $331/332$ for $3$-regular graphs~\cite{Berman1999}. For random $3$-regular graphs, one can bound the maximum achievable cut fraction using the replica method from statistical physics, leading to a bound of $0.92386$ almost surely over random choice of the graph~\cite{harangi2025rsbboundsmaximumcut}.

\subsection{Classical solvers}\label{sec:res_classical_solvers}
The Goemans--Williamson (GW) algorithm~\cite{goemans1995} approximates the solution to Max-Cut in polynomial time using a convex semidefinite programming (SDP) relaxation. By performing randomized rounding on the SDP optimal solution, one obtains a cut with worst-case approximation ratio $\approx 0.8786$. This is optimal under the Unique Games Conjecture~\cite{Khot2007}. The SDP relaxation can be solved using Interior Point Methods (IPMs)~\cite{wolkowicz2012handbook}. These algorithms solve large, dense linear systems exactly at each iteration, and thus become prohibitively expensive at intermediate problem sizes; in our experiments solving the SDP to optimality becomes challenging for $N \geq 2{,}000$ (see runtime benchmarks in \cref{fig:compare_runtime}). For $3$-regular graphs, the Halperin--Livnat--Zwick (HLZ) algorithm~\cite{Halperin2004} tightens the SDP relaxation by adding structure-aware inequalities and performing a local improvement step, raising the provable worst-case approximation ratio to $0.9326$. For more details, we refer the reader to \cref{sec:classical_solvers}.
 
Beyond provable algorithms, there exist many classical heuristics that work well. Benchmarking shows that Burer--Monteiro (BM)~\cite{burer2002rank} is the best-performing method for 3-regular graphs (see \cref{fig:compare_methods_overN} and \cref{sec:compare_classical_solvers}). BM addresses the scalability limitations faced when solving SDPs with second-order methods like IPMs by using low-rank change of variables. This approach ensures the solution is positive semidefinite, and reduces the number of variables from $\mathcal{O}(N^2)$ to $\mathcal{O}(kN)$ with $k \ll N$. One can solve the low-rank SDP formulation using first order methods (e.g., augmented Lagrangian), but the problem at hand is nonconvex. For the rank-2 case used here for Max-Cut, each vertex is mapped to an angle $\theta_i$ on the unit circle, and the non-convex objective is optimized with gradient-based methods; the best cut is then extracted by a deterministic sweep (see \cref{sec:classical_solvers} for details). 
The overall algorithmic flowchart of BM, and GW or HLZ is summarized in \cref{fig:algorithmic_pipline}A. Their solutions obtained through the rounding can be further improved through a local search step, while HLZ has its own local search rules and a general greedy local search (see \cref{sec:local_search}) works for BM and GW solutions.

BM is similar to GW and HLZ in that it first solves a continuous relaxation of the Max-Cut problem, followed by rounding and local search. The performance of BM depends on two tunable parameters: (i)~the number of independent random initializations (multistarts) $M$, since the rank-2 landscape is highly non-convex; and (ii)~the stopping criterion for iterative search, namely the number of sequential perturbation rounds $K$, in which each rounded bitstring is mapped back to angles, randomly perturbed, and re-optimized. In \cref{fig:algorithmic_pipline}A, it is represented by a closed loop from the refined solution to the nonconvex solving. Increasing either $M$ or $K$ yields monotonic improvements at proportional computational cost. 

We find that the C++ implementation of BM (named \texttt{BURER2002}) in \texttt{MQLib} is the most performant off-the-shelf solver for Max-Cut. We additionally introduce a modification, denoted \texttt{MQLib+}, where we parallelize BM over random initializations in step (i). Parallelization enables running BM with high $K$ within a modest time budget, leading to large performance gains over \texttt{MQLib} (see \cref{fig:bm_rsb_bound}).

\begin{figure*}[th]
    \centering
    \includegraphics[width=\textwidth]{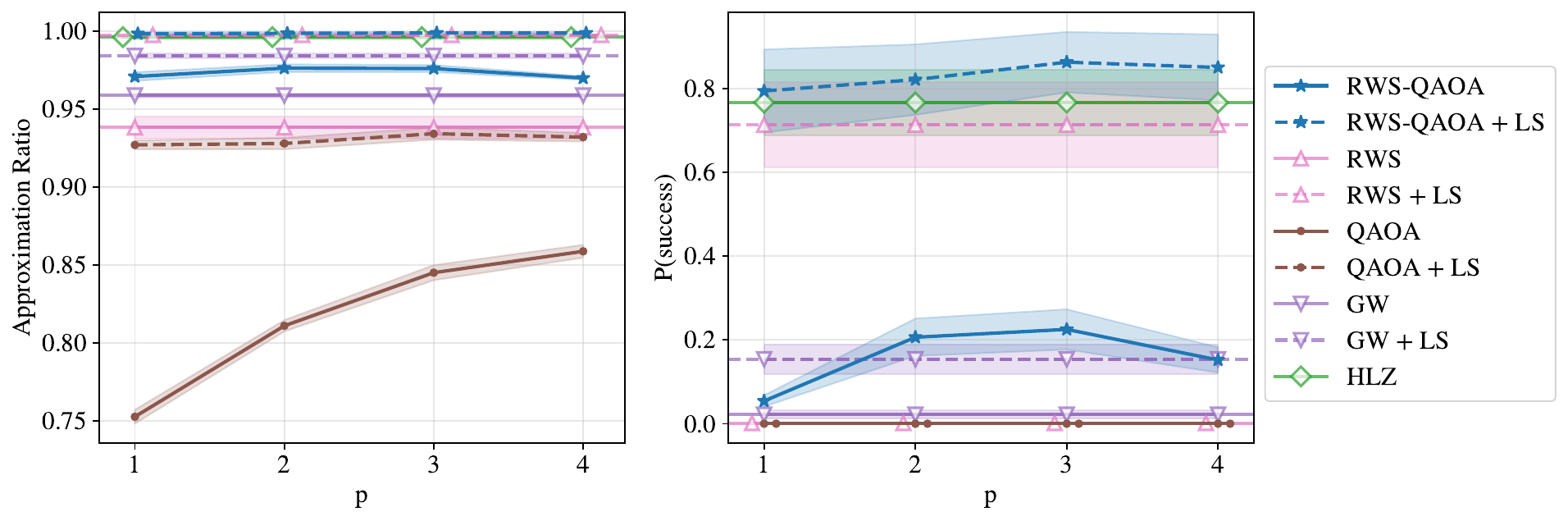}
    \caption{
    \textbf{Regularized Warm-started QAOA outperforms algorithms with best provable guarantees on hardware.} 
    Experimental results of Max-Cut on five $N=96$ $3$-regular graphs executed on the Quantinuum trapped-ion device \texttt{Helios}. RWS‑QAOA, with and without a single‑step local search (LS), outperforms Goemans--Williamson (GW), Halperin--Livnat--Zwick (HLZ), and standard QAOA (each with and without LS) in both approximation ratio and success probability. HLZ natively includes a local search step. Shaded areas denote standard errors.
    }\label{fig:helios_result}
\end{figure*}
\subsection{RWS-QAOA}
We now introduce our RWS-QAOA algorithm. The intuition is as follows. The best classical heuristic, BM, proceeds by first solving a relaxed continuous problem, followed by rounding and local improvement. A natural quantum extension is to load the solution of the classical relaxation onto qubits as a quantum state, quantumly improve upon it using QAOA evolution before measuring and post-processing using classical local search. Below, we introduce this approach in detail and discuss some challenges that we overcome to make it practical.

Recall that QAOA~\cite{farhi2014quantumapproximateoptimizationalgorithm,hogg2000quantumoptimization,Hogg2000} prepares a parameterized quantum state $\ket{\psi(\boldsymbol{\gamma}, \boldsymbol{\beta})}$
\begin{equation}\label{eq:standard_qaoa}
    \ket{\psi(\boldsymbol{\gamma}, \boldsymbol{\beta})} = \prod_{t = 1}^p e^{-i\beta_t \mathbf{H}_M}e^{-i\gamma_t \mathbf{H}_C} \ket{\psi_0},
\end{equation}
where $\mathbf{H}_C$ is the cost Hamiltonian, $\mathbf{H}_M$ is the mixer, 
and $\boldsymbol{\gamma}, \boldsymbol{\beta}$ are tunable parameters. In standard QAOA, $\ket{\psi_0} = \ket{+}^{\otimes N}$ and $\mathbf{H}_M = \sum_i \mathbf{X}_i$.

In warm-started QAOA~\cite{tate2023warm,augustino2024strategies,egger2021warm}, the initial state encodes a classically obtained approximate solution. One simple method is applying a ${R_y}(\theta_i)$ gate on each qubit, leading to a parameterized initial state $\ket{\psi_0(\boldsymbol{\theta})}$
\begin{equation}
    \ket{\psi_0(\boldsymbol{\theta})} = {\left[ \cos(\theta_i /2) \ket{0} + \sin(\theta_i /2) \ket{1} \right]}^{\otimes N},
\end{equation}
where $\boldsymbol{\theta} = (\theta_1, \ldots, \theta_N)$ is determined from the solution of the relaxed optimization problem. In this state the probability of obtaining outcome $1$ on the i-th qubit is given by $p_i = \sin^2(\theta_i/2)$.  As performance of QAOA improves by selecting the mixing Hamiltonian so that $\ket{\psi_0(\boldsymbol{\theta})}$ is its ground state (up to sign choice)~\cite{He2023}, we ensure that this is the case by picking the mixing Hamiltonian to be $\mathbf{H}_M(\boldsymbol{\theta}):=\sum_{i=1}^N \sin(\theta_i) \mathbf{X}_i + \cos(\theta_i) \mathbf{Z}_i$.
Thus, the warm-started QAOA state is given by:
\begin{equation}\label{eq:warm_started_state}
\ket{\psi(\boldsymbol{\theta}, \boldsymbol{\gamma}, \boldsymbol{\beta})} = \prod_{t = 1}^p e^{-i\beta_t \mathbf{H}_M(\boldsymbol{\theta})}e^{-i\gamma_t \mathbf{H}_C} \ket{\psi_0(\boldsymbol{\theta})}.
\end{equation}
When $\theta_i = \pi/2$ for all qubits, the warm-started QAOA reduces to the standard version, \cref{eq:standard_qaoa}.

A natural idea is to choose $\boldsymbol{\theta}$ so that the initial state maximizes the expected energy $$\mathbb{E}_{\mathbf{x} \sim \ket{\psi_0(\boldsymbol{\theta})}}\left[ \mathbf{x}^\top \mathbf{Q} \mathbf{x} \right] \equiv \braket{\psi_0(\boldsymbol{\theta}) \vert \mathbf{H}_C \vert \psi_0(\boldsymbol{\theta})}$$ 
with $\mathbf{H}_C = \sum_{(i,j) \in E }\frac{1}{2}(1-\mathbf{Z}_i\mathbf{Z}_j)$, i.e., set $\boldsymbol{\theta}$ according to
\begin{equation}\label{eq:direct_p0}
   \boldsymbol{\theta}_{C} = \arg\max_{\boldsymbol{\theta} \in {[0,2\pi)}^N}  \mathcal{L(\boldsymbol{\theta})} = \mathbb{E}_{\mathbf{x} \sim \ket{\psi_0(\boldsymbol{\theta})}}\left[ \mathbf{x}^\top \mathbf{Q} \mathbf{x} \right].
\end{equation}
However, the resulting state $\ket{\psi_0(\boldsymbol{\theta}_{C})}$ often becomes a superposition dominated by a small number of classical bitstrings, or even a single bitstring. If the mixing Hamiltonian is chosen so that $\ket{\psi_0(\boldsymbol{\theta}_{C} )}$ is its ground state, then this initial state is a common eigenstate of both the cost and mixing Hamiltonians. As a result, the state remains invariant under QAOA evolution and the algorithm gets stuck without exploring new solutions, which still holds even if the mixing Hamiltonian is not initial state dependent~\cite{cain2022qaoa}. This phenomenon limits the effectiveness of warm starts that are too ``classical'' or sharply peaked.

Furthermore, while full-rank SDP based warm start methods~\cite{augustino2024strategies,egger2021warm} can provide high-quality initial states, they are computationally expensive and scale poorly for large problem instances. This motivates the need for efficient and scalable warm start strategies that avoid these pitfalls and enable meaningful quantum evolution. 

To avoid this stalling while keeping preprocessing lightweight, we introduce a regularized warm starting approach for determining the initial state $\ket{\psi_0(\boldsymbol{\theta})}$ that explicitly encourages the starting state to have similar probability for $0$ and $1$ on each qubit. If the probability of observing $1$ on the i-th qubit is $p_i$ then the simplest such regularization term is $\sum_{i=1}^N (p_i - 1/2)^2$ which is equivalent to $\sum_{i=1}^N \sin^2{\theta_i}$ up to an overall constant.

The objective function for regularized warm start (RWS) is as follows:
\begin{equation}\label{eq:rws_objective}
    \mathcal{L_{\lambda}(\mathbf{p})}
    \equiv \mathbb{E}_{\mathbf{x} \sim \ket{\psi_0(\boldsymbol{\theta})}}\left[ \mathbf{x}^\top \mathbf{Q} \mathbf{x} \right] - 4 \lambda \sum_{i=1}^N p_i (1 - p_i),
\end{equation}
where $\mathbf{p} = (p_1, \ldots, p_N)$, $\lambda > 0$ is the regularization hyper-parameter and the factor of $4$ is chosen for convenience. When $\lambda \rightarrow 0$, our approach reduces to \eqref{eq:direct_p0}, while when $\lambda \rightarrow \infty$, the solution is trivially $\pi/2$ and the warm-started QAOA reduces to standard QAOA. Throughout this paper, we set $\lambda = 0.6$ for Max-Cut on 3-regular graphs. A detailed discussion on setting this hyperparameter $\lambda$ can be found in \cref{sec:set_regularization_strength}. 

In addition, we propose a protocol to get a set of fixed high-quality parameters $\boldsymbol{\gamma}$ and $\boldsymbol{\beta}$ for regular graphs in \cref{sec:fix_qaoa_para}. 
Thus, for any 3-regular graph, we have a non-variational circuit to prepare the RWS-QAOA state as described in \cref{eq:warm_started_state}. 

RWS-QAOA shares a similar relax-and-round structure of classical solvers but augments it with a quantum evolution step. The overall algorithmic flowchart is shown in \cref{fig:algorithmic_pipline}B. 
Related quantum approaches to relaxation and rounding have been explored in the literature.
Dupont et al.~\cite{dupont2024extending, dupont2025benchmarking} treat relaxation and rounding as a post-processing step applied to quantum outcomes and generalize the idea to quantum preconditioning~\cite{dupont2025optimization}, while Fuller et al.~\cite{fuller2024approximate} and He et al.~\cite{he2025non} relax the problem formulation via quantum random access codes.
Conversely, \v{C}epait\.{e} et al.~\cite{vcepaite2025quantum} use quantum solutions to warm-start classical solvers.

\subsection{Comparison with classical algorithms with provable guarantees on hardware}

We randomly select five 3-regular graphs with $96$ nodes and solve them using RWS-QAOA, standard QAOA, the regularized warm start alone, GW~\cite{goemans1995}, and HLZ~\cite{Halperin2004}. Experiments are executed on Quantinuum's trapped ion quantum processor, \texttt{Helios}~\cite{Helios}, using $96$ physical qubits. We run RWS-QAOA and QAOA at depths up to $p=4$, yielding circuits with up to $576$ two-qubit gates. Both quantum methods are executed non-variationally: RWS-QAOA uses the fixed parameter schedule described in \cref{sec:fix_qaoa_para}, while standard QAOA uses the tree parameters~\cite{wurtz2021fixed}. After collecting outputs from all solvers, including both measured bitstrings from the quantum devices and rounded bitstrings from the SDP relaxations, we apply a single step of local search to each solution (see \cref{sec:local_search}). HLZ natively includes the same local search.

Results are summarized in \cref{fig:helios_result}. We report both approximation ratio and p(success), where p(success) for GW and HLZ is defined as the fraction of rounded bitstrings that achieve the optimal cut over the total number of roundings. Overall, RWS-QAOA on hardware outperforms the state-of-the-art provable algorithms.

Across all experiments, RWS-QAOA consistently outperforms the standard QAOA. The performance of the standard QAOA increases with $p$, whereas RWS-QAOA achieves its best performance at $p=3$ and then declines slightly due to hardware noise. Without local search, RWS-QAOA exceeds GW in both approximation ratio and p(success), and attains an approximation ratio close to HLZ. Because the approximation ratio of RWS-QAOA is already high, local search yields limited additional gains in that metric; however, it substantially increases p(success), so that RWS-QAOA plus local search surpasses HLZ. This pattern suggests RWS-QAOA frequently produces near-optimal bitstrings that can be reliably elevated to optimality via local search, a behavior which is less pronounced for standard QAOA. An analysis of the fidelity of the circuit execution is provided in \cref{sec:hardware_error}.

\begin{figure*}[th]
    \centering
    \includegraphics[width=1\linewidth]{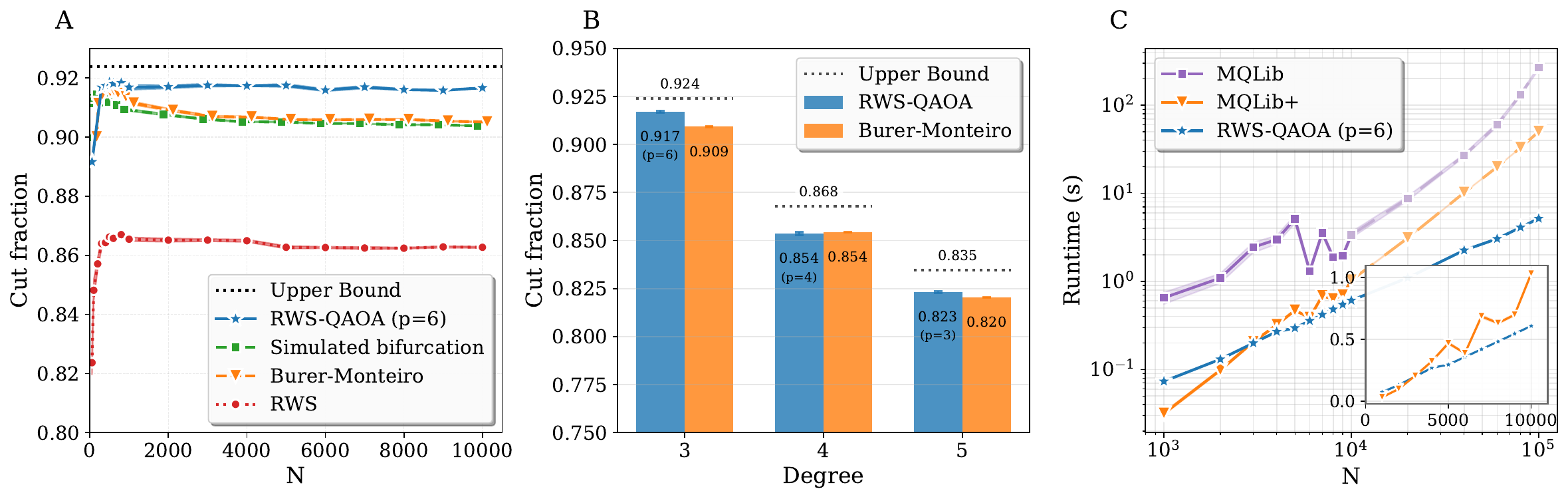}
    \caption{ 
    \textbf{RWS-QAOA outperforms classical heuristics on large graphs and conditions for runtime crossover.}
    (A) Average cut fraction of fixed-depth RWS-QAOA ($p=6$) versus classical heuristics on 3-regular graphs, where local-search post-processing and, where applicable, iterative refinement are disabled for all solvers (see text for justification).
    (B) Fixed-depth RWS-QAOA consistently performs as well as or better than the restricted Burer--Monteiro algorithm on ten $N=2{,}000$ random graphs with different degrees. RWS-QAOA is simulated using our best known parameters (see \cref{tab:fixed_parameters}).  
    (C) Estimated runtime for unrestricted Burer--Monteiro (\texttt{MQLib} and parallelized \texttt{MQLib+}) to match the solution quality of RWS-QAOA with $p=6$. For $N \leq 10{,}000$, the target cut fraction is obtained from tensor-network simulation; for $N > 10{,}000$ (faint and dashed lines), the target is held fixed at the $N{=}10{,}000$ value ($\approx 0.9167$). The RWS-QAOA runtime considers the overhead of fault-tolerant execution on a superconducting quantum computer. 
    Shaded areas and error bars denote standard error over graph instances.
    }
    \label{fig:compare_methods_overN}
\end{figure*}
\subsection{Comparison with restricted classical heuristics in simulation}
While RWS-QAOA outperforms provable algorithms on hardware at small scale, the comparison with state-of-the-art heuristics must be performed at much larger scale where the instances cannot be easily solved to optimality. In the near term, the scope of such large-scale studies is limited by the capabilities of classical simulators. Specifically, while local expectations can be computed for shallow circuits on a large number of qubits, sampling from such circuits is not possible with existing techniques (see \cref{fig:classical_simulation}). Consequently, while we are able to predict the \emph{average} cut fraction of the bitstrings produced by QAOA, we cannot obtain the bitstrings and ascertain the impact of local improvements. To make the comparison in this Section fair, we similarly disable the local improvement step in classical heuristics. Furthermore, since the sequential perturbation step in BM requires iteratively rounding the solutions, we also disable this iterative refinement (the closed loop in \cref{fig:algorithmic_pipline}).

In \cref{fig:compare_methods_overN}A, for 3-regular graphs with up to $10{,}000$ nodes, we compare RWS-QAOA at a fixed depth $p=6$ against representative classical heuristics, including simulated bifurcation (SB)~\cite{Goto2019} and the Burer--Monteiro (BM) approach~\cite{burer2002rank}. We use the open-source implementation of SB in \cite{Ageron_Simulated_Bifurcation_SB_2023} and BM in \texttt{MQLib}~\cite{Dunning2018} by setting the number of sequential perturbation rounds $K=1$. 
BM and SB results are reported as the average over $20$ random instances at each size. A detailed comparative study of the classical solvers is included in \cref{sec:compare_classical_solvers}. 

In RWS-QAOA, we optimize the initial state for each instance by maximizing \eqref{eq:rws_objective}, simulate the RWS-QAOA at $p=6$ using the fixed parameters (see \cref{sec:fix_qaoa_para}), and, for each $N$, randomly select $6$ problem instances. To enable comparisons at this scale, we compute the energy of the RWS-QAOA state via exact tensor-network contraction (see \cref{sec:mps_simulation}). 

Unless otherwise specified, RWS-QAOA uses a best-of-$M$ strategy with $M=100$ multistarts (initializations): we select $\boldsymbol{\theta}^\star$ as the maximizer of \cref{eq:rws_objective} over $100$ trials. For fairness, BM and SB also use $100$ multistarts. As shown in \cref{sec:stability_of_warm_start}, performance of all solvers remains stable with $100$ multistarts. Note that in BM and SB, each initialization yields a single solution bitstring, so ``best-of-$M$'' selects the best bitstring directly. In contrast, in RWS-QAOA, ``best-of-$M$'' selects the warm-start parameters that maximize \cref{eq:rws_objective}, not a bitstring. This setup favors classical solvers, as we compare the expected value of the RWS-QAOA state against the best solution found by the classical solvers, but we still refer to it as the \emph{fair comparison}.
Under this fair but classical-favored setup, the fixed-depth RWS-QAOA with $p=6$ outperforms the classical heuristics in \cref{fig:compare_methods_overN}A. We report the standard errors across instances as shaded regions, but they are too small to be visible.  

Additionally, we show the Max-Cut results for higher degree regular graphs in \cref{fig:compare_methods_overN}B. Given the computational cost of optimizing RWS-QAOA parameters, we compare our best-known results of RWS-QAOA ($p=4$ for degree-4 and $p=3$ for degree-5) against BM in a fair comparison setup. RWS-QAOA consistently performs as well as or better than BM, demonstrating that RWS-QAOA generalizes across graph degrees. For all cases, there still exists a performance gap between the most performant algorithm and the RSB upper bound of the cut fraction~\cite{harangi2025rsb} (see \cref{sec:rsb_converge}). The gap indicates room for the algorithms to improve toward global optimality, which could be addressed by future exploration of quantum algorithms.

\begin{figure*}[t]
    \centering
    \includegraphics[width=1\linewidth]{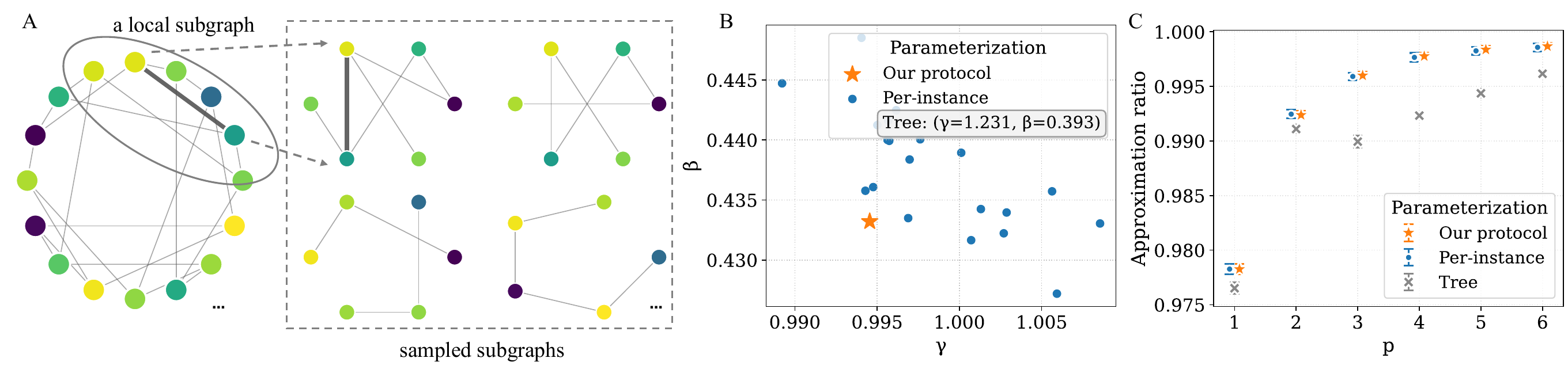}
    \caption{\textbf{QAOA parameter setting in RWS-QAOA.} 
    (A) The protocol for obtaining fixed parameters. We solve the regularized initial states for multiple graphs. Given a target $p$, we extract the subgraphs from the original graphs and sample a subset to conduct the parameter optimization. 
    (B) Visualization of the fixed parameters and instance-wise optimized parameters at $p=1$. The per-instance parameters are numerically optimized for $N=1{,}000$ instances, while the fixed parameters are obtained through the proposed protocol. 
    (C) Performance of RWS-QAOA for five $N=500$ instances under different parameter schedules up to $p=6$. The fixed parameter schedule and the per-instance optimized parameters perform very similarly, and both outperform the tree parameters. Error bars represent standard errors.} %
    \label{fig:qaoa_para_setup}
\end{figure*}

\subsection{Conditions for outperforming unrestricted classical heuristics}

We now remove all restrictions on classical heuristics and identify conditions for RWS-QAOA to outperform them in terms of runtime. We remark that the comparison is unfair to the quantum side since RWS-QAOA is still restricted to no local search and since local search clearly helps RWS-QAOA at small scale (\cref{fig:helios_result}). Therefore the conditions presented in this Section give a worst-case analysis of quantum-classical crossover, and in practice we expect better quantum performance.

We begin by noting that the best available implementation of BM (namely, \texttt{MQLib}) is suboptimal. Specifically, local search from each of $M$ initializations of BM can be run in parallel. This also enables setting the duration of sequential iteration $K$ much higher. 
We implement this parallelized implementation and denote it \texttt{MQLib+}. Both \texttt{MQLib} and \texttt{MQLib+} are executed on an x$86\_64$ system with $96$ vCPUs (AMD EPYC 7R13 Processor, 2.7~GHz) and 186~GB RAM. Note that these changes cannot impact the performance of the ``restricted'' BM considered in the previous Section.

To compute the crossover point, we measure how long classical solvers take to match the average cut fraction achieved by RWS-QAOA with $p=6$ at each size $N$ and compare this runtime to an estimate of RWS-QAOA runtime. For $N \le 10{,}000$, we take the empirical cut fraction values as reported in \cref{fig:compare_methods_overN}A. As the classical simulation of $p=6$ RWS-QAOA for larger graphs is computationally costly and the performance is very stable, for $N > 10{,}000$, we use the average cut fraction obtained at $N = 10{,}000$, $\approx 0.9167$, as the target crossover for \texttt{MQLib} and \texttt{MQLib+} (denoted as faint and dashed lines in \cref{fig:compare_methods_overN}C).  

On the classical side, we report the average time of solving $100$ graph instances at each size $N$. In \texttt{MQLib+}, we parallelize the solver for each instance with $M=100$ initialization trials. The runtime of each graph instance is taken as the minimum crossover time over all $M$ trials. 

On the quantum side, we estimate the runtime of RWS-QAOA with $p=6$ on a surface-code-based fault-tolerant quantum computer with $10^{-3}$ physical error rate and $1\mu$s cycle time~\cite{omanakuttan2025threshold}. We target 90\% fidelity of running the RWS-QAOA circuit within a superconducting surface code architecture. The estimation is supported through the optimal Clifford + $T$ decomposition of the rotation gates and high-quality magic-state factories through magic state cultivation \cite{gidney2024magic}. The detailed quantum runtime estimation can be found in \cref{sec:resource_est}.

\cref{fig:compare_methods_overN}C shows that the runtime, defined as the one-shot execution of the algorithm including all the error correction primitives, which includes magic state cultivation and decoding, and parallel operation of resource factories, is on the order of a few seconds.
RWS-QAOA achieves a runtime crossover at $N=3{,}000$, and the crossover time is below $0.2$ seconds with fewer than $1.3$ million physical qubits, a threshold that is considered achievable for future quantum processors~\cite{mohseni2024build}.

\subsection{Protocol for optimizing RWS-QAOA parameters}\label{sec:fix_qaoa_para}

We now present the protocol used to obtain the fixed parameters in the experiments above. When applying standard QAOA for regular graphs, there exist fixed values for the QAOA parameters that are independent of the specific graph instance and only depend on the graph degree $D$ and QAOA depth $p$, yielding high-quality QAOA performance~\cite{wurtz2021fixed, wurtz2021maxcut}. These parameters are numerically optimized over the worst-case tree subgraphs, thus called ``tree parameters''~\cite{wurtz2021fixed}. 

Since the mixer acts vertex by vertex, the support of an observable $\mathbf{Z}_i \mathbf{Z}_j$ on edge $(i,j)$ grows by at most 2 after each application of the cost function unitary $e^{-i \gamma \mathbf{H}_C}$. Thus, for depth-$p$ QAOA the support of an observable spans at most $2p+1$ vertices (including the edge itself). 
Let $G^p_{(i,j)}$ be the subgraph corresponding to the edge $(i,j)$ and $\mathcal{S}$ be the set of all these subgraphs. Since the cost function is a sum of such edge observables, by regularity the QAOA expectation value can be evaluated on their associated subgraphs
\begin{align}
     F^G_p(\boldsymbol{\gamma},\boldsymbol{\beta}) & = \sum_{(i,j) \in E} \braket{\psi(\boldsymbol{\gamma},\boldsymbol{\beta})\vert \frac{1}{2}(1-\mathbf{Z}_i\mathbf{Z}_j)\vert \psi(\boldsymbol{\gamma},\boldsymbol{\beta})} \\
     & \equiv \sum_{G^p_{(i,j)} \in \mathcal{S}} g_{G^p_{(i,j)}}(\boldsymbol{\gamma},\boldsymbol{\beta}).
\end{align}
Thus, one can optimize $\boldsymbol{\gamma}, \boldsymbol{\beta}$ over the summation of sub-graph energies~\cite{wurtz2021fixed, Basso2022, farhi2025, apte2026}. Many other techniques for efficient setting standard QAOA parameters exist~\cite{hao2024end, ICCAD_qaoapara}.

When it comes to RWS-QAOA, the local subgraphs continue to exhibit similar structures. However, the initial state of each subgraph may vary depending on the specific $\boldsymbol{\theta}$ assigned to each node. Empirically, we observe that while the optimal QAOA parameters for individual subgraphs can differ, the distribution of these parameters remains consistent across subgraphs of varying sizes and instances. This consistency suggests that a set of fixed, high-quality RWS-QAOA parameters can still be identified and applied effectively. This makes RWS-QAOA a non-variational algorithm, while other warm-started methods usually require additional numerical optimization of the QAOA parameters~\cite{yu2025warm, egger2021warm, tate2023warm}.

We introduce the following protocol for obtaining fixed RWS-QAOA parameters $\boldsymbol{\gamma}$ and $\boldsymbol{\beta}$. First, we collect a diverse set of graph instances $\mathcal{G}$ from the same graph family (3-regular graph here). We solve for the warm-started $\boldsymbol{\theta}$ values for each graph in $\mathcal{G}$ using a fixed regularization strength $\lambda$. For a fixed QAOA depth $p$, we then construct the local subgraph $G^{p}_{i,j}$ associated with each edge in a graph $G$, incorporating the corresponding $\boldsymbol{\theta}$ values. This process yields a pool of subgraphs representative of the problem space. From this pool, we sample a subset of subgraphs $\mathcal{S}_{\text{sample}}$ and perform optimization of the RWS-QAOA parameters for the total energies of these sampled subgraphs
\begin{equation}
    \max_{\boldsymbol{\gamma},\boldsymbol{\beta}} \sum_{G^{p}_{i,j} \in \mathcal{S}_{\text{sample}}} g_{G^{p}_{i,j}}(\boldsymbol{\gamma},\boldsymbol{\beta}).
\end{equation}
As the number of sampled subgraphs increases, the optimized parameters tend to concentrate around fixed values. When these concentrated parameters are applied to previously unseen graphs, they closely approximate the parameters that would be obtained by direct optimization on those graphs, resulting in consistently strong QAOA performance. It also indicates that the WS-subgraphs across different instances follow similar distributions, although $\boldsymbol{\theta}$ is clearly instance dependent. The resulting parameters and the RWS-QAOA performance under different parameter setting can be found in \cref{fig:qaoa_para_setup}. A table containing the optimized parameters is presented in \cref{sec:lut_paras}. 

\section{Discussion}

Combining classical and quantum algorithms in a hybrid workflow has long been hypothesized to be the most likely path to improving upon classical state-of-the-art for optimization. Our work shows that constant-depth quantum circuits may be sufficient for such a hybrid algorithm to outperform purely classical ones. 

The RWS-QAOA introduced in this work can be used as a better performing drop-in replacement for QAOA in many hybrid schemes that use QAOA as a subproblem solver~\cite{Sciorilli2025,UshijimaMwesigwa2021,2503.12551,Kotil2025}. We hope that the simplicity of using RWS-QAOA and the availability of pre-optimized parameters (\cref{tab:fixed_parameters}) encourages the community to leverage it to improve the performance of hybrid algorithms.

Our results focus on the performance of classical and quantum algorithms on finite-sized instances and within finite total runtime. The lack of asymptotic guarantees makes our results sensitive to properties of future quantum computers such as logical clock rate. 
Nonetheless, the logical two-qubit gate depth of our circuit is problem-size independent and scales only as $\mathcal{O}(Dp)$ for $D$-regular graphs. Moreover, given a fixed QAOA depth, the physical non-Clifford depth grows approximately linearly with problem size (see \cref{fig:qaoa_resource_estimation}C), leading to a close-to-linear FT runtime. This makes it possible to outperform classical solvers even on quantum computers with slow clock rates, provided that resource state preparation can be parallelized.

Our fault-tolerant runtime estimate assumes that a single-shot measurement of the RWS-QAOA state yields a bitstring whose cut value is close to the expected energy. Even without a formal concentration guarantee, at least half of the probability mass lies at or above the mean energy, so a constant number of independent circuit executions (which can be run in parallel) suffices to obtain a bitstring at or above the mean with high probability, at the cost of a small constant-factor overhead in qubit count. Concentration results, which have been proven for certain QAOA settings~\cite{1812.04170,farhi2014quantumapproximateoptimizationalgorithm, Basso2022, boulebnane2025quantum}, would further strengthen this by ensuring that nearly all measurement outcomes lie close to the mean, reducing the overhead to a single copy. Establishing a formal concentration bound for warm-started QAOA remains an interesting open question, but its absence affects our resource estimates by at most a small constant factor in space.

\section{Methods}

\begin{figure*}[t]
    \centering
    \includegraphics[width=1\linewidth]{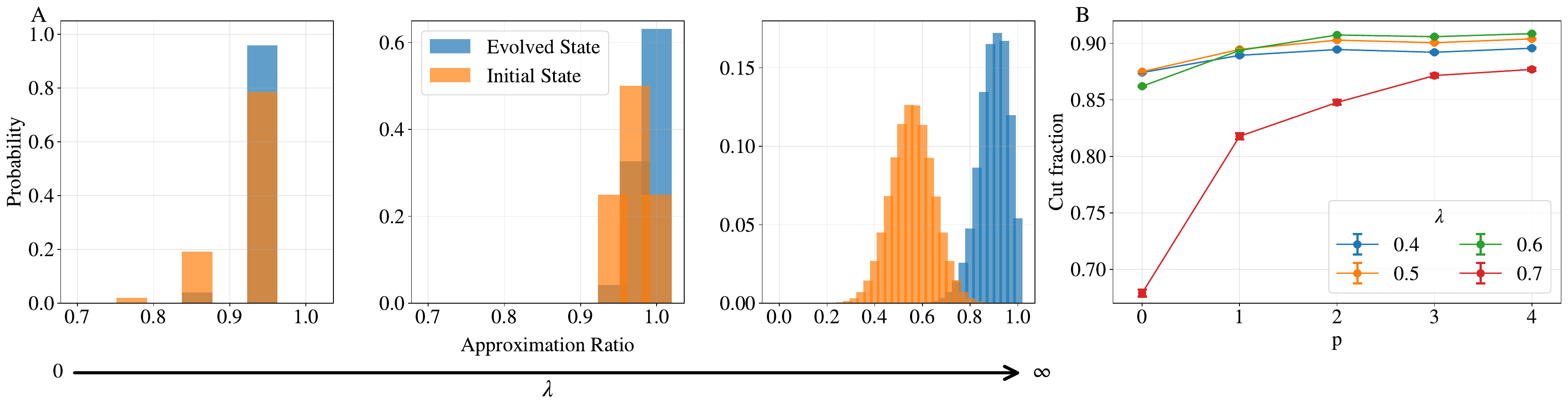}
    \caption{
    \textbf{Regularization strength in RWS-QAOA.} 
    (A) An illustrative example showing the impact of $\lambda$. With a proper choice of $\lambda$, the evolved quantum state concentrates towards the high-quality state quickly.  
    (B) Cut fraction of five $N=2{,}000$ $3$-regular graphs under different $\lambda$ for depths up to $p=4$; $p=0$ represents the initial state only. Here, the RWS-QAOA parameters are fixed to be the tree parameters~\cite{wurtz2021fixed}. Error bars represent standard errors. Empirically, $\lambda = 0.6$ performs well. 
    }
    \label{fig:lambda_setup}
\end{figure*}
\subsection{Solving for the regularized warm start}

The regularized warm starting problem~\cref{eq:rws_objective} can be efficiently solved. 
For the initial product state the measurement outcomes are independent Bernoulli random variables
\begin{equation}
x_i\sim \mathrm{Bernoulli}(p_i),\quad
p_i \equiv \Pr[x_i=1] = \sin^2\!\left(\frac{\theta_i}{2}\right), 
\end{equation}
and therefore $\mathbb{E}[x_i]=p_i$ is the expectation of the Boolean variable $x_i$ and $\mathbb{E}[x_i x_j]=p_i p_j$ for $i\neq j$, is the expectation of the product $x_i x_j$. Hence the expectation of the initial state energy is
\begin{align}
\mathbb{E}[\mathbf{x}^\top \mathbf{Q} \mathbf{x}] 
& = \sum_{i=1}^N \mathbf{Q}_{ii} \mathbb{E}[x_i]
  + \sum_{\substack{i,j=1\\i\neq j}}^N \mathbf{Q}_{ij} \mathbb{E}[x_i x_j]\\ 
& = \sum_{i=1}^N \mathbf{Q}_{ii} p_i
  + \sum_{i \neq j} \mathbf{Q}_{ij} p_i p_j ~. 
\label{eq:expected_energy_qubo_prob}
\end{align}
Therefore, \cref{eq:rws_objective} can be rewritten as 
\begin{align}\label{eq:objective_prob}
\mathcal{L}_{\lambda} (\mathbf{p}) & = \sum_{i=1}^N \mathbf{Q}_{ii} p_i
  + \sum_{i \neq j} \mathbf{Q}_{ij} p_i p_j - 4 \lambda \sum_{i=1}^N p_i (1 - p_i)~. 
\end{align}
The optimization is performed over the vector $\mathbf{p} = (p_1, \ldots, p_N)$, where each variable is constrained to the unit interval
$0 \leq p_i \leq 1$ and thus the feasible set is the unit hypercube. Since the objective function $\mathcal{L}_\lambda(\mathbf{p})$ is continuous and the domain is compact, the Weierstrass extreme value theorem guarantees that a global minimum exists for any choice of $\mathbf{Q}$ and $\lambda$.

Differentiating the above and using the fact that $\mathbf{Q}^{\top} = \mathbf{Q}$, we obtain the first and second derivatives 
\begin{align}
\frac{\partial \mathcal{L}_{\lambda}}{\partial p_i}
& = \mathbf{Q}_{ii} -  4 \lambda + 2 \sum_{i \neq j} \mathbf{Q}_{ij} p_j + 8 \lambda p_i~,\\
 \frac{\partial^2 \mathcal{L}_{\lambda}}{\partial p_i \partial p_j}
& = 2 (1 - \delta_{ij}) \mathbf{Q}_{ij} + 8 \lambda \delta_{ij} \coloneqq \mathbf{H}_{ij}~.
\label{eq:grad_prob_theta_coord}
\end{align}
Importantly, the Hessian matrix $\mathbf{H}_{ij}$ is constant and thus if the minimum eigenvalue exceeds zero then the minimum is unique and attained at the stationary point. Since the derivative can be calculated analytically the objective function can be quickly minimized by any gradient-based optimization methods, such as gradient descent or Adam~\cite{adam}. This produces the initial set of $\mathbf{p}$ or equivalently the vector of rotation angles $\boldsymbol{\theta}$. Since the optimization objective is non-convex in practice we run many parallel optimizations starting with different initial points. Running the optimization for a large but fixed number of steps leads to convergence within a numerical tolerance threshold. As the gradient computation is linear in the number of non-zero terms of QUBO matrix $\textrm{nnz}(\mathbf{Q})$, the overall run-time for the warm start part of RWS-QAOA is also linear in $\textrm{nnz}(\mathbf{Q})$. 

\subsection{Choice of regularization strength}\label{sec:set_regularization_strength}

One of the important hyperparameters in RWS-QAOA is the strength of the regularization $\lambda$ in warm start; an illustrative figure on the effect of $\lambda$ can be found in \cref{fig:lambda_setup}A. When $\lambda$ is too large, RWS-QAOA reduces to standard QAOA as the initial state is enforced to the uniform superposition. When $\lambda$ is too small, the quantum state can get stuck to a non-optimal solution. In contrast, a proper choice of $\lambda$ enables fast convergence toward the optimal solutions. 

Recall from \cref{eq:grad_prob_theta_coord} that the Hessian matrix of the RWS objective is $\mathbf{H}_{ij} = 2 (1 - \delta_{ij}) \mathbf{Q}_{ij} + 8 \lambda \delta_{ij}$, and the presence of $1 - \delta_{ij}$ essentially kills the diagonal part of $\mathbf{Q}$. Specializing to Max-Cut on regular graphs, we note that $\mathbf{Q} = -\mathbf{L}$ and thus the off diagonal part is simply negative of the adjacency matrix $-\mathbf{A}$. Therefore, the Hessian is $2\cdot(4 \lambda \mathbf{I} - \mathbf{A})$. Note that in this case $\mathbf{p} = (1/2, \ldots, 1/2)$ is always a stationary point since the terms with and without $\lambda$ in the gradient as written in \cref{eq:grad_prob_theta_coord} vanish 
\begin{equation}
    8\lambda p_i = 4\lambda~,\quad 2 \sum_{i \neq j} \mathbf{Q}_{ij} p_j = \sum_{i \neq j} \mathbf{Q}_{ij} =  D = -\mathbf{Q}_{ii}~.
\end{equation}
For $D$-regular graphs, the maximum eigenvalue of $\mathbf{A}$ is $D$ with eigenvector $(1, \ldots, 1)$~\cite{Brouwer2012}. Therefore, the minimum eigenvalue of the Hessian is $2\cdot(4 \lambda  - D)$ which is positive if $\lambda > D/4$ with the unique global minimum being $\mathbf{p} = (1/2, \ldots, 1/2)$. As $\lambda$ increases beyond $D/4$ the uniform superposition state $\ket{+}^{\otimes N}$ is chosen at the initial state and thus warm start is no longer as effective. The ideal value of $\lambda$ lies in the interval $(0, D/4)$ and in practice as shown in \cref{fig:lambda_setup}B, we have observed that a fixed choice of $\lambda = 0.6$ produces very good states after QAOA evolution while performance remains stable for $\lambda \in [0.4, 0.6]$.
Throughout this paper we set $\lambda = 0.6$, which empirically yields the best performance for $3$-regular graphs, though it is not provably optimal. 

When the number of non-zero terms $\textrm{nnz}(\mathbf{Q})$ scales as $\mathcal{O}(N^2)$ instead of linearly as is the case of Max-Cut on $D$ regular graphs it is useful to normalize $\mathbb{E}[\mathbf{x}^\top \mathbf{Q} \mathbf{x}]$ by a factor of $N$ so that it is comparable in contribution to $\sum_{i=1}^N p_i (1 - p_i) =\mathcal{O}(N)$. Without this scaling, the regularization term becomes ineffective as $N \to \infty$.

\subsection{Resource estimation on FT implementation}\label{sec:resource_est}
In this section, we outline the details of the resource estimates of the fault tolerant implementation of  RWS-QAOA using planar rotated surface code.
First, we count the total number of single-qubit rotation gates required for the RWS-QAOA circuits in \cref{fig:algorithmic_pipline}B.
For MaxCut of a $D$-regular graph, the total number of rotations for QAOA depth $p$ is $p\times\left(\frac{D}{2}N + N\right) + N$, where $N$ denotes the number of logical qubits. 
Specifically, $\frac{D}{2}N$ rotation gates are required for the phase operator and $N$ rotation gates per QAOA depth are required for the mixer, and an additional $N$ gates are required for the initialization of the warm-started state. 
This initialization overhead reflects the complexity of preparing a non-trivial quantum state informed by classical optimization.

Given the total number of gates and a target circuit fidelity ($0.9$ in our setting), we first determine the required accuracy for each rotation gate $R_z$ in the algorithm. 
For a specified rotation accuracy and a chosen decomposition that maps arbitrary single-qubit rotations to Clifford + $T$ circuits~\cite{ross2016optimal, Bocharov_efficient_synthesis_2015}, we select the optimal decomposition error $\delta_{\mathrm{opt}}$ and the minimum $T$ gate infidelity $\epsilon_T$ required to achieve the target, balancing circuit accuracy against $T$-gate overhead as described in \cite{omanakuttan2025threshold}.

Once we determine the minimum $T$-gate infidelity, we find the target surface-code distance that achieves this error rate as:
\begin{equation}
    d_{\mathrm{tar}} = \left\lceil 2\,\frac{\log(\epsilon_T)}{\log\!\left(p_{\mathrm{ph}} / p_{\mathrm{th}}\right)} \right\rceil.
    \label{eq:distance_surface_code}
\end{equation}
Here, $p_{\mathrm{ph}}$ is the maximum error rate of the physical components and $p_{\mathrm{th}}$ is the surface-code threshold. The value of $\epsilon_T$ also determines the magic-state preparation scheme for the algorithm.

For example, for the $p=6$ RWS-QAOA circuits for $100{,}000$-node 3-regular graphs, as detailed in \cref{fig:qaoa_resource_estimation}, we require magic states with infidelity $< 10^{-9}$.
From \cref{eq:distance_surface_code}, to achieve an error rate of $10^{-9}$, we need a surface code of distance $d=17$ for a physical error rate of $10^{-3}$.
At these error rates, magic-state cultivation developed by Gidney et al.~\cite{gidney2024magic} can be used and requires approximately $2d^2 = 578$ physical qubits in a time of $173$ code cycles (where code cycle $\tau$ is the time for a single syndrome extraction) from Fig.~1 in Ref.~\cite{gidney2024magic}. Similarly, we can obtain the required magic state infidelity, code distance, and code cycles for different number of logical qubits.

This yields the total number of physical qubits required for the fault-tolerant, RWS-QAOA:
\begin{equation}
    N_{\text{phys}} = 2d^2 (N + N_{\mathrm{fac}}),
\end{equation}
where $N$ is the number of logical qubits, $d$ is the surface-code distance, and $N_{\mathrm{fac}}$ is the number of resource factories used to generate the $T$-gates that balances the parallelization and physical qubit overhead.
Here, we assume a constant $N_{\mathrm{fac}} = 1200$ for different sized problems, which consumes $\approx 1.2\%$ percent of the total physical qubits at $N=100{,}000$. 

The total quantum runtime $t_q$ of the RWS-QAOA is estimated as 
\begin{equation}\label{eq:quantum_runtime_eq}
    t_q = \frac{\#~\text{code cycle} }{d}\times\frac{N_T}{N_{\mathrm{fac}}}\, t_{\mathrm{LC}},
\end{equation}
where $N_T$ is the total number of magic states required for the algorithm. $t_{\mathrm{LC}} = d \times \tau$ is logical cycle time for a surface code with distance $d$ with $\tau = 1 \, \mu s$ being the time for 1 round of syndrome extraction for superconducting systems \cite{focus_beyond_quadratic}. 
One limitation here is that we do not account for routing time; this overhead can be reduced by deploying additional factories \cite{PRXQuantum.3.020342}.

\subsection{Classical simulation of RWS-QAOA}\label{sec:mps_simulation}

\subsubsection{Exact tensor network contraction}
To compute energy expectation values exactly, we contract tensor networks within the causal lightcones of the corresponding local observables. Given a QAOA circuit and a Hamiltonian decomposed into local terms (e.g., $\mathbf{Z}_i \mathbf{Z}_j$ for each edge $(i,j)$), we extract, for each observable, the subset of gates that can influence it by tracing backward through the circuit to form its lightcone. On graphs of maximum degree $D$ and depth $p$, the number of leaves in the lightcone scales as $\mathcal{O}((D-1)^p)$. 
Each local expectation value can be contracted independently with memory cost $2^{\mathcal{O}(t_W)}$, where $t_W$ is the treewidth of the subgraph within the causal lightcone. This cost is independent of the total system size $N$ beyond the lightcone radius. The time complexity of calculating the total energy is linear in the number of local observables, $\mathcal{O}(DN)$ in this case and the time complexity of computing a single local observable is $\mathcal{O}(p(D-1)^p 2^{\mathcal{O}(t_W)})$~\cite{doi:10.1137/050644756}, where $\mathcal{O}(p(D-1)^p)$ is the number of edges in the contraction tree.
In the worst case $t_W = \mathcal{O}\qty((D-1)^p)$, and in the more often tree-like best case, especially for a problem that is high-girth, $t_W = \mathcal{O}\qty(2p)$~\cite{farhi2025}. 
This approach yields exact results for each term without approximations, and parallelization across terms is straightforward. We implement this for graph degree up to $D=5$, QAOA depth up to $p=6$, and system size up to $N=10{,}000$ using \texttt{quimb}~\cite{gray2018quimb}. We leverage the Aurora supercomputer~\cite{allen2025} at the Argonne Leadership Computing Facility to evaluate many terms in parallel using up to $7{,}500$ Intel Data Center GPU Max Series devices (each with two GPU tiles) at a time~\cite{auroraoverview}. %
Recent complexity-theoretic results clarify the limits of this approach. For general graphs and any fixed QAOA depth, not only is exactly evaluating the expectation value of QAOA at prescribed angles NP-hard, but so is approximating this expectation within additive error $2^{-\mathcal{O}(N)}$ in the number of vertices~\cite{wang2025unifiedcomplexityalgorithmaccountconstantround}.

\begin{figure}[t]
    \centering
    \includegraphics[width=\linewidth]{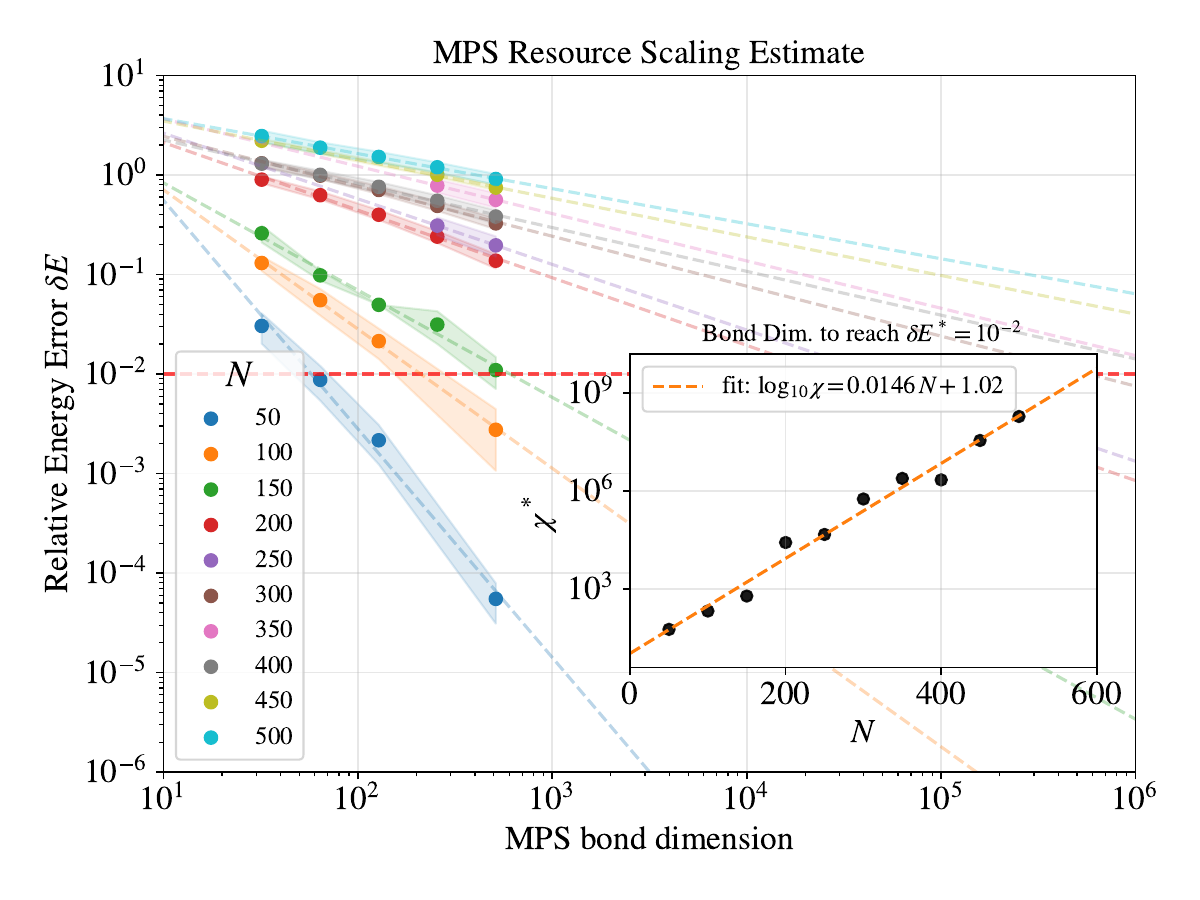}
    \caption{
    \textbf{Classical simulation cost of MPS scales exponentially with system size.} The approximate relative energy error $\delta E$ of the MPS simulation as a function of the MPS bond dimension for various system sizes $N$, averaged over 3 graph instances. For each system size, we fit a power law scaling, and estimate the bond dimension $\chi^*$ needed to achieve an error of $\delta E^* = 10^{-2}$ (red line). The inset shows the bond dimension $\chi^*$ vs.\ system size $N$; we fit an exponential curve, giving a goodness of fit $R^2 \simeq 0.98$. 
    }
    \label{fig:classical_simulation}
\end{figure}

\subsubsection{Approximate MPS sampling with fixed bond dimension}
For sampling from the final QAOA state under realistic resource constraints, we adopt a matrix product state (MPS) approach with fixed bond dimension. We first compile the QAOA circuit (a sequence of one- and two-qubit unitaries) into a sequence of matrix product operators (MPOs) and apply them sequentially to the MPS. After each MPO application, we perform an SVD-based truncation to cap the bond dimension at $\chi$, discarding singular values $\{\lambda_i\}_{i>\chi}$ at each cut. Let $\ket{\psi}$ denote the exact QAOA state and $\ket{\bar{\psi}}$ the truncated MPS. Define the truncation vector $\ket{\delta} \equiv \ket{\psi} - \ket{\bar{\psi}}$ and the truncation error
$$
\epsilon \equiv \sqrt{\left|\braket{\delta|\delta}\right|} \;\leq\; \sqrt{\sum_{\text{all cuts}}\sum_{i>\chi}\lambda_i^2},
$$
where the right-hand side is the standard cumulative discarded weight across all truncations~\cite{Verstraete_2006}. The resulting relative energy error obeys
\begin{align}
    \delta E\;&=\; \left|\bra{\psi}\mathbf{H}\ket{{\psi}} - \bra{\bar{\psi}}\mathbf{H}\ket{\bar{\psi}} \right|/\|\mathbf{H}\| \\
    &= \left| \bra{\delta}\mathbf{H}\ket{\bar{\psi}} 
    + \bra{\bar{\psi}}\mathbf{H}\ket{\delta}
    + \bra{\delta}\mathbf{H}\ket{{\delta}} \right| /\|\mathbf{H}\|\\
    &\leq 2\epsilon\,\|\bar{\psi}\| +\epsilon^2,
\end{align}
where $\|\mathbf{H}\|$ is the operator norm of the Hamiltonian $\mathbf{H}$ and $\|\bar{\psi}\|$ is the norm of the unnormalized MPS after truncation. 

We perform MPS simulations using the ITensor library~\cite{ITensor} with NVIDIA A100 GPUs for system sizes up to $N=500$ and QAOA depths up to $p=6$. We track the cumulative discarded weight at each truncation step, obtaining $\epsilon$. The tradeoff between cost (bond dimension) and accuracy (relative energy error) of this MPS sampling approach is summarized in \cref{fig:classical_simulation}, where the approximate relative energy error $\delta E$ decreases with increasing bond dimension $\chi$. To see what bond dimension $\chi^*$ is needed to achieve a desired accuracy $\delta E^*$, we fit a line to $\log \delta E$ vs. $\log \chi$ for each system size $N$ and extrapolate to the point $(\chi^*, \delta E^*)$. The inset of \cref{fig:classical_simulation} shows $\chi^*$ vs $N$ for $\delta E^*=10^{-2}$. Fitting a line to $\log\chi^*$ vs $N$ gives a goodness of fit of $R^2 \simeq 0.98$, showing that the computational cost scales exponentially with system size for a fixed accuracy. 
Warm starts can shape the initial MPS spectrum and can reduce the bond dimension needed at shallow depths, but as $N$ grows and graph structure becomes more complex, the entanglement generated across layers typically forces $\chi$ to scale rapidly, producing the observed exponential trend in the inset of \cref{fig:classical_simulation}. This aligns with broader evidence that sampling from certain quantum circuit families is computationally hard in general~\cite{farhi2019quantumsupremacyquantumapproximate}, whereas localized expectation values can remain tractable when effective treewidth is controlled~\cite{wang2025unifiedcomplexityalgorithmaccountconstantround}.

\subsection{Classical algorithms for Max-Cut}\label{sec:classical_solvers}

In this subsection, we will review classical algorithms for solving the Max-Cut problem. The methods we discuss are those providing provably good approximation ratios or those that have shown strong empirical performance on Max-Cut benchmarks~\cite{Dunning2018}. It will be useful to recast \cref{e:max_cut} using variables $\mathbf{y} \in \{\pm 1\}^{N}$ via the transformation $y_i = 2x_i - 1 \in \{\pm 1\}$. This yields
\begin{equation} \label{e:max_cut_ising}
    \begin{aligned}
\mathsf{OPT}(G) &= \frac{1}{2}\max_{\mathbf{y}  \in \{\pm 1\}^N} \sum_{(i,j) \in E} (1 - y_i y_j) \\
    &= \frac{1}{2} \max_{\mathbf{y}  \in \{\pm 1\}^N} \mathbf{y}^{\top} \mathbf{L} \mathbf{y} \\
    &= \frac{1}{2} \max_{\mathbf{y}  \in \{\pm 1\}^N} \operatorname{Tr}( \mathbf{L} \mathbf{y}\mathbf{y}^{\top}) ,
\end{aligned}
\end{equation}
where $\mathbf{L}$ is the graph Laplacian.

\subsubsection{Goemans--Williamson algorithm}
\Cref{e:max_cut_ising} is a challenging non-convex optimization problem. The key idea behind the Goemans--Williamson algorithm~\cite{goemans1995} is to first replace the discrete variables $y_i \in \{\pm 1 \}$ with continuous unit vectors $\mathbf{v}_i \in \mathbb{S}^N$ with $\|\mathbf{v}_i\|_2 = 1$ and $y_{i} y_j =\mathbf{v}_i \cdot \mathbf{v}_j$, where $\cdot$ denotes the Euclidean inner product on $\mathbb{R}^N$. Defining the Gram matrix $\mathbf{Y} \in \mathbb{R}^{N \times N}$ with entries $\mathbf{Y}_{ij} = \mathbf{v}_i \cdot \mathbf{v}_j$ and letting $\mathcal{S}^N_{+} := \{ M \in \mathbb{R}^{N \times N} : M \succeq 0\}$ denote the cone of $N \times N$ positive semidefinite matrices, \eqref{e:max_cut_ising} is equivalent to
\begin{equation}\label{e:max_cut_SDP_nonconvex}
\mathsf{OPT}(G) = \max_{\left\{ \mathbf{Y} \in \mathcal{S}^N_{+} : \operatorname{rank}(\mathbf{Y}) =1, ~ \operatorname{diag}(\mathbf{Y}) = \mathbf{e} \right\} } \frac{1}{2} \operatorname{Tr}(\mathbf{L} \mathbf{Y}).
\end{equation}
The rank constraint ensures that any optimal solution to \eqref{e:max_cut_SDP_nonconvex} satisfies $\mathbf{Y}_{\star} = \mathbf{y}_{\star} \mathbf{y}_{\star}^{\top}$ where $\mathbf{y}_{\star} \in \{ \pm 1\}^N$ is an optimal solution of \eqref{e:max_cut_ising}, but is non-convex. Upon dropping it, we obtain a \textit{convex} relaxation of \eqref{e:max_cut_ising}
\begin{equation}\label{e:max_cut_SDP}
\textsf{SDP}(G) := \max_{\left\{ \mathbf{Y} \in \mathcal{S}^N_{+} : \operatorname{diag}(\mathbf{Y}) = \mathbf{e} \right\} } \frac{1}{2} \operatorname{Tr}(\mathbf{L} \mathbf{Y}).
\end{equation}
Since we dropped the rank constraint, one generally has $\textsf{SDP}(G) \geq \mathsf{OPT}(G)$ and $\mathbf{Y}_{\star} \neq \mathbf{y}_{\star} \mathbf{y}_{\star}^{\top}$. However, interior point algorithms can solve \eqref{e:max_cut_SDP} in polynomial time. 

After obtaining an optimal solution $\mathbf{Y}_{\star}$ to \eqref{e:max_cut_SDP}, we compute the factorization $\mathbf{Y}_{\star} = \mathbf{V} \mathbf{V}^{\top}$ where the $i$-th row of $\mathbf{V}$ is $\mathbf{v}_i^{\top}$. Then, sample a random vector $\mathbf{r} \sim \mathcal{N}(0, \mathbf{I}_N)$ and assign vertex $i$ to a side via $y_i = \operatorname{sign}(\mathbf{v}_i \cdot \mathbf{r})$. For any edge $(i,j)$, the probability of being cut is given by 
\begin{equation}
    \Pr[y_i \neq y_j] = \frac{\arccos\!\big(\mathbf{v}_i \cdot \mathbf{v}_j\big)}{\pi},
\end{equation}
which also equals the expected cut value for this edge. On the other hand, the contribution of this edge to SDP objective value is simply $(1 - \mathbf{v}_i \cdot \mathbf{v}_j)/2$. The ratio of these two quantities is lower bounded by 
\begin{align}
\frac{\arccos(t)}{\pi} &\;\ge\; \alpha \cdot \frac{1 - t}{2}
&\forall ~ t \in [-1,1],
\end{align}
where 
$$\alpha \;=\; \min_{\theta \in [0,\pi]} \frac{2\theta/\pi}{1 - \cos\theta} \;\approx\; 0.87856.$$
By the linearity of the Max-Cut objective over the edges, Goemans--Williamson achieves an approximation ratio of at least $\alpha$:
\begin{equation}
\mathrm{SDP} \geq \mathrm{OPT} \geq \mathrm{GW}
\ge \alpha \cdot \mathrm{OPT}~.  
\end{equation}
Although there are instances which saturate the lower bound, i.e., $\mathrm{GW}
= \alpha \cdot \mathrm{OPT}$~\cite{karloff1996good, alon2001constructing}, performance on regular graphs exceeds the lower bound in practice, as shown in \cref{fig:compare_methods_overN}. 

Using \emph{interior-point methods} (IPMs), the SDP \eqref{e:max_cut_SDP} can be solved to precision $\varepsilon \in (0,1)$ in time $\mathcal{O}\!\left(N^{3.5}\log(1/\varepsilon)\right)$. We benchmark the runtime of the method in \cref{fig:compare_runtime}, and the observed scaling makes it quite expensive to run for graphs with $N \geq 2000$. 

\subsubsection{Halperin-Livnat-Zwick algorithm}

For unweighted graphs with maximum degree of 3, it is possible to tighten the relaxation \eqref{e:max_cut_SDP} by imposing additional constraints that account for the graph structure. This leads to the following program:  
\begin{align} 
& \max \sum_{(i,j) \in E} \frac{1 - \mathbf{v}_i \cdot \mathbf{v}_j}{2}~,\nonumber \\
& \text{subject to} ~ \|\mathbf{v}_i\|=1 \in \mathbb{R}^{N}~, \label{e:HLZ} \\ 
& \mathbf{v}_i \cdot \mathbf{v}_j + \mathbf{v}_i \cdot \mathbf{v}_k + \mathbf{v}_j \cdot \mathbf{v}_k \ge -1,~\forall\, 1 \le i,j,k \le N, \nonumber  \\
& \mathbf{v}_i \cdot \mathbf{v}_j - \mathbf{v}_i \cdot \mathbf{v}_k - \mathbf{v}_j \cdot \mathbf{v}_k \ge -1,~ \forall\, 1 \le i,j,k \le N, \nonumber \\
& \mathbf{v}_i \cdot \mathbf{v}_j + \mathbf{v}_i \cdot \mathbf{v}_k + \mathbf{v}_j \cdot \mathbf{v}_k = -1,~\forall\, (i,j),(j,k) \in E.  \nonumber 
\end{align} 
Solving this program and applying the previously described rounding procedure yields a candidate cut. 

The subsequent local improvement phase applies three types of moves iteratively until none increases the cut value.  Let $V_3$ denote vertices with all three neighbors on the same side of the cut, and $V_2$ denote vertices with exactly two neighbors on the same side.
Move (a) flips a vertex in $V_3$ with the fewest $V_3$-neighbors, guaranteed to increase the cut by at least one edge. 
Move (b) identifies a path $u \rightarrow v_1 \rightarrow \ldots \rightarrow v_k \rightarrow w$ of unsatisfied edges where internal vertices $v_i \in V_2$ and endpoints $u,w \notin V_2$, then flips alternate internal vertices ($v_i$ for odd $i$), satisfying all path edges.
Move (c) handles cycles $v_1 \rightarrow v_2 \rightarrow \ldots \rightarrow v_k \rightarrow v_1$ of unsatisfied edges with all $v_i \in V_2$, flipping alternate vertices ($v_i$ for even $i$) to satisfy the entire cycle. 

After the local improvement step, the resulting Halperin-Livnat-Zwick (HLZ) algorithm achieves a provable approximation ratio of at least $0.9326$~\cite{Halperin2004}. However, this approach does not generalize to other graph degrees, and just as in the case of Goemans-Williamson, the SDP~\eqref{e:HLZ} can be solved in time $\mathcal{O}\!\left(N^{3.5}\log(1/\varepsilon)\right)$, using IPMs. %

\subsubsection{Burer--Monteiro algorithm}
Second-order methods for solving~\eqref{e:max_cut_SDP} (like IPMs) incur $\mathcal{O}(N^3)$ cost per iteration, as they require solving large linear systems. The Burer--Monteiro method~\cite{burer2002rank} overcomes scalability limitations of standard second-order methods for SDPs by employing the low rank change of variables $\mathbf{Y} = \mathbf{U} \mathbf{U}^\top$, where $\mathbf{U} \in \mathbb{R}^{k \times N}$ and $k \ll N$. This reduces the number of variables from $\mathcal{O}(N^2)$ to $\mathcal{O}(k N)$, while ensuring that the solution is positive semidefinite by construction. Upon lifting the constraints to the objective through the use of Lagrange multipliers, the resulting (non-convex) optimization problem over $\mathbb{R}^{k \times N}$ can be solved using first-order methods, making this approach favorable for large graphs. 

For the rank-2 case ($k=2$), each vector $\mathbf{v}_i$ is constrained to lie on the unit circle in $\mathbb{R}^2$. That is, 
$\mathbf{v}_i = (\cos \theta_i, \sin \theta_i)$ for some angle $\theta_i \in [0, 2\pi)$. The inner product between two such vectors is simply $\mathbf{v}_i^\top \mathbf{v}_j = \cos(\theta_i - \theta_j)$. Substituting this into the Max-Cut objective, the rank-2 Burer--Monteiro relaxation becomes:
\begin{equation}\label{e:bm_rank2}
\max_{\{\theta_1, \dots, \theta_N \}} \frac{1}{2} \sum_{(i,j) \in E} \left(1 - \cos(\theta_i - \theta_j)\right)~.
\end{equation}
\Cref{e:bm_rank2} is non-convex in the angles $\{\theta_i\}_{i \in [N]}$, but the gradient of the objective function is easy to compute and can be evaluated using $\mathcal{O}(|E|)$ operations. In practice, the rank-2 Burer--Monteiro method typically converges in a fixed number of steps (often a few thousand), meaning that the practical runtime also scales as $\mathcal{O}(|E|)$. This becomes $\mathcal{O}(N)$ for regular graphs, and we numerically verify this in \cref{fig:compare_runtime}. Since the optimization landscape is non-convex, a randomized multi-start strategy is generally employed. 

After solving for the angles $\theta_1, \dots, \theta_N$, each vertex is represented by a point on the unit circle. To obtain a cut, one can ``sweep" a line through the origin at angle $\phi$ and assign each vertex $i$ to one side of the cut according to $\text{sign}(\cos(\theta_i - \phi))$. As $\phi$ varies from $0$ to $2\pi$, the cut value changes only when the line passes through a vector $\mathbf{v}_i$. By checking all $N$ possible transitions, one can efficiently determine the best cut induced by the solution. The rounding flowchart is summarized in \cref{sec:more_alg_charts}.

\subsubsection{Simulated bifurcation}

Simulated Bifurcation is a physics-inspired algorithm for solving QUBO problems, such as Max-Cut~\cite{Goto2019}. The method models the optimization process as the evolution of a dynamical system, where bifurcations correspond to transitions between candidate solutions. This approach is particularly well-suited for problems with quadratic objectives, but does not easily generalize to higher-order or non-quadratic formulations.

A key advantage of Simulated Bifurcation is its computational efficiency, which depends on the sparsity of the underlying problem. For sparse graphs (e.g., regular graphs) where each node has a bounded degree, the QUBO formulation contains only $\mathcal{O}(N)$ nonzero terms. This results in a runtime that scales at most linearly with the number of vertices as verified in our runtime benchmarks \cref{fig:compare_runtime}. In contrast, the runtime becomes quadratic in $N$ for dense graphs with $\mathcal{O}(N^2)$ edges. Although it is very difficult to prove bounds on the performance of simulated bifurcation, it was one of the best-performing classical algorithms in the comparison presented in \cref{fig:compare_methods_overN}. 

Our implementation of Simulated Bifurcation is based on~\cite{Ageron_Simulated_Bifurcation_SB_2023}, which supports GPU-acceleration of matrix-vector products required to evolve the state of the candidate solutions at any given step.

\subsubsection{Local search post-processing}\label{sec:local_search}
Local search is a standard post-processing technique for improving solutions to Max-Cut. Starting from any candidate solution, such as those produced by SDP rounding, simulated bifurcation, or QAOA, the local search algorithm iteratively flips bits whenever such moves increase the cut value. 

The post-processing consists of two greedy improvement passes applied to an initial binary assignment $\textbf{x}$. First, we visit each vertex $v_i$, and flip $v_i$ only if this strictly increases the cut. Then, we visit each edge $(i, j)$ and flip the pair only if this strictly increases the cut. These passes can be repeated until no improving single or double flip remains, yielding a labeling that is locally optimal with respect to one and two-bit moves.

The run-time for single flips is $\mathcal{O}(N)$ while that for double flips is $\mathcal{O}(|E|)$, which simplifies to $\mathcal{O}(N)$ for a single pass on regular graphs. The gains from post-processing are quite substantial for solutions obtained from the Burer--Monteiro algorithm. Note that since a candidate solution is needed for post-processing, we cannot apply this procedure to large-scale simulations in which we only compute the expectation value using tensor contraction without sampling from the output of the QAOA circuit.

\section*{Acknowledgments}
We thank Enrico Fontana for the initial exploration of classical simulation. 
We thank Soorya Rethinasamy for reviewing the paper draft. 
We thank Rob Otter for the executive support of the work and invaluable feedback on this project.
We thank the colleagues at the Global Technology Applied Research center of JPMorganChase
for support.
An award of computer time was provided by the INCITE program. This research used resources of the Argonne Leadership Computing Facility, which is a DOE Office of Science User Facility supported under Contract DE-AC02-06CH11357.

\section*{Author Contributions}
Z.H. and R.S. conceptualized the project. Z.H. and A.A. proposed the regularized warm start method. Z.H. conducted the hardware experiments. A.K. and Z.H. implemented and performed the large-scale tensor network simulations. B.A. and A.A. implemented the SDP-based algorithms. A.B., A.A., Z.H., and R.S. benchmarked classical solvers and analyzed the data. S.O. performed the fault-tolerant resource estimation. All authors contributed to the writing of the manuscript and shaping of the project.

\section*{Disclaimer}
This paper was prepared for informational purposes with contributions from the Global Technology Applied Research center of JPMorgan Chase \& Co. This paper is not a product of the Research Department of JPMorgan Chase \& Co. or its affiliates. Neither JPMorgan Chase \& Co. nor any of its affiliates makes any explicit or implied representation or warranty and none of them accept any liability in connection with this paper, including, without limitation, with respect to the completeness, accuracy, or reliability of the information contained herein and the potential legal, compliance, tax, or accounting effects thereof. This document is not intended as investment research or investment advice, or as a recommendation, offer, or solicitation for the purchase or sale of any security, financial instrument, financial product or service, or to be used in any way for evaluating the merits of participating in any transaction.

\putbib[main.bib]

\end{bibunit}
\clearpage
\newpage

\begin{bibunit}[apsrev4-2-author-truncate]

\onecolumngrid

\begin{center}
    \textbf{\large Supplementary Material for: \\ Regularized Warm-Started Quantum Approximate Optimization and Conditions for Surpassing Classical Solvers on the Max-Cut Problem}
\end{center}

\section{Fixed parameters for RWS-QAOA}\label{sec:lut_paras}
The proposed protocol for optimizing RWS-QAOA parameters is summarized in Algorithm~\ref{alg:fixed_para}. 

\begin{algorithm}[H]
\caption{Protocol for Obtaining Fixed RWS-QAOA Parameters}\label{alg:fixed_para}
\KwIn{Set of graph instances $\mathcal{G}$, QAOA depth $p$, number of subgraphs to sample $K$, regularization parameter $\lambda$}
\KwOut{Fixed RWS-QAOA parameters $(\boldsymbol{\gamma}^*, \boldsymbol{\beta}^*)$}
\textbf{Warm Start:} For each graph $G \in \mathcal{G}$, with a fixed $\lambda$, compute warm-started $\boldsymbol{\theta}$ values for all nodes\;
Initialize an empty pool of subgraphs $\mathcal{S}$\;
For each graph, construct local subgraphs of depth $p$ around its edges and add them to $\mathcal{S}$\;
Randomly sample $K$ subgraphs from pool $\mathcal{S}$ to form set $\mathcal{S}_{\text{sample}}$\;
\textbf{Optimization:} Optimize QAOA parameters $(\boldsymbol{\gamma}, \boldsymbol{\beta})$ for the total energy of the sampled subgraphs $\mathcal{S}_{\text{sample}}$\;
As $K$ increases, obtain concentrated parameters $(\boldsymbol{\gamma}^*, \boldsymbol{\beta}^*)$\;
\end{algorithm}

We report the optimized RWS-QAOA parameters for $D$-regular graphs with $D=3,4,5$ in \cref{tab:fixed_parameters}. The hyperparameter $\lambda$ is selected by sweeping $\lambda \in [0, \frac{D}{4}]$ with a step size of 0.1, i.e., $0.1, 0.2, \ldots$, and choosing the value that yields the best performance at a relatively high depth $p$. Specifically, if we target small-$p$ performance, a smaller $\lambda$ will be preferred as it could lead to higher cut values in the initial state ($p=0$). The hyperparameter $\lambda$ and parameters $\boldsymbol{\gamma}$ and $\boldsymbol{\beta}$ reported here are not guaranteed to be optimal, but empirically perform fairly well. 
\begin{table}[h]
\centering
\small
\setlength{\tabcolsep}{4pt}
\begin{tabular}{@{}c c c l l@{}}
\toprule
Graph & $\lambda$ & $p$ & $\boldsymbol{\gamma}$ & $\boldsymbol{\beta}$ \\
\midrule
\multirow{6}{*}{degree-3} & \multirow{6}{*}{0.6} & 1 & [0.9945] & [0.4332] \\
\cmidrule(lr){3-5}
                          &                       & 2 & [0.9996, 2.1510] & [0.4638, 0.2564] \\
\cmidrule(lr){3-5}
                          &                       & 3 & [0.5551, 1.4950, 2.3055] & [0.6161, 0.3287, 0.1795] \\
\cmidrule(lr){3-5}
                          &                       & 4 & [0.4141, 0.9741, 1.9294, 2.2130] & [0.8026, 0.4948, 0.2468, 0.1517] \\
\cmidrule(lr){3-5}
                          &                       & 5 & [0.3175, 0.7904, 1.3684, 2.0624, 2.1531] & [0.7858, 0.6179, 0.3380, 0.1995, 0.1207] \\
\cmidrule(lr){3-5}
                          &                       & 6 & [0.2793, 0.6845, 1.1566, 1.5702, 2.1050, 2.2143] & [0.7811, 0.6177, 0.4271, 0.2699, 0.1650, 0.1007] \\
\midrule
\multirow{4}{*}{degree-4} & \multirow{4}{*}{0.7} & 1 & [0.9267] & [0.4377] \\
\cmidrule(lr){3-5}
                          &                       & 2 & [0.5816, 1.3216] & [0.5426, 0.2652] \\
\cmidrule(lr){3-5}
                          &                       & 3 & [0.4047, 0.9305, 1.5222] & [0.6798, 0.4075, 0.1802] \\
\cmidrule(lr){3-5}
                          &                       & 4 & [0.3154, 0.7349, 1.1747, 1.6615] & [0.7634, 0.5736, 0.3214, 0.1589] \\
\midrule
\multirow{3}{*}{degree-5} & \multirow{3}{*}{0.7} & 1 & [0.9899] & [0.4373] \\
\cmidrule(lr){3-5}
                          &                       & 2 & [0.9870, 2.0958] & [0.4860, 0.2606] \\
\cmidrule(lr){3-5}
                          &                       & 3 & [0.5976, 1.4999, 2.1220] & [0.5470, 0.3253, 0.1977] \\
\bottomrule
\end{tabular}
\caption{Optimized parameters for regular graphs with different degrees: $\boldsymbol{\gamma}$ and $\boldsymbol{\beta}$ for varying $p$.}
\label{tab:fixed_parameters}
\end{table}

\section{More results of RWS-QAOA}

\subsection{Error analysis of hardware result}\label{sec:hardware_error}

We now study the expected fidelity of the RWS-QAOA circuit run on the \texttt{Helios} trapped ion quantum computer without error detection or correction. Although our fidelity estimate will be tailored to this device based on the gate errors as reported in~\cite{ransford2025}, the analysis will hold for all devices with all-to-all connectivity. Note that for hardware with nearest-neighbor connectivity such as superconducting qubits~\cite{devoret2004superconductingqubitsshortreview}, the fidelity can be substantially different since the unitary for cost Hamiltonian $e^{-i\gamma_t \mathbf{H}_C}$ has to be decomposed into nearest neighbor gates. 

A phenomenological error model for fidelity was developed in~\cite{ransford2025}, which takes into account the number of two-qubit gates and state preparation and measurement (SPAM) errors, while neglecting single-qubit gate errors since the single-qubit gate error rate is much lower than either two-qubit gate error or SPAM error. The estimate for fidelity for a circuit with $|\mathcal{Q}_2|$ two-qubit gates and $N$ qubits is:
\begin{equation}\label{e:fidelity_est}
    F(N, \mathcal{Q}_2) \simeq (1-p_{\text{spam}})^N \bigg(1 - \frac{5}{4} \varepsilon_{\mathcal{Q}_2}\bigg)^{|\mathcal{Q}_2|}~,
\end{equation}
where $p_{\text{spam}}$ is the effective SPAM error and $\varepsilon_{\mathcal{Q}_2}$ is the effective average two-qubit error rate. The estimated values for \texttt{Helios} are $p_{\text{spam}} = 5.3 \times 10^{-4}$ and $\varepsilon_{\mathcal{Q}_2} = 2.0 \times 10^{-3}$. For the RWS-QAOA circuit, the mixer unitary including the basis change has only single-qubit gates. Thus, it suffices to count the number of two-qubit gates required to implement the unitaries $e^{-i\gamma_t \mathbf{H}_C}$ across the $p$ layers. The total number of such gates in each layer simply equals the number of edges, as each edge contributes a term coupling the qubits corresponding to vertices. Therefore, for a graph $G$ with number of edges $|E|$ the fidelity is
\begin{equation}\label{e:fidelity_graph}
    F(N, |E|, p) \simeq (1-p_{\text{spam}})^N \bigg(1 - \frac{5}{4} \varepsilon_{\mathcal{Q}_2}\bigg)^{|E|\cdot p}~.
\end{equation}

\begin{figure}[ht!]
    \centering
   \includegraphics[width=0.5\linewidth]{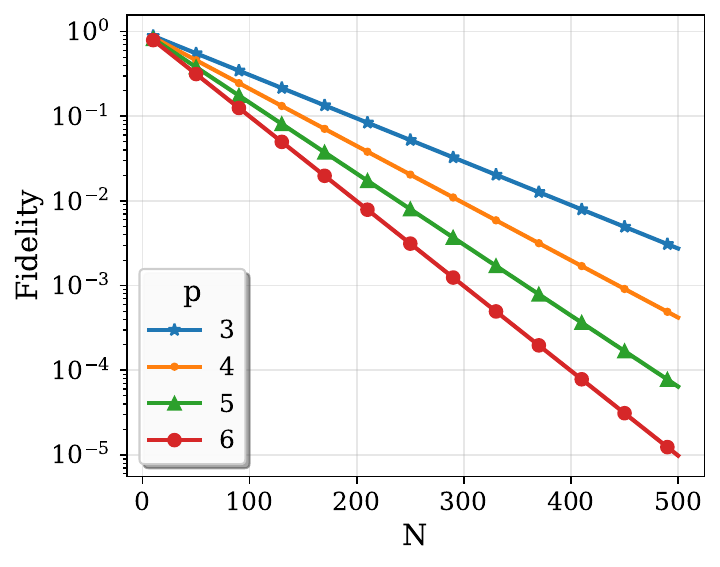}
    \caption{
     Estimated fidelity versus number of vertices $N$ for 3-regular graphs at various circuit depths $p$, using the phenomenological error model with \texttt{Helios} parameters ($p_{\text{spam}} = 5.3 \times 10^{-4}$, $\varepsilon_{\mathcal{Q}_2} = 2.0 \times 10^{-3}$). The log-linear plot shows exponential decay with slope proportional to $p$.
    }
    \label{fig:fidelity}
\end{figure}

For regular graphs, the number of edges is $|E| = \frac{ND}{2}$ and therefore the fidelity decays exponentially in $N$, $D$, and $p$. By plugging in the values of $p_{\text{spam}}$ and $\varepsilon_{\mathcal{Q}_2}$, the fidelity estimate at $p = 4$ for a degree-3 graph on $96$ vertices is $F \simeq 0.22$. As shown in \cref{fig:fidelity}, the fidelity decreases exponentially with the number of vertices $N$, with slope proportional to $p$, underscoring the need for quantum error correction when solving larger graph instances with RWS-QAOA. 

\subsection{Comparison over other warm-start methods}
We compare RWS-QAOA with warm-started QAOA variants commonly referred to as warmest-QAOA~\cite{tate2023warm}, specifically those initialized from the Goemans--Williamson (GW) and Burer--Monteiro (BM) solutions, denoted GW-WS-QAOA and BM-WS-QAOA, respectively. 
While there exist many variants in mapping the classical SDP or BM solutions to the quantum state~\cite{bhattacharyya2025solving, tate2023warm}, we prepared the initial state of GW-WS-QAOA by randomly projecting the SDP solutions to the 2-dimensional space and using a rank-2 approximation for BM. 
Warmest-QAOA typically requires instance-wise parameter optimization; however, \cite{augustino2024strategies} reports that warmest QAOA initialized from the GW solution performs well with fixed tree parameters. Because RWS-QAOA uses fixed parameters determined by our protocol, we compare it against warmest-QAOA under the same fixed schedule. For this reason, we omit warm-start approaches that entail heavy instance-specific parameter optimization, such as \cite{yu2025warm, egger2021warm}.

\begin{figure}[ht]
    \centering
   \includegraphics[width=0.5\linewidth]{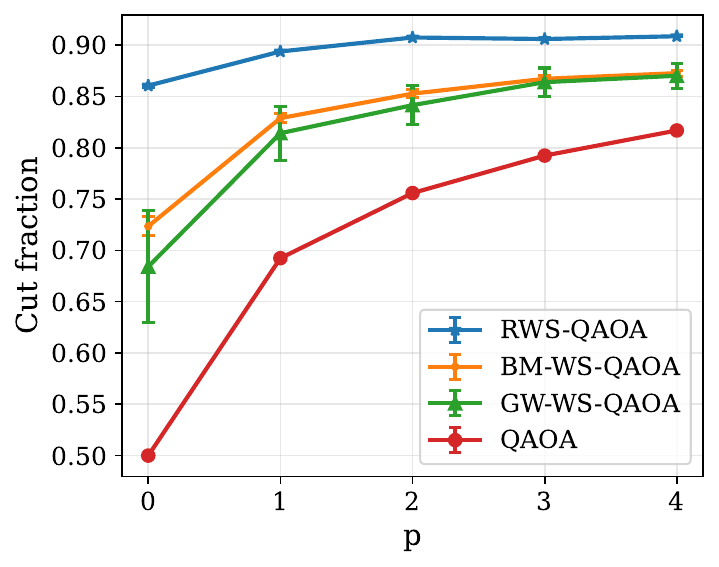}
    \caption{
        Comparison of warm-started QAOA methods on five $2{,}000$-node $3$-regular graph instances. Under our fixed parameter schedules, RWS-QAOA achieves the best performance across depths $p$. GW-WS-QAOA and BM-WS-QAOA use the tree parameter schedule. Error bars represent standard errors. 
    }
    \label{fig:compare_ws_method}
\end{figure}

As shown in \cref{fig:compare_ws_method}, we randomly selected five $3$-regular graphs with $N=2{,}000$ and simulated state energies using tensor-network contraction. Across depths $p$, RWS-QAOA consistently outperforms GW-WS-QAOA, BM-WS-QAOA, and standard QAOA. Here, $p=0$ denotes the warm-start initial state (no QAOA evolution). 

\subsection{Stability of solvers with respect to multistarts}\label{sec:stability_of_warm_start}

\begin{figure*}[th]
    \centering
    \includegraphics[width=1\linewidth]{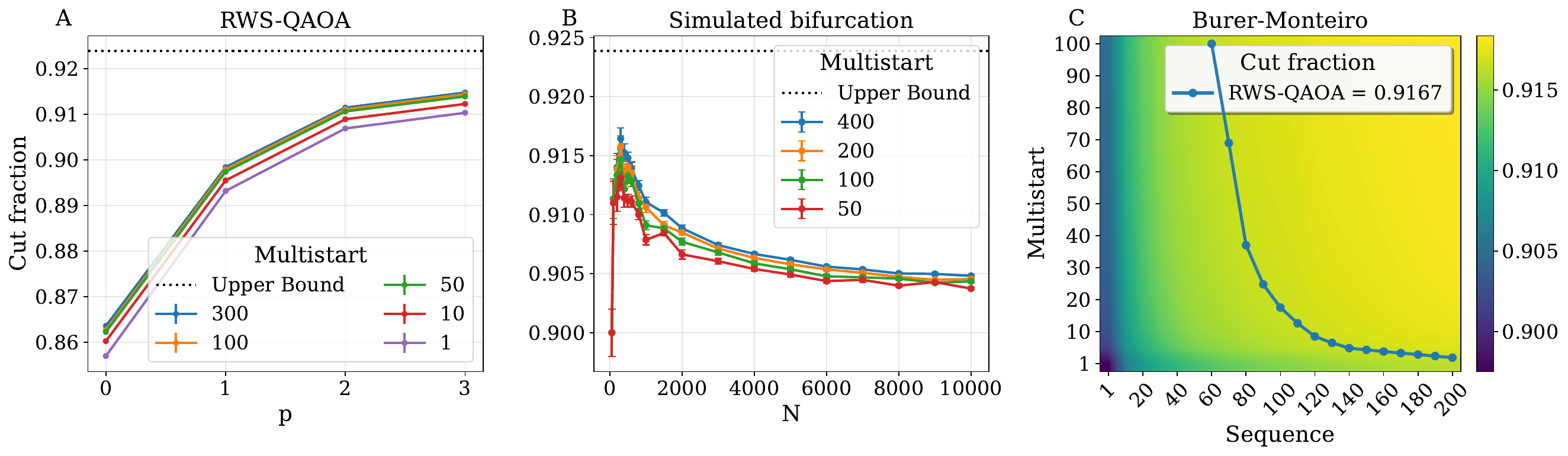}
    \caption{
    Algorithmic performance under different numbers of multistarts.
    (A) RWS-QAOA. More initializations of the random rotation angles improve performance with a marginally decreasing effect. Error bars represent standard errors. 
    (B) Simulated Bifurcation. Cut fraction achieved by the algorithm as a function of the number of multistarts evaluated in parallel. The performance varies by a very small amount as the number of multistarts is increased.
    (C) Burer--Monteiro. The cut fraction on twenty $N=10{,}000$ graphs with varying numbers of initializations and sequential perturbations. The blue line highlights the performance crossover with RWS-QAOA at $p=6$.
    }
    \label{fig:res_multiple_ini}
\end{figure*}

\textbf{RWS-QAOA}. In RWS-QAOA, we first maximize \eqref{eq:objective_prob} to obtain parameters $\boldsymbol{\theta}^\star$ for preparing the warm-start initial state. Because \cref{eq:objective_prob} is highly nonconvex, the quality of $\boldsymbol{\theta}^\star$ depends on the initialization $\boldsymbol{\theta}_0$. To assess robustness, we conduct a stability analysis by varying the number of random initializations $M$ and selecting $\boldsymbol{\theta}^\star$ via a best-of-$M$ strategy (i.e., the maximizer of the objective over $M$ starts). Although a higher objective value does not necessarily yield a better cut in the initial state or superior post-evolution performance, \cref{fig:res_multiple_ini}A empirically shows that increasing $M$ improves RWS-QAOA performance at both $p=0$ (warm start only) and higher depths $p$, with diminishing marginal gains as $M$ grows. Balancing accuracy and cost, we therefore report $M=100$ initializations in the main text and use this setting as the reference for comparisons against classical heuristics.

\begin{figure}[htb!]
    \centering
    \includegraphics[width=1\linewidth]{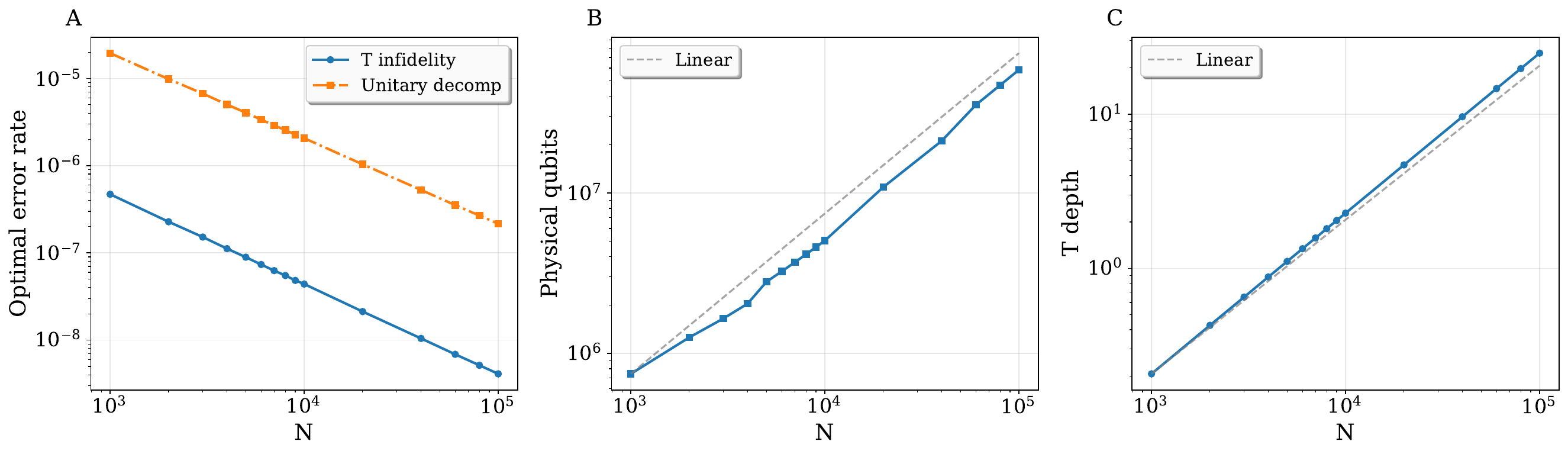}
    \caption{ 
    Additional resource estimation details for RWS-QAOA with $p=6$ on $3$-regular graphs.
    (A) The optimal $T$-gate infidelity and optimal decomposition error. (B) The number of physical qubits grows almost linearly with the number of logical qubits. The slight deviation is due to the increasing $T$-gate fidelity requirement and the correspondingly larger code distance. (C) The $T$-gate depth grows approximately linearly as the number of logical qubits increases. The slight deviation is due to the higher unitary decomposition accuracy required. 
    }
    \label{fig:qaoa_resource_estimation}
\end{figure}
\textbf{Simulated Bifurcation}. The Simulated Bifurcation algorithm is highly parallelizable because it relies only on matrix-vector multiplication. Therefore, one can perform a multi-agent search of the optimal solution by evolving several spin vectors in parallel. The performance of the Simulated Bifurcation algorithm depends on the number of multistarts. However, the dependence is quite mild and we see an increase of only one part in a thousand as the number of multistarts is increased from 50 to 400, as shown in \cref{fig:res_multiple_ini}B.  

\textbf{Burer--Monteiro}. Intuitively, BM exhibits improved performance under both iterative execution and multistarts. 
We further quantify its dependence on the number of random initializations $M$ and the number of sequential perturbation iterations 
$K$ in \cref{fig:res_multiple_ini}C. 
For small $M$, substantially larger 
$K$ is required to achieve performance parity with RWS‑QAOA; in particular, when 
$M<5$, the required $K$ increases markedly. Conversely, the incremental gains from additional multi‑starts diminish as $M$ grows, indicating diminishing marginal returns. Across experiments in \cref{fig:compare_methods_overN}A and B, we set $M=100$ to match the RWS-QAOA configuration, which provides stable and reproducible BM results.

\subsection{Details of resource estimation}
Here we provide additional details of the resource estimation for RWS-QAOA with $p=6$ on $3$-regular graphs discussed in \cref{sec:resource_est}.
In \cref{fig:qaoa_resource_estimation}A, we plot the optimal error rates of the $T$-gate and unitary synthesis for RWS-QAOA as a function of the number of logical qubits, based on the method in~\cite{omanakuttan2025threshold}.
For the full range of logical qubit counts considered (up to $10^5$), the T-gate infidelity is below $10^{-9}$, so these states can be prepared with low resource overhead as studied in \cite{gidney2024magic}. 
Consequently, magic-state preparation is not the bottleneck for the successful implementation of these algorithms.
As the number of logical qubits grows, both the number of physical qubits (\cref{fig:qaoa_resource_estimation}B) and the $T$-gate depth, defined as $\frac{N_T}{N_{\mathrm{fac}}}$ in \cref{eq:quantum_runtime_eq} (\cref{fig:qaoa_resource_estimation}C), increase almost linearly. These results demonstrate the scalability of RWS-QAOA when accounting for fault-tolerant overhead.

\section{More discussion on classical solvers}
\subsection{Comprehensive comparison among classical solvers}\label{sec:compare_classical_solvers}
\begin{figure}[htb!]
    \centering
   \includegraphics[width=0.7\linewidth]{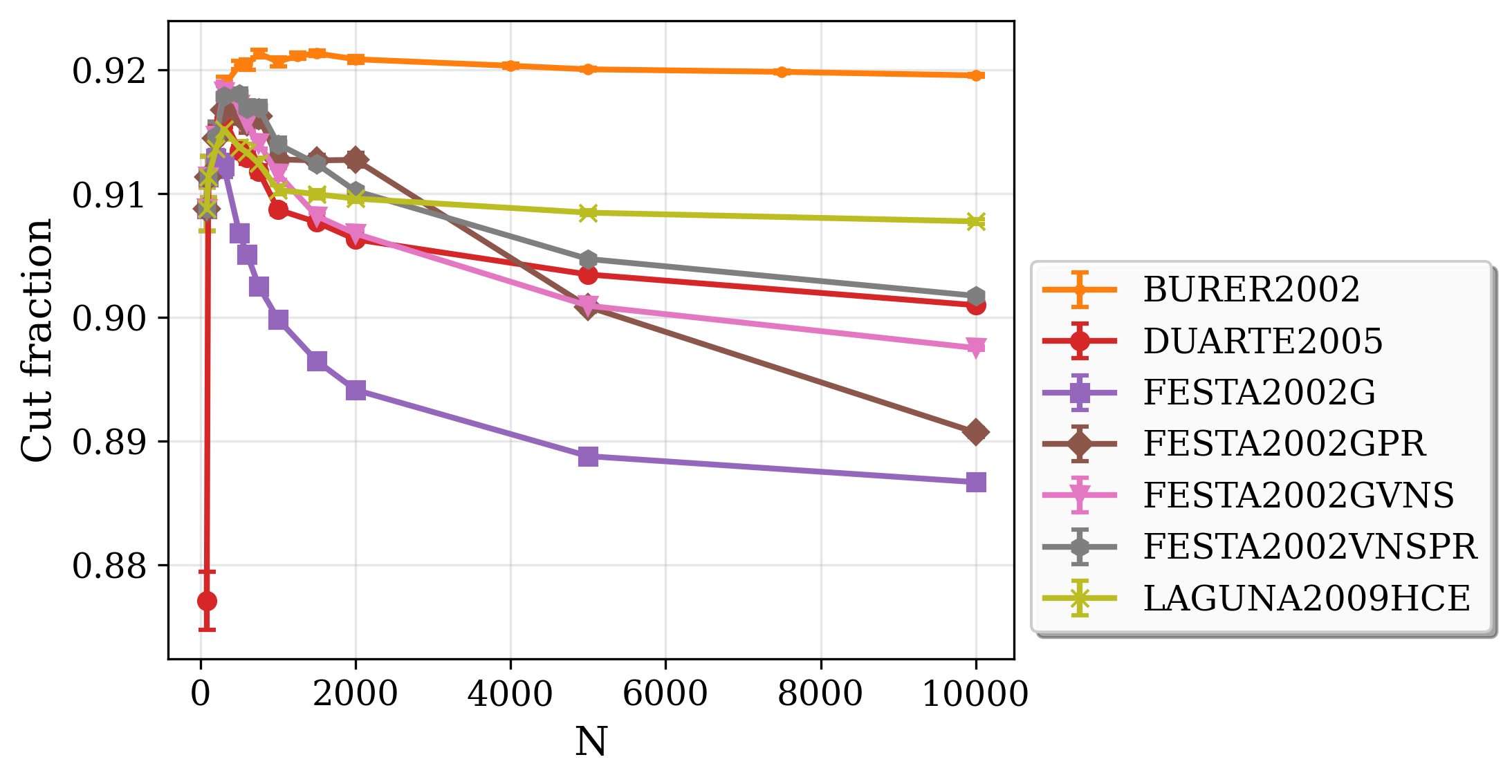}
    \caption{Cut fractions on $20$ degree-$3$ random graphs for top-performing \texttt{MQLib} Max-Cut heuristics with varying numbers of nodes. \texttt{BURER2002} (Burer--Monteiro with local search) is the strongest and is chosen for like-for-like comparison with RWS-QAOA. Error bars represent standard errors.
    Note that for RWS-QAOA comparisons in other figures we exclude local search to ensure fairness and label the solver as ``Burer--Monteiro''. 
    }
    \label{fig:more_classical_solvers}
\end{figure}

We benchmarked the classical Max-Cut heuristics available in \texttt{MQLib}~\cite{Dunning2018} with their default setups on random degree-$3$ graphs. As shown in \cref{fig:more_classical_solvers}, \texttt{BURER2002} achieves the highest cut fractions among all evaluated methods. The \texttt{MQLib} implementation of \texttt{BURER2002} augments the Burer--Monteiro approach with an explicit local search step (described in \cref{sec:local_search}). We therefore adopt \texttt{BURER2002} as the primary classical baseline for large-scale comparison with RWS-QAOA.

\subsection{Convergence of Burer--Monteiro towards RSB bound}\label{sec:rsb_converge}

\begin{figure*}[htb!]
    \centering
    \includegraphics[width=1\linewidth]{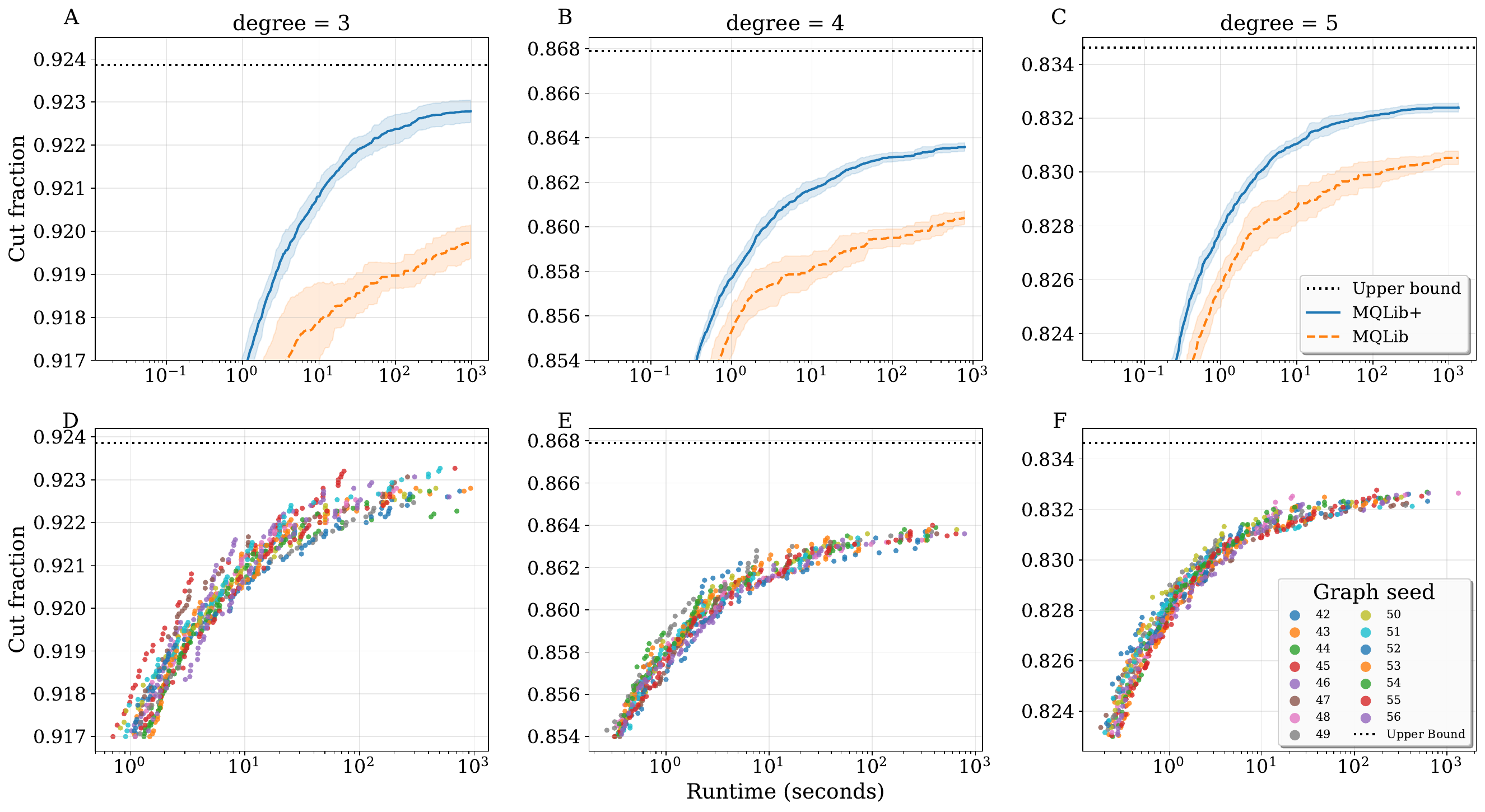}
    \caption{Cut fractions for Max-Cut on random degree-$3$, $4$, and $5$ regular graphs with $10{,}000$ nodes (left to right) using the parallelized \texttt{MQLib+} and default \texttt{MQLib} implementation of \texttt{BURER2002}. In \texttt{MQLib+}, we set the maximum number of sequences to $10{,}000$, run $100$ parallel initializations, and set a sufficiently large number of rounds of sequential perturbations. 
    (A--C): \texttt{MQLib+} finds better cuts than the default \texttt{MQLib} within the same time. The shaded regions represent the standard deviation. 
    (D--F): The instance-wise performance under \texttt{MQLib+}. 
    As the runtime increases, a notable gap persists between the cut found by \texttt{BURER2002} and the upper bound.
    }
    \label{fig:bm_rsb_bound}
\end{figure*}

Theoretical upper bounds on the achievable cut fraction for random regular graphs can be obtained using techniques from statistical physics, particularly the replica method. These bounds are believed to be sharp in the large system limit and serve as important benchmarks for both classical and quantum algorithms~\cite{Dembo2017}. For random $D$-regular graphs, the best known upper bounds on the cut fraction as a function of degree $D$ are summarized for various degrees in~\cite{harangi2025rsbboundsmaximumcut}.

In \cref{fig:bm_rsb_bound}, we compare the performance of \texttt{BURER2002} under the default \texttt{MQLib} and our optimized \texttt{MQLib+}. In the default \texttt{MQLib}, the whole program is executed sequentially. The program restarts with a new random initialization after $10$ rounds of sequential perturbations yield no improvement. In \texttt{MQLib+}, we use $100$ multistarts and set a total of $500$ sequential perturbations. We report the runtime of \texttt{MQLib+} reaching a certain cut fraction by taking the minimum over all trials.

Under \texttt{BURER2002}, increasing the runtime yields monotonic improvements in the cut fraction up to a point and additional time produces diminishing returns. Consequently, the residual gap between the best achieved cut fraction by BM even under \texttt{MQLib+} and the RSB bound persists as shown in \cref{fig:bm_rsb_bound}.

\subsection{Classical runtime comparison}

We now compare the runtime of three prominent classical algorithms for Max-Cut discussed in \cref{sec:classical_solvers}, in addition to the time required to compute the angles $\boldsymbol{\theta} = (\theta_1, \theta_2, \ldots , \theta_N)$ for RWS-QAOA. Since the Goemans--Williamson algorithm involves solving an SDP with an $N \times N$ matrix, the runtime per step is $\Theta(N^3)$, leading to an overall runtime that is at least cubic in $N$. For the Simulated Bifurcation algorithm, each step requires a matrix-vector product which is linear in the number of non-zero entries of the $\mathbf{Q}$ matrix for a QUBO problem. Running for a constant number of steps independent of the problem size $N$ suffices for convergence, and in practice this is set to tens of thousands of steps. This leads to a linear runtime $\mathcal{O}(N)$ for Max-Cut on regular graphs, as the Laplacian of the graph is the $\mathbf{Q}$ matrix when Max-Cut is written as a QUBO with a number of non-zero terms equal to $|V| + 2 |E|$.
\begin{figure}[h]
    \centering
    \includegraphics[width=0.5\linewidth]{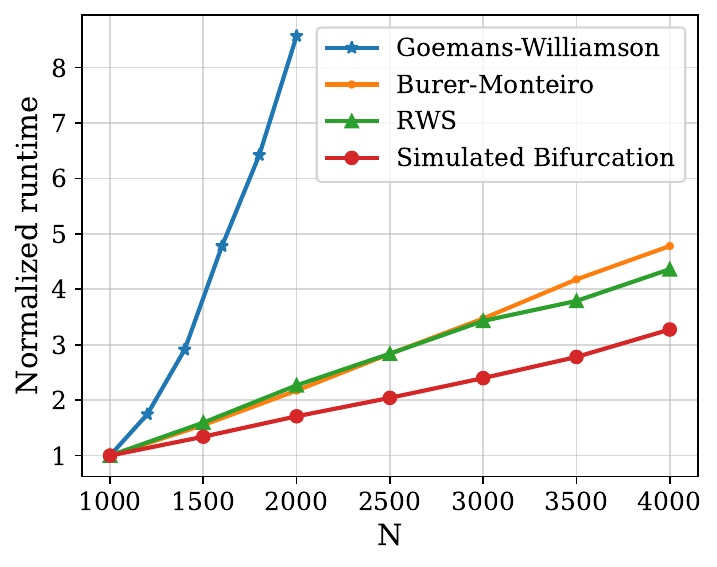}
    \caption{
    Runtime scaling of classical solvers and the RWS-QAOA warm-start angle computation as a function of the number of vertices $N$, averaged over $20$ random $3$-regular graphs per size. Runtimes are normalized by the value at $N=1{,}000$ to account for differences in software implementations. All methods scale nearly linearly except Goemans--Williamson, which exhibits super-cubic scaling.
    }
    \label{fig:compare_runtime}
\end{figure}

For both Burer--Monteiro and the warm-start computation of RWS-QAOA, the computation of the objective and gradient takes time linear in the number of edges. The gradient-descent-based optimizer (e.g., Adam) converges in a fixed number of steps independent of the problem size. For regular graphs, this leads to an overall linear runtime in the number of vertices. These expectations are confirmed in \cref{fig:compare_runtime}, which shows linear runtime for all methods except the Goemans--Williamson algorithm.

\subsection{Detailed algorithms for Burer--Monteiro rounding and classical post-processing}\label{sec:more_alg_charts}
In addition to the runtime of classical solving, another advantage of Burer--Monteiro over SDP-based solvers is that its rounding step is deterministic, which is summarized in Algorithm~\ref{alg:bm_det_rounding}. 

The local search step described in \cref{sec:local_search} can be summarized in Algorithm~\ref{alg:local_search}. It is used for the GW, BM, and RWS-QAOA described in \cref{fig:algorithmic_pipline} while HLZ has a specialized local search as introduced in \cref{sec:classical_solvers}. 

\begin{algorithm}[H]
\caption{Deterministic rounding in Burer--Monteiro~\cite{burer2002rank}}\label{alg:bm_det_rounding}
\KwIn{Angles $\boldsymbol{\theta} = (\theta_1, \ldots, \theta_N)$}
\KwOut{Best solution $x^*$}
\SetKwInOut{Initialize}{Initialize}
\Initialize{
    Set $\phi \gets 0$, $f^\star \gets -\infty$, $i \gets 1$\;
    Let $j$ be the smallest index such that $\theta_j > \pi$; if none exists, set $j \gets N+1$\;
    Set $\theta_{N+1} \gets 2\pi$\;
}
\While{$\phi \leq \pi$}{
    Generate cut $\textbf{x}$ by:
    $
    x_i = 
    \begin{cases}
        +1, & \text{if } \theta_i \in [\phi, \phi + \pi) \\
        -1, & \text{otherwise}
    \end{cases}
    $
    and compute its performance value $f(\textbf{x})$: \\
    \If{$f(\textbf{x}) > f^\star$}{
        Set $f^\star \gets f(\textbf{x})$\;
        Set $\textbf{x}^\star \gets \textbf{x}$\;
    }
    \eIf{$\theta_i \leq \theta_j - \pi$}{
        Set $\phi \gets \theta_i$\;
        Increment $i$ by $1$\;
    }{
        Set $\phi \gets \theta_j - \pi$\;
        Increment $j$ by $1$\;
    }
}
\Return{$\textbf{x}^\star$}
\end{algorithm}

\begin{algorithm}[H]
\caption{Local search post-processing}\label{alg:local_search}
\KwIn{Graph $G$, initial solution $\textbf{x}$}
\KwOut{Refined solution $\textbf{x}$}
\textbf{Phase 1: Single flips}\;
\For{each vertex $i$}{
    \If{flipping bit $x_i$ improves cut}{
        Flip bit $x_i$\;
    }
}
\textbf{Phase 2: Double flips}\;
\For{each edge $(i,j)$}{
    \If{flipping bits $x_i$, $x_j$ improves cut}{
        Flip bits $x_i$, $x_j$\;
    }
}
\Return{$\textbf{x}$}
\end{algorithm}

\putbib[main]
\end{bibunit}


\begin{thebibliography}{96}%
\makeatletter
\providecommand \@ifxundefined [1]{%
 \@ifx{#1\undefined}
}%
\providecommand \@ifnum [1]{%
 \ifnum #1\expandafter \@firstoftwo
 \else \expandafter \@secondoftwo
 \fi
}%
\providecommand \@ifx [1]{%
 \ifx #1\expandafter \@firstoftwo
 \else \expandafter \@secondoftwo
 \fi
}%
\providecommand \natexlab [1]{#1}%
\providecommand \enquote  [1]{``#1''}%
\providecommand \bibnamefont  [1]{#1}%
\providecommand \bibfnamefont [1]{#1}%
\providecommand \citenamefont [1]{#1}%
\providecommand \href@noop [0]{\@secondoftwo}%
\providecommand \href [0]{\begingroup \@sanitize@url \@href}%
\providecommand \@href[1]{\@@startlink{#1}\@@href}%
\providecommand \@@href[1]{\endgroup#1\@@endlink}%
\providecommand \@sanitize@url [0]{\catcode `\\12\catcode `\$12\catcode `\&12\catcode `\#12\catcode `\^12\catcode `\_12\catcode `\%12\relax}%
\providecommand \@@startlink[1]{}%
\providecommand \@@endlink[0]{}%
\providecommand \url  [0]{\begingroup\@sanitize@url \@url }%
\providecommand \@url [1]{\endgroup\@href {#1}{\urlprefix }}%
\providecommand \urlprefix  [0]{URL }%
\providecommand \Eprint [0]{\href }%
\providecommand \doibase [0]{https://doi.org/}%
\providecommand \selectlanguage [0]{\@gobble}%
\providecommand \bibinfo  [0]{\@secondoftwo}%
\providecommand \bibfield  [0]{\@secondoftwo}%
\providecommand \translation [1]{[#1]}%
\providecommand \BibitemOpen [0]{}%
\providecommand \bibitemStop [0]{}%
\providecommand \bibitemNoStop [0]{.\EOS\space}%
\providecommand \EOS [0]{\spacefactor3000\relax}%
\providecommand \BibitemShut  [1]{\csname bibitem#1\endcsname}%
\let\auto@bib@innerbib\@empty
\bibitem [{\citenamefont {{Google Quantum AI}}(2025)}]{quantumecho2025}%
  \BibitemOpen
  \bibfield  {author} {\bibinfo {author} {\bibnamefont {{Google Quantum AI}}},\ }\bibfield  {title} {\bibinfo {title} {Observation of constructive interference at the edge of quantum ergodicity},\ }\href {https://doi.org/10.1038/s41586-025-09526-6} {\bibfield  {journal} {\bibinfo  {journal} {Nature}\ }\textbf {\bibinfo {volume} {646}},\ \bibinfo {pages} {825–830} (\bibinfo {year} {2025})}\BibitemShut {NoStop}%
\bibitem [{\citenamefont {Liu}\ \emph {et~al.}(2025{\natexlab{a}})\citenamefont {Liu}, \citenamefont {Niroula}, \citenamefont {DeCross}, \citenamefont {Foreman}, \citenamefont {Kon}, \citenamefont {Primaatmaja}, \citenamefont {Allman}, \citenamefont {Campora~III}, \citenamefont {Isanaka}, \citenamefont {Singhal} \emph {et~al.}}]{liu2025certified}%
  \BibitemOpen
  \bibfield  {author} {\bibinfo {author} {\bibfnamefont {M.}~\bibnamefont {Liu}}, \bibinfo {author} {\bibfnamefont {P.}~\bibnamefont {Niroula}}, \bibinfo {author} {\bibfnamefont {M.}~\bibnamefont {DeCross}}, \emph {et~al.},\ }\bibfield  {title} {\bibinfo {title} {Certified randomness amplification by dynamically probing remote random quantum states},\ }\href {https://arxiv.org/abs/2511.03686} {\bibfield  {journal} {\bibinfo  {journal} {arXiv preprint arXiv:2511.03686}\ } (\bibinfo {year} {2025}{\natexlab{a}})}\BibitemShut {NoStop}%
\bibitem [{\citenamefont {Niroula}\ \emph {et~al.}(2026)\citenamefont {Niroula}, \citenamefont {Liu}, \citenamefont {Omanakuttan}, \citenamefont {Amaro}, \citenamefont {Chakrabarti}, \citenamefont {Ghosh}, \citenamefont {He}, \citenamefont {Jin}, \citenamefont {Kaleoglu}, \citenamefont {Kordonowy} \emph {et~al.}}]{niroula2026digital}%
  \BibitemOpen
  \bibfield  {author} {\bibinfo {author} {\bibfnamefont {P.}~\bibnamefont {Niroula}}, \bibinfo {author} {\bibfnamefont {M.}~\bibnamefont {Liu}}, \bibinfo {author} {\bibfnamefont {S.}~\bibnamefont {Omanakuttan}}, \emph {et~al.},\ }\bibfield  {title} {\bibinfo {title} {Digital signatures with classical shadows on near-term quantum computers},\ }\href {https://arxiv.org/abs/2602.04859} {\bibfield  {journal} {\bibinfo  {journal} {arXiv preprint arXiv:2602.04859}\ } (\bibinfo {year} {2026})}\BibitemShut {NoStop}%
\bibitem [{\citenamefont {Liu}\ \emph {et~al.}(2025{\natexlab{b}})\citenamefont {Liu}, \citenamefont {Shaydulin}, \citenamefont {Niroula}, \citenamefont {DeCross}, \citenamefont {Hung}, \citenamefont {Kon}, \citenamefont {Cervero-Martín}, \citenamefont {Chakraborty}, \citenamefont {Amer}, \citenamefont {Aaronson}, \citenamefont {Acharya}, \citenamefont {Alexeev}, \citenamefont {Berg}, \citenamefont {Chakrabarti}, \citenamefont {Curchod}, \citenamefont {Dreiling}, \citenamefont {Erickson}, \citenamefont {Foltz}, \citenamefont {Foss-Feig}, \citenamefont {Hayes}, \citenamefont {Humble}, \citenamefont {Kumar}, \citenamefont {Larson}, \citenamefont {Lykov}, \citenamefont {Mills}, \citenamefont {Moses}, \citenamefont {Neyenhuis}, \citenamefont {Eloul}, \citenamefont {Siegfried}, \citenamefont {Walker}, \citenamefont {Lim},\ and\ \citenamefont {Pistoia}}]{Liu2025}%
  \BibitemOpen
  \bibfield  {author} {\bibinfo {author} {\bibfnamefont {M.}~\bibnamefont {Liu}}, \bibinfo {author} {\bibfnamefont {R.}~\bibnamefont {Shaydulin}}, \bibinfo {author} {\bibfnamefont {P.}~\bibnamefont {Niroula}}, \emph {et~al.},\ }\bibfield  {title} {\bibinfo {title} {Certified randomness using a trapped-ion quantum processor},\ }\href {https://doi.org/10.1038/s41586-025-08737-1} {\bibfield  {journal} {\bibinfo  {journal} {Nature}\ }\textbf {\bibinfo {volume} {640}},\ \bibinfo {pages} {343–348} (\bibinfo {year} {2025}{\natexlab{b}})}\BibitemShut {NoStop}%
\bibitem [{\citenamefont {King}\ \emph {et~al.}(2025)\citenamefont {King}, \citenamefont {Nocera}, \citenamefont {Rams}, \citenamefont {Dziarmaga}, \citenamefont {Wiersema}, \citenamefont {Bernoudy}, \citenamefont {Raymond}, \citenamefont {Kaushal}, \citenamefont {Heinsdorf}, \citenamefont {Harris}, \citenamefont {Boothby}, \citenamefont {Altomare}, \citenamefont {Asad}, \citenamefont {Berkley}, \citenamefont {Boschnak}, \citenamefont {Chern}, \citenamefont {Christiani}, \citenamefont {Cibere}, \citenamefont {Connor}, \citenamefont {Dehn}, \citenamefont {Deshpande}, \citenamefont {Ejtemaee}, \citenamefont {Farre}, \citenamefont {Hamer}, \citenamefont {Hoskinson}, \citenamefont {Huang}, \citenamefont {Johnson}, \citenamefont {Kortas}, \citenamefont {Ladizinsky}, \citenamefont {Lanting}, \citenamefont {Lai}, \citenamefont {Li}, \citenamefont {MacDonald}, \citenamefont {Marsden}, \citenamefont {McGeoch}, \citenamefont {Molavi}, \citenamefont {Oh}, \citenamefont {Neufeld}, \citenamefont {Norouzpour},
  \citenamefont {Pasvolsky}, \citenamefont {Poitras}, \citenamefont {Poulin-Lamarre}, \citenamefont {Prescott}, \citenamefont {Reis}, \citenamefont {Rich}, \citenamefont {Samani}, \citenamefont {Sheldan}, \citenamefont {Smirnov}, \citenamefont {Sterpka}, \citenamefont {Trullas~Clavera}, \citenamefont {Tsai}, \citenamefont {Volkmann}, \citenamefont {Whiticar}, \citenamefont {Whittaker}, \citenamefont {Wilkinson}, \citenamefont {Yao}, \citenamefont {Yi}, \citenamefont {Sandvik}, \citenamefont {Alvarez}, \citenamefont {Melko}, \citenamefont {Carrasquilla}, \citenamefont {Franz},\ and\ \citenamefont {Amin}}]{King2025}%
  \BibitemOpen
  \bibfield  {author} {\bibinfo {author} {\bibfnamefont {A.~D.}\ \bibnamefont {King}}, \bibinfo {author} {\bibfnamefont {A.}~\bibnamefont {Nocera}}, \bibinfo {author} {\bibfnamefont {M.~M.}\ \bibnamefont {Rams}}, \emph {et~al.},\ }\bibfield  {title} {\bibinfo {title} {Beyond-classical computation in quantum simulation},\ }\href {https://doi.org/10.1126/science.ado6285} {\bibfield  {journal} {\bibinfo  {journal} {Science}\ }\textbf {\bibinfo {volume} {388}},\ \bibinfo {pages} {199–204} (\bibinfo {year} {2025})}\BibitemShut {NoStop}%
\bibitem [{\citenamefont {Haghshenas}\ \emph {et~al.}(2025)\citenamefont {Haghshenas}, \citenamefont {Chertkov}, \citenamefont {Mills}, \citenamefont {Kadow}, \citenamefont {Lin}, \citenamefont {Chen}, \citenamefont {Cade}, \citenamefont {Niesen}, \citenamefont {Begu{\v{s}}i{\'c}}, \citenamefont {Rudolph} \emph {et~al.}}]{haghshenas2025digital}%
  \BibitemOpen
  \bibfield  {author} {\bibinfo {author} {\bibfnamefont {R.}~\bibnamefont {Haghshenas}}, \bibinfo {author} {\bibfnamefont {E.}~\bibnamefont {Chertkov}}, \bibinfo {author} {\bibfnamefont {M.}~\bibnamefont {Mills}}, \emph {et~al.},\ }\bibfield  {title} {\bibinfo {title} {Digital quantum magnetism at the frontier of classical simulations},\ }\href {https://arxiv.org/abs/2503.20870} {\bibfield  {journal} {\bibinfo  {journal} {arXiv preprint arXiv:2503.20870}\ } (\bibinfo {year} {2025})}\BibitemShut {NoStop}%
\bibitem [{\citenamefont {Jordan}\ \emph {et~al.}(2025)\citenamefont {Jordan}, \citenamefont {Shutty}, \citenamefont {Wootters}, \citenamefont {Zalcman}, \citenamefont {Schmidhuber}, \citenamefont {King}, \citenamefont {Isakov}, \citenamefont {Khattar},\ and\ \citenamefont {Babbush}}]{Jordan2025}%
  \BibitemOpen
  \bibfield  {author} {\bibinfo {author} {\bibfnamefont {S.~P.}\ \bibnamefont {Jordan}}, \bibinfo {author} {\bibfnamefont {N.}~\bibnamefont {Shutty}}, \bibinfo {author} {\bibfnamefont {M.}~\bibnamefont {Wootters}}, \emph {et~al.},\ }\bibfield  {title} {\bibinfo {title} {Optimization by decoded quantum interferometry},\ }\href {https://doi.org/10.1038/s41586-025-09527-5} {\bibfield  {journal} {\bibinfo  {journal} {Nature}\ }\textbf {\bibinfo {volume} {646}},\ \bibinfo {pages} {831–836} (\bibinfo {year} {2025})}\BibitemShut {NoStop}%
\bibitem [{\citenamefont {Chakrabarti}\ \emph {et~al.}(2024)\citenamefont {Chakrabarti}, \citenamefont {Herman}, \citenamefont {Ozgul}, \citenamefont {Zhu}, \citenamefont {Augustino}, \citenamefont {Hao}, \citenamefont {He}, \citenamefont {Shaydulin},\ and\ \citenamefont {Pistoia}}]{chakrabartiShortPath}%
  \BibitemOpen
  \bibfield  {author} {\bibinfo {author} {\bibfnamefont {S.}~\bibnamefont {Chakrabarti}}, \bibinfo {author} {\bibfnamefont {D.}~\bibnamefont {Herman}}, \bibinfo {author} {\bibfnamefont {G.}~\bibnamefont {Ozgul}}, \emph {et~al.},\ }\bibfield  {title} {\bibinfo {title} {Generalized short path algorithms: Towards super-quadratic speedup over {Markov} chain search for combinatorial optimization},\ }\href {https://arxiv.org/abs/2410.23270} {\bibfield  {journal} {\bibinfo  {journal} {arXiv preprint arXiv:2410.23270}\ } (\bibinfo {year} {2024})}\BibitemShut {NoStop}%
\bibitem [{\citenamefont {Dalzell}\ \emph {et~al.}(2023)\citenamefont {Dalzell}, \citenamefont {Pancotti}, \citenamefont {Campbell},\ and\ \citenamefont {Brandão}}]{Dalzell2023}%
  \BibitemOpen
  \bibfield  {author} {\bibinfo {author} {\bibfnamefont {A.~M.}\ \bibnamefont {Dalzell}}, \bibinfo {author} {\bibfnamefont {N.}~\bibnamefont {Pancotti}}, \bibinfo {author} {\bibfnamefont {E.~T.}\ \bibnamefont {Campbell}},\ and\ \bibinfo {author} {\bibfnamefont {F.~G.}\ \bibnamefont {Brandão}},\ }\bibfield  {title} {\bibinfo {title} {Mind the gap: Achieving a super-{Grover} quantum speedup by jumping to the end},\ }in\ \href {https://doi.org/10.1145/3564246.3585203} {\emph {\bibinfo {booktitle} {Proceedings of the 55th Annual ACM Symposium on Theory of Computing}}},\ \bibinfo {series and number} {STOC ’23}\ (\bibinfo  {publisher} {ACM},\ \bibinfo {year} {2023})\ p.\ \bibinfo {pages} {1131–1144}\BibitemShut {NoStop}%
\bibitem [{\citenamefont {Augustino}\ \emph {et~al.}(2026)\citenamefont {Augustino}, \citenamefont {Herman}, \citenamefont {Ozgul}, \citenamefont {Watkins}, \citenamefont {Acharya}, \citenamefont {Fontana}, \citenamefont {Kim},\ and\ \citenamefont {Chakrabarti}}]{2602.13494}%
  \BibitemOpen
  \bibfield  {author} {\bibinfo {author} {\bibfnamefont {B.}~\bibnamefont {Augustino}}, \bibinfo {author} {\bibfnamefont {D.}~\bibnamefont {Herman}}, \bibinfo {author} {\bibfnamefont {G.}~\bibnamefont {Ozgul}}, \emph {et~al.},\ }\bibfield  {title} {\bibinfo {title} {Quantum speedups for group relaxations of integer linear programs},\ }\href {https://arxiv.org/abs/2602.13494} {\bibfield  {journal} {\bibinfo  {journal} {arXiv preprint arXiv:2602.13494}\ } (\bibinfo {year} {2026})}\BibitemShut {NoStop}%
\bibitem [{\citenamefont {Herman}\ \emph {et~al.}(2025)\citenamefont {Herman}, \citenamefont {Ozgul}, \citenamefont {Apte}, \citenamefont {Kim}, \citenamefont {Prakash}, \citenamefont {Shen},\ and\ \citenamefont {Chakrabarti}}]{2510.03385}%
  \BibitemOpen
  \bibfield  {author} {\bibinfo {author} {\bibfnamefont {D.}~\bibnamefont {Herman}}, \bibinfo {author} {\bibfnamefont {G.}~\bibnamefont {Ozgul}}, \bibinfo {author} {\bibfnamefont {A.}~\bibnamefont {Apte}}, \emph {et~al.},\ }\bibfield  {title} {\bibinfo {title} {Mechanisms for quantum advantage in global optimization of nonconvex functions},\ }\href {https://arxiv.org/abs/2510.03385} {\bibfield  {journal} {\bibinfo  {journal} {arXiv preprint arXiv:2510.03385}\ } (\bibinfo {year} {2025})}\BibitemShut {NoStop}%
\bibitem [{\citenamefont {Augustino}\ \emph {et~al.}(2023)\citenamefont {Augustino}, \citenamefont {Leng}, \citenamefont {Nannicini}, \citenamefont {Terlaky},\ and\ \citenamefont {Wu}}]{2311.03977}%
  \BibitemOpen
  \bibfield  {author} {\bibinfo {author} {\bibfnamefont {B.}~\bibnamefont {Augustino}}, \bibinfo {author} {\bibfnamefont {J.}~\bibnamefont {Leng}}, \bibinfo {author} {\bibfnamefont {G.}~\bibnamefont {Nannicini}}, \emph {et~al.},\ }\bibfield  {title} {\bibinfo {title} {A quantum central path algorithm for linear optimization},\ }\href {https://arxiv.org/abs/2311.03977} {\bibfield  {journal} {\bibinfo  {journal} {arXiv preprint arXiv:2311.03977}\ } (\bibinfo {year} {2023})}\BibitemShut {NoStop}%
\bibitem [{\citenamefont {Leng}\ \emph {et~al.}(2023)\citenamefont {Leng}, \citenamefont {Hickman}, \citenamefont {Li},\ and\ \citenamefont {Wu}}]{2303.01471}%
  \BibitemOpen
  \bibfield  {author} {\bibinfo {author} {\bibfnamefont {J.}~\bibnamefont {Leng}}, \bibinfo {author} {\bibfnamefont {E.}~\bibnamefont {Hickman}}, \bibinfo {author} {\bibfnamefont {J.}~\bibnamefont {Li}},\ and\ \bibinfo {author} {\bibfnamefont {X.}~\bibnamefont {Wu}},\ }\bibfield  {title} {\bibinfo {title} {Quantum {Hamiltonian} descent},\ }\href {https://arxiv.org/abs/2303.01471} {\bibfield  {journal} {\bibinfo  {journal} {arXiv preprint arXiv:2303.01471}\ } (\bibinfo {year} {2023})}\BibitemShut {NoStop}%
\bibitem [{\citenamefont {Abbas}\ \emph {et~al.}(2024)\citenamefont {Abbas}, \citenamefont {Ambainis}, \citenamefont {Augustino}, \citenamefont {B\"{a}rtschi}, \citenamefont {Buhrman}, \citenamefont {Coffrin}, \citenamefont {Cortiana}, \citenamefont {Dunjko}, \citenamefont {Egger}, \citenamefont {Elmegreen}, \citenamefont {Franco}, \citenamefont {Fratini}, \citenamefont {Fuller}, \citenamefont {Gacon}, \citenamefont {Gonciulea}, \citenamefont {Gribling}, \citenamefont {Gupta}, \citenamefont {Hadfield}, \citenamefont {Heese}, \citenamefont {Kircher}, \citenamefont {Kleinert}, \citenamefont {Koch}, \citenamefont {Korpas}, \citenamefont {Lenk}, \citenamefont {Marecek}, \citenamefont {Markov}, \citenamefont {Mazzola}, \citenamefont {Mensa}, \citenamefont {Mohseni}, \citenamefont {Nannicini}, \citenamefont {O’Meara}, \citenamefont {Tapia}, \citenamefont {Pokutta}, \citenamefont {Proissl}, \citenamefont {Rebentrost}, \citenamefont {Sahin}, \citenamefont {Symons}, \citenamefont {Tornow}, \citenamefont {Valls},
  \citenamefont {Woerner}, \citenamefont {Wolf-Bauwens}, \citenamefont {Yard}, \citenamefont {Yarkoni}, \citenamefont {Zechiel}, \citenamefont {Zhuk},\ and\ \citenamefont {Zoufal}}]{Abbas2024}%
  \BibitemOpen
  \bibfield  {author} {\bibinfo {author} {\bibfnamefont {A.}~\bibnamefont {Abbas}}, \bibinfo {author} {\bibfnamefont {A.}~\bibnamefont {Ambainis}}, \bibinfo {author} {\bibfnamefont {B.}~\bibnamefont {Augustino}}, \emph {et~al.},\ }\bibfield  {title} {\bibinfo {title} {Challenges and opportunities in quantum optimization},\ }\href {https://doi.org/10.1038/s42254-024-00770-9} {\bibfield  {journal} {\bibinfo  {journal} {Nature Reviews Physics}\ }\textbf {\bibinfo {volume} {6}},\ \bibinfo {pages} {718–735} (\bibinfo {year} {2024})}\BibitemShut {NoStop}%
\bibitem [{\citenamefont {Omanakuttan}\ \emph {et~al.}(2025)\citenamefont {Omanakuttan}, \citenamefont {He}, \citenamefont {Zhang}, \citenamefont {Hao}, \citenamefont {Babakhani}, \citenamefont {Boulebnane}, \citenamefont {Chakrabarti}, \citenamefont {Herman}, \citenamefont {Sullivan}, \citenamefont {Perlin} \emph {et~al.}}]{omanakuttan2025threshold}%
  \BibitemOpen
  \bibfield  {author} {\bibinfo {author} {\bibfnamefont {S.}~\bibnamefont {Omanakuttan}}, \bibinfo {author} {\bibfnamefont {Z.}~\bibnamefont {He}}, \bibinfo {author} {\bibfnamefont {Z.}~\bibnamefont {Zhang}}, \emph {et~al.},\ }\bibfield  {title} {\bibinfo {title} {Threshold for fault-tolerant quantum advantage with the quantum approximate optimization algorithm},\ }\href {https://arxiv.org/abs/2504.01897} {\bibfield  {journal} {\bibinfo  {journal} {arXiv preprint arXiv:2504.01897}\ } (\bibinfo {year} {2025})}\BibitemShut {NoStop}%
\bibitem [{\citenamefont {Farhi}\ \emph {et~al.}(2014)\citenamefont {Farhi}, \citenamefont {Goldstone},\ and\ \citenamefont {Gutmann}}]{farhi2014quantumapproximateoptimizationalgorithm}%
  \BibitemOpen
  \bibfield  {author} {\bibinfo {author} {\bibfnamefont {E.}~\bibnamefont {Farhi}}, \bibinfo {author} {\bibfnamefont {J.}~\bibnamefont {Goldstone}},\ and\ \bibinfo {author} {\bibfnamefont {S.}~\bibnamefont {Gutmann}},\ }\bibfield  {title} {\bibinfo {title} {A quantum approximate optimization algorithm},\ }\href {https://arxiv.org/abs/1411.4028} {\bibfield  {journal} {\bibinfo  {journal} {arXiv preprint arXiv:1411.4028}\ } (\bibinfo {year} {2014})}\BibitemShut {NoStop}%
\bibitem [{\citenamefont {Hogg}\ and\ \citenamefont {Portnov}(2000)}]{hogg2000quantumoptimization}%
  \BibitemOpen
  \bibfield  {author} {\bibinfo {author} {\bibfnamefont {T.}~\bibnamefont {Hogg}}\ and\ \bibinfo {author} {\bibfnamefont {D.}~\bibnamefont {Portnov}},\ }\bibfield  {title} {\bibinfo {title} {Quantum optimization},\ }\href {https://arxiv.org/abs/quant-ph/0006090} {\bibfield  {journal} {\bibinfo  {journal} {arXiv preprint arXiv:quant-ph/0006090}\ } (\bibinfo {year} {2000})}\BibitemShut {NoStop}%
\bibitem [{\citenamefont {Hogg}(2000)}]{Hogg2000}%
  \BibitemOpen
  \bibfield  {author} {\bibinfo {author} {\bibfnamefont {T.}~\bibnamefont {Hogg}},\ }\bibfield  {title} {\bibinfo {title} {Quantum search heuristics},\ }\href {https://doi.org/10.1103/physreva.61.052311} {\bibfield  {journal} {\bibinfo  {journal} {Physical Review A}\ }\textbf {\bibinfo {volume} {61}},\ \bibinfo {pages} {052311} (\bibinfo {year} {2000})}\BibitemShut {NoStop}%
\bibitem [{\citenamefont {Shaydulin}\ and\ \citenamefont {Pistoia}(2023)}]{Shaydulin2023npgeq}%
  \BibitemOpen
  \bibfield  {author} {\bibinfo {author} {\bibfnamefont {R.}~\bibnamefont {Shaydulin}}\ and\ \bibinfo {author} {\bibfnamefont {M.}~\bibnamefont {Pistoia}},\ }\bibfield  {title} {\bibinfo {title} {{QAOA} with $n\cdot p\geq 200$},\ }in\ \href {https://doi.org/10.1109/qce57702.2023.00121} {\emph {\bibinfo {booktitle} {2023 IEEE International Conference on Quantum Computing and Engineering (QCE)}}}\ (\bibinfo  {publisher} {IEEE},\ \bibinfo {year} {2023})\ p.\ \bibinfo {pages} {1074–1077}\BibitemShut {NoStop}%
\bibitem [{\citenamefont {Pelofske}\ \emph {et~al.}(2023)\citenamefont {Pelofske}, \citenamefont {B\"{a}rtschi},\ and\ \citenamefont {Eidenbenz}}]{Pelofske2023}%
  \BibitemOpen
  \bibfield  {author} {\bibinfo {author} {\bibfnamefont {E.}~\bibnamefont {Pelofske}}, \bibinfo {author} {\bibfnamefont {A.}~\bibnamefont {B\"{a}rtschi}},\ and\ \bibinfo {author} {\bibfnamefont {S.}~\bibnamefont {Eidenbenz}},\ }\bibfield  {title} {\bibinfo {title} {Quantum annealing vs. {QAOA}: 127 qubit higher-order ising problems on~{NISQ} computers},\ }in\ \href {https://doi.org/10.1007/978-3-031-32041-5_13} {\emph {\bibinfo {booktitle} {Lecture Notes in Computer Science}}}\ (\bibinfo  {publisher} {Springer Nature Switzerland},\ \bibinfo {year} {2023})\ pp.\ \bibinfo {pages} {240--258}\BibitemShut {NoStop}%
\bibitem [{\citenamefont {Pelofske}\ \emph {et~al.}(2024)\citenamefont {Pelofske}, \citenamefont {B\"{a}rtschi}, \citenamefont {Cincio}, \citenamefont {Golden},\ and\ \citenamefont {Eidenbenz}}]{Pelofske2024}%
  \BibitemOpen
  \bibfield  {author} {\bibinfo {author} {\bibfnamefont {E.}~\bibnamefont {Pelofske}}, \bibinfo {author} {\bibfnamefont {A.}~\bibnamefont {B\"{a}rtschi}}, \bibinfo {author} {\bibfnamefont {L.}~\bibnamefont {Cincio}}, \emph {et~al.},\ }\bibfield  {title} {\bibinfo {title} {Scaling whole-chip {QAOA} for higher-order {Ising} spin glass models on heavy-hex graphs},\ }\bibfield  {journal} {\bibinfo  {journal} {npj Quantum Information}\ }\textbf {\bibinfo {volume} {10}},\ \href {https://doi.org/10.1038/s41534-024-00906-w} {10.1038/s41534-024-00906-w} (\bibinfo {year} {2024})\BibitemShut {NoStop}%
\bibitem [{\citenamefont {He}\ \emph {et~al.}(2025{\natexlab{a}})\citenamefont {He}, \citenamefont {Amaro}, \citenamefont {Shaydulin},\ and\ \citenamefont {Pistoia}}]{he2024performance}%
  \BibitemOpen
  \bibfield  {author} {\bibinfo {author} {\bibfnamefont {Z.}~\bibnamefont {He}}, \bibinfo {author} {\bibfnamefont {D.}~\bibnamefont {Amaro}}, \bibinfo {author} {\bibfnamefont {R.}~\bibnamefont {Shaydulin}},\ and\ \bibinfo {author} {\bibfnamefont {M.}~\bibnamefont {Pistoia}},\ }\bibfield  {title} {\bibinfo {title} {Performance of quantum approximate optimization with quantum error detection},\ }\href {https://doi.org/10.1038/s42005-025-02136-8} {\bibfield  {journal} {\bibinfo  {journal} {Communications Physics}\ }\textbf {\bibinfo {volume} {8}},\ \bibinfo {pages} {217} (\bibinfo {year} {2025}{\natexlab{a}})}\BibitemShut {NoStop}%
\bibitem [{\citenamefont {Boulebnane}\ and\ \citenamefont {Montanaro}(2024{\natexlab{a}})}]{boulebnane2022solving}%
  \BibitemOpen
  \bibfield  {author} {\bibinfo {author} {\bibfnamefont {S.}~\bibnamefont {Boulebnane}}\ and\ \bibinfo {author} {\bibfnamefont {A.}~\bibnamefont {Montanaro}},\ }\bibfield  {title} {\bibinfo {title} {Solving boolean satisfiability problems with the quantum approximate optimization algorithm},\ }\href {https://doi.org/10.1103/PRXQuantum.5.030348} {\bibfield  {journal} {\bibinfo  {journal} {PRX Quantum}\ }\textbf {\bibinfo {volume} {5}},\ \bibinfo {pages} {030348} (\bibinfo {year} {2024}{\natexlab{a}})}\BibitemShut {NoStop}%
\bibitem [{\citenamefont {Shaydulin}\ \emph {et~al.}(2024{\natexlab{a}})\citenamefont {Shaydulin}, \citenamefont {Li}, \citenamefont {Chakrabarti}, \citenamefont {DeCross}, \citenamefont {Herman}, \citenamefont {Kumar}, \citenamefont {Larson}, \citenamefont {Lykov}, \citenamefont {Minssen}, \citenamefont {Sun} \emph {et~al.}}]{shaydulin2023evidence}%
  \BibitemOpen
  \bibfield  {author} {\bibinfo {author} {\bibfnamefont {R.}~\bibnamefont {Shaydulin}}, \bibinfo {author} {\bibfnamefont {C.}~\bibnamefont {Li}}, \bibinfo {author} {\bibfnamefont {S.}~\bibnamefont {Chakrabarti}}, \emph {et~al.},\ }\bibfield  {title} {\bibinfo {title} {Evidence of scaling advantage for the quantum approximate optimization algorithm on a classically intractable problem},\ }\href {https://doi.org/10.1126/sciadv.adm6761} {\bibfield  {journal} {\bibinfo  {journal} {Science Advances}\ }\textbf {\bibinfo {volume} {10}},\ \bibinfo {pages} {eadm6761} (\bibinfo {year} {2024}{\natexlab{a}})}\BibitemShut {NoStop}%
\bibitem [{\citenamefont {Boulebnane}\ \emph {et~al.}(2024)\citenamefont {Boulebnane}, \citenamefont {Ciudad-Alañón}, \citenamefont {Mineh}, \citenamefont {Montanaro},\ and\ \citenamefont {Vaishnav}}]{Vaishnav2024}%
  \BibitemOpen
  \bibfield  {author} {\bibinfo {author} {\bibfnamefont {S.}~\bibnamefont {Boulebnane}}, \bibinfo {author} {\bibfnamefont {M.}~\bibnamefont {Ciudad-Alañón}}, \bibinfo {author} {\bibfnamefont {L.}~\bibnamefont {Mineh}}, \emph {et~al.},\ }\bibfield  {title} {\bibinfo {title} {Applying the quantum approximate optimization algorithm to general constraint satisfaction problems},\ }\href {https://arxiv.org/abs/2411.17442} {\bibfield  {journal} {\bibinfo  {journal} {arXiv preprint arXiv:2411.17442}\ } (\bibinfo {year} {2024})}\BibitemShut {NoStop}%
\bibitem [{\citenamefont {Apte}\ \emph {et~al.}(2025)\citenamefont {Apte}, \citenamefont {Sureshbabu}, \citenamefont {Shaydulin}, \citenamefont {Boulebnane}, \citenamefont {He}, \citenamefont {Herman}, \citenamefont {Sud},\ and\ \citenamefont {Pistoia}}]{apte2025iterative}%
  \BibitemOpen
  \bibfield  {author} {\bibinfo {author} {\bibfnamefont {A.}~\bibnamefont {Apte}}, \bibinfo {author} {\bibfnamefont {S.~H.}\ \bibnamefont {Sureshbabu}}, \bibinfo {author} {\bibfnamefont {R.}~\bibnamefont {Shaydulin}}, \emph {et~al.},\ }\bibfield  {title} {\bibinfo {title} {Iterative interpolation schedules for quantum approximate optimization algorithm},\ }\href {https://arxiv.org/abs/2504.01694} {\bibfield  {journal} {\bibinfo  {journal} {arXiv preprint arXiv:2504.01694}\ } (\bibinfo {year} {2025})}\BibitemShut {NoStop}%
\bibitem [{\citenamefont {Montanaro}\ and\ \citenamefont {Zhou}(2024)}]{Zhou2024}%
  \BibitemOpen
  \bibfield  {author} {\bibinfo {author} {\bibfnamefont {A.}~\bibnamefont {Montanaro}}\ and\ \bibinfo {author} {\bibfnamefont {L.}~\bibnamefont {Zhou}},\ }\bibfield  {title} {\bibinfo {title} {Quantum speedups in solving near-symmetric optimization problems by low-depth {QAOA}},\ }\href {https://arxiv.org/abs/2411.04979} {\bibfield  {journal} {\bibinfo  {journal} {arXiv preprint arXiv:2411.04979}\ } (\bibinfo {year} {2024})}\BibitemShut {NoStop}%
\bibitem [{\citenamefont {Shaydulin}\ \emph {et~al.}(2024{\natexlab{b}})\citenamefont {Shaydulin}, \citenamefont {Li}, \citenamefont {Chakrabarti}, \citenamefont {DeCross}, \citenamefont {Herman}, \citenamefont {Kumar}, \citenamefont {Larson}, \citenamefont {Lykov}, \citenamefont {Minssen}, \citenamefont {Sun} \emph {et~al.}}]{shaydulin2024evidence}%
  \BibitemOpen
  \bibfield  {author} {\bibinfo {author} {\bibfnamefont {R.}~\bibnamefont {Shaydulin}}, \bibinfo {author} {\bibfnamefont {C.}~\bibnamefont {Li}}, \bibinfo {author} {\bibfnamefont {S.}~\bibnamefont {Chakrabarti}}, \emph {et~al.},\ }\bibfield  {title} {\bibinfo {title} {Evidence of scaling advantage for the quantum approximate optimization algorithm on a classically intractable problem},\ }\href {https://doi.org/10.1126/sciadv.adm6761} {\bibfield  {journal} {\bibinfo  {journal} {Science Advances}\ }\textbf {\bibinfo {volume} {10}},\ \bibinfo {pages} {eadm6761} (\bibinfo {year} {2024}{\natexlab{b}})}\BibitemShut {NoStop}%
\bibitem [{\citenamefont {Boulebnane}\ and\ \citenamefont {Montanaro}(2024{\natexlab{b}})}]{boulebnane2024solving}%
  \BibitemOpen
  \bibfield  {author} {\bibinfo {author} {\bibfnamefont {S.}~\bibnamefont {Boulebnane}}\ and\ \bibinfo {author} {\bibfnamefont {A.}~\bibnamefont {Montanaro}},\ }\bibfield  {title} {\bibinfo {title} {Solving boolean satisfiability problems with the quantum approximate optimization algorithm},\ }\href {https://doi.org/10.1103/PRXQuantum.5.030348} {\bibfield  {journal} {\bibinfo  {journal} {PRX Quantum}\ }\textbf {\bibinfo {volume} {5}},\ \bibinfo {pages} {030348} (\bibinfo {year} {2024}{\natexlab{b}})}\BibitemShut {NoStop}%
\bibitem [{\citenamefont {Bhattacharyya}\ \emph {et~al.}(2025)\citenamefont {Bhattacharyya}, \citenamefont {Capriotti},\ and\ \citenamefont {Tate}}]{bhattacharyya2025solving}%
  \BibitemOpen
  \bibfield  {author} {\bibinfo {author} {\bibfnamefont {B.}~\bibnamefont {Bhattacharyya}}, \bibinfo {author} {\bibfnamefont {M.}~\bibnamefont {Capriotti}},\ and\ \bibinfo {author} {\bibfnamefont {R.}~\bibnamefont {Tate}},\ }\bibfield  {title} {\bibinfo {title} {Solving general {QUBOs} with warm-start {QAOA} via a reduction to {Max-Cut}},\ }\href {https://arxiv.org/abs/2504.06253} {\bibfield  {journal} {\bibinfo  {journal} {arXiv preprint arXiv:2504.06253}\ } (\bibinfo {year} {2025})}\BibitemShut {NoStop}%
\bibitem [{\citenamefont {Tate}\ \emph {et~al.}(2023)\citenamefont {Tate}, \citenamefont {Moondra}, \citenamefont {Gard}, \citenamefont {Mohler},\ and\ \citenamefont {Gupta}}]{tate2023warm}%
  \BibitemOpen
  \bibfield  {author} {\bibinfo {author} {\bibfnamefont {R.}~\bibnamefont {Tate}}, \bibinfo {author} {\bibfnamefont {J.}~\bibnamefont {Moondra}}, \bibinfo {author} {\bibfnamefont {B.}~\bibnamefont {Gard}}, \emph {et~al.},\ }\bibfield  {title} {\bibinfo {title} {Warm-started {QAOA} with custom mixers provably converges and computationally beats {Goemans-Williamson}'s max-cut at low circuit depths},\ }\href {https://doi.org/10.22331/q-2023-09-26-1121} {\bibfield  {journal} {\bibinfo  {journal} {Quantum}\ }\textbf {\bibinfo {volume} {7}},\ \bibinfo {pages} {1121} (\bibinfo {year} {2023})}\BibitemShut {NoStop}%
\bibitem [{\citenamefont {Augustino}\ \emph {et~al.}(2024)\citenamefont {Augustino}, \citenamefont {Cain}, \citenamefont {Farhi}, \citenamefont {Gupta}, \citenamefont {Gutmann}, \citenamefont {Ranard}, \citenamefont {Tang},\ and\ \citenamefont {Van~Kirk}}]{augustino2024strategies}%
  \BibitemOpen
  \bibfield  {author} {\bibinfo {author} {\bibfnamefont {B.}~\bibnamefont {Augustino}}, \bibinfo {author} {\bibfnamefont {M.}~\bibnamefont {Cain}}, \bibinfo {author} {\bibfnamefont {E.}~\bibnamefont {Farhi}}, \emph {et~al.},\ }\bibfield  {title} {\bibinfo {title} {Strategies for running the {QAOA} at hundreds of qubits},\ }\href {https://arxiv.org/abs/2410.03015} {\bibfield  {journal} {\bibinfo  {journal} {arXiv preprint arXiv:2410.03015}\ } (\bibinfo {year} {2024})}\BibitemShut {NoStop}%
\bibitem [{\citenamefont {Egger}\ \emph {et~al.}(2021)\citenamefont {Egger}, \citenamefont {Mare{\v{c}}ek},\ and\ \citenamefont {Woerner}}]{egger2021warm}%
  \BibitemOpen
  \bibfield  {author} {\bibinfo {author} {\bibfnamefont {D.~J.}\ \bibnamefont {Egger}}, \bibinfo {author} {\bibfnamefont {J.}~\bibnamefont {Mare{\v{c}}ek}},\ and\ \bibinfo {author} {\bibfnamefont {S.}~\bibnamefont {Woerner}},\ }\bibfield  {title} {\bibinfo {title} {Warm-starting quantum optimization},\ }\href {https://doi.org/10.22331/q-2021-06-17-479} {\bibfield  {journal} {\bibinfo  {journal} {Quantum}\ }\textbf {\bibinfo {volume} {5}},\ \bibinfo {pages} {479} (\bibinfo {year} {2021})}\BibitemShut {NoStop}%
\bibitem [{\citenamefont {Yu}\ \emph {et~al.}(2025)\citenamefont {Yu}, \citenamefont {Wang}, \citenamefont {Shannon},\ and\ \citenamefont {Joynt}}]{yu2025warm}%
  \BibitemOpen
  \bibfield  {author} {\bibinfo {author} {\bibfnamefont {Y.}~\bibnamefont {Yu}}, \bibinfo {author} {\bibfnamefont {X.-B.}\ \bibnamefont {Wang}}, \bibinfo {author} {\bibfnamefont {N.}~\bibnamefont {Shannon}},\ and\ \bibinfo {author} {\bibfnamefont {R.}~\bibnamefont {Joynt}},\ }\bibfield  {title} {\bibinfo {title} {Warm-start adaptive-bias quantum approximate optimization algorithm},\ }\href {https://doi.org/10.1103/nt3w-j4mj} {\bibfield  {journal} {\bibinfo  {journal} {Physical Review A}\ }\textbf {\bibinfo {volume} {112}},\ \bibinfo {pages} {012422} (\bibinfo {year} {2025})}\BibitemShut {NoStop}%
\bibitem [{\citenamefont {Yuan}\ \emph {et~al.}(2025)\citenamefont {Yuan}, \citenamefont {Yang},\ and\ \citenamefont {Barnes}}]{yuan2025iterative}%
  \BibitemOpen
  \bibfield  {author} {\bibinfo {author} {\bibfnamefont {H.}~\bibnamefont {Yuan}}, \bibinfo {author} {\bibfnamefont {S.}~\bibnamefont {Yang}},\ and\ \bibinfo {author} {\bibfnamefont {C.~H.}\ \bibnamefont {Barnes}},\ }\bibfield  {title} {\bibinfo {title} {Iterative quantum optimisation with a warm-started quantum state},\ }\href {https://arxiv.org/abs/2502.09704} {\bibfield  {journal} {\bibinfo  {journal} {arXiv preprint arXiv:2502.09704}\ } (\bibinfo {year} {2025})}\BibitemShut {NoStop}%
\bibitem [{\citenamefont {Nguyen}\ and\ \citenamefont {Kieferov{\'a}}(2025)}]{nguyen2025theoretical}%
  \BibitemOpen
  \bibfield  {author} {\bibinfo {author} {\bibfnamefont {T.}~\bibnamefont {Nguyen}}\ and\ \bibinfo {author} {\bibfnamefont {M.}~\bibnamefont {Kieferov{\'a}}},\ }\bibfield  {title} {\bibinfo {title} {Theoretical guarantees of variational quantum algorithm with guiding states},\ }\href {https://arxiv.org/abs/2510.06764} {\bibfield  {journal} {\bibinfo  {journal} {arXiv preprint arXiv:2510.06764}\ } (\bibinfo {year} {2025})}\BibitemShut {NoStop}%
\bibitem [{\citenamefont {Puig}\ \emph {et~al.}(2025)\citenamefont {Puig}, \citenamefont {Drudis}, \citenamefont {Thanasilp},\ and\ \citenamefont {Holmes}}]{PRXQuantum.6.010317}%
  \BibitemOpen
  \bibfield  {author} {\bibinfo {author} {\bibfnamefont {R.}~\bibnamefont {Puig}}, \bibinfo {author} {\bibfnamefont {M.}~\bibnamefont {Drudis}}, \bibinfo {author} {\bibfnamefont {S.}~\bibnamefont {Thanasilp}},\ and\ \bibinfo {author} {\bibfnamefont {Z.}~\bibnamefont {Holmes}},\ }\bibfield  {title} {\bibinfo {title} {Variational quantum simulation: A case study for understanding warm starts},\ }\href {https://doi.org/10.1103/PRXQuantum.6.010317} {\bibfield  {journal} {\bibinfo  {journal} {PRX Quantum}\ }\textbf {\bibinfo {volume} {6}},\ \bibinfo {pages} {010317} (\bibinfo {year} {2025})}\BibitemShut {NoStop}%
\bibitem [{\citenamefont {He}\ \emph {et~al.}(2023)\citenamefont {He}, \citenamefont {Shaydulin}, \citenamefont {Chakrabarti}, \citenamefont {Herman}, \citenamefont {Li}, \citenamefont {Sun},\ and\ \citenamefont {Pistoia}}]{He2023}%
  \BibitemOpen
  \bibfield  {author} {\bibinfo {author} {\bibfnamefont {Z.}~\bibnamefont {He}}, \bibinfo {author} {\bibfnamefont {R.}~\bibnamefont {Shaydulin}}, \bibinfo {author} {\bibfnamefont {S.}~\bibnamefont {Chakrabarti}}, \emph {et~al.},\ }\bibfield  {title} {\bibinfo {title} {Alignment between initial state and mixer improves {QAOA} performance for constrained optimization},\ }\bibfield  {journal} {\bibinfo  {journal} {npj Quantum Information}\ }\textbf {\bibinfo {volume} {9}},\ \href {https://doi.org/10.1038/s41534-023-00787-5} {10.1038/s41534-023-00787-5} (\bibinfo {year} {2023})\BibitemShut {NoStop}%
\bibitem [{\citenamefont {Cain}\ \emph {et~al.}(2022)\citenamefont {Cain}, \citenamefont {Farhi}, \citenamefont {Gutmann}, \citenamefont {Ranard},\ and\ \citenamefont {Tang}}]{cain2022qaoa}%
  \BibitemOpen
  \bibfield  {author} {\bibinfo {author} {\bibfnamefont {M.}~\bibnamefont {Cain}}, \bibinfo {author} {\bibfnamefont {E.}~\bibnamefont {Farhi}}, \bibinfo {author} {\bibfnamefont {S.}~\bibnamefont {Gutmann}}, \emph {et~al.},\ }\bibfield  {title} {\bibinfo {title} {The {QAOA} gets stuck starting from a good classical string},\ }\href {https://arxiv.org/abs/2207.05089} {\bibfield  {journal} {\bibinfo  {journal} {arXiv preprint arXiv:2207.05089}\ } (\bibinfo {year} {2022})}\BibitemShut {NoStop}%
\bibitem [{\citenamefont {Ransford}\ \emph {et~al.}(2025)\citenamefont {Ransford}, \citenamefont {Allman}, \citenamefont {Arkinstall}, \citenamefont {Campora~III}, \citenamefont {Cooper}, \citenamefont {Delaney}, \citenamefont {Dreiling}, \citenamefont {Estey}, \citenamefont {Figgatt}, \citenamefont {Hall} \emph {et~al.}}]{ransford2025}%
  \BibitemOpen
  \bibfield  {author} {\bibinfo {author} {\bibfnamefont {A.}~\bibnamefont {Ransford}}, \bibinfo {author} {\bibfnamefont {M.~S.}\ \bibnamefont {Allman}}, \bibinfo {author} {\bibfnamefont {J.}~\bibnamefont {Arkinstall}}, \emph {et~al.},\ }\bibfield  {title} {\bibinfo {title} {{Helios}: A 98-qubit trapped-ion quantum computer},\ }\href {https://arxiv.org/abs/2511.05465} {\bibfield  {journal} {\bibinfo  {journal} {arXiv preprint arXiv:2511.05465}\ } (\bibinfo {year} {2025})}\BibitemShut {NoStop}%
\bibitem [{\citenamefont {Goemans}\ and\ \citenamefont {Williamson}(1995)}]{goemans1995}%
  \BibitemOpen
  \bibfield  {author} {\bibinfo {author} {\bibfnamefont {M.~X.}\ \bibnamefont {Goemans}}\ and\ \bibinfo {author} {\bibfnamefont {D.~P.}\ \bibnamefont {Williamson}},\ }\bibfield  {title} {\bibinfo {title} {Improved approximation algorithms for maximum cut and satisfiability problems using semidefinite programming},\ }\href {https://doi.org/10.1145/227683.227684} {\bibfield  {journal} {\bibinfo  {journal} {Journal of the ACM}\ }\textbf {\bibinfo {volume} {42}},\ \bibinfo {pages} {1115–1145} (\bibinfo {year} {1995})}\BibitemShut {NoStop}%
\bibitem [{\citenamefont {Halperin}\ \emph {et~al.}(2004)\citenamefont {Halperin}, \citenamefont {Livnat},\ and\ \citenamefont {Zwick}}]{Halperin2004}%
  \BibitemOpen
  \bibfield  {author} {\bibinfo {author} {\bibfnamefont {E.}~\bibnamefont {Halperin}}, \bibinfo {author} {\bibfnamefont {D.}~\bibnamefont {Livnat}},\ and\ \bibinfo {author} {\bibfnamefont {U.}~\bibnamefont {Zwick}},\ }\bibfield  {title} {\bibinfo {title} {{MAX CUT} in cubic graphs},\ }\href {https://doi.org/10.1016/j.jalgor.2004.06.001} {\bibfield  {journal} {\bibinfo  {journal} {Journal of Algorithms}\ }\textbf {\bibinfo {volume} {53}},\ \bibinfo {pages} {169–185} (\bibinfo {year} {2004})}\BibitemShut {NoStop}%
\bibitem [{\citenamefont {Kitaev}(2006)}]{kitaev2006anyons}%
  \BibitemOpen
  \bibfield  {author} {\bibinfo {author} {\bibfnamefont {A.}~\bibnamefont {Kitaev}},\ }\bibfield  {title} {\bibinfo {title} {Anyons in an exactly solved model and beyond},\ }\href {https://doi.org/10.1016/j.aop.2005.10.005} {\bibfield  {journal} {\bibinfo  {journal} {Annals of Physics}\ }\textbf {\bibinfo {volume} {321}},\ \bibinfo {pages} {2} (\bibinfo {year} {2006})}\BibitemShut {NoStop}%
\bibitem [{\citenamefont {{Google Quantum AI and Collaborators}}(2025)}]{google2025quantum}%
  \BibitemOpen
  \bibfield  {author} {\bibinfo {author} {\bibnamefont {{Google Quantum AI and Collaborators}}},\ }\bibfield  {title} {\bibinfo {title} {Quantum error correction below the surface code threshold},\ }\href {https://doi.org/10.1038/s41586-024-08449-y} {\bibfield  {journal} {\bibinfo  {journal} {Nature}\ }\textbf {\bibinfo {volume} {638}},\ \bibinfo {pages} {920} (\bibinfo {year} {2025})}\BibitemShut {NoStop}%
\bibitem [{\citenamefont {Gidney}(2025)}]{gidney2025factor}%
  \BibitemOpen
  \bibfield  {author} {\bibinfo {author} {\bibfnamefont {C.}~\bibnamefont {Gidney}},\ }\bibfield  {title} {\bibinfo {title} {How to factor 2048 bit {RSA} integers with less than a million noisy qubits},\ }\href {https://arxiv.org/abs/2505.15917} {\bibfield  {journal} {\bibinfo  {journal} {arXiv preprint arXiv:2505.15917}\ } (\bibinfo {year} {2025})}\BibitemShut {NoStop}%
\bibitem [{\citenamefont {West}\ \emph {et~al.}(2001)\citenamefont {West} \emph {et~al.}}]{west2001introduction}%
  \BibitemOpen
  \bibfield  {author} {\bibinfo {author} {\bibfnamefont {D.~B.}\ \bibnamefont {West}} \emph {et~al.},\ }\href@noop {} {\emph {\bibinfo {title} {Introduction to graph theory}}},\ Vol.~\bibinfo {volume} {2}\ (\bibinfo  {publisher} {Prentice hall Upper Saddle River},\ \bibinfo {year} {2001})\BibitemShut {NoStop}%
\bibitem [{\citenamefont {Barahona}\ \emph {et~al.}(1988)\citenamefont {Barahona}, \citenamefont {Gr\"{o}tschel}, \citenamefont {J\"{u}nger},\ and\ \citenamefont {Reinelt}}]{Barahona1988}%
  \BibitemOpen
  \bibfield  {author} {\bibinfo {author} {\bibfnamefont {F.}~\bibnamefont {Barahona}}, \bibinfo {author} {\bibfnamefont {M.}~\bibnamefont {Gr\"{o}tschel}}, \bibinfo {author} {\bibfnamefont {M.}~\bibnamefont {J\"{u}nger}},\ and\ \bibinfo {author} {\bibfnamefont {G.}~\bibnamefont {Reinelt}},\ }\bibfield  {title} {\bibinfo {title} {An application of combinatorial optimization to statistical physics and circuit layout design},\ }\href {https://doi.org/10.1287/opre.36.3.493} {\bibfield  {journal} {\bibinfo  {journal} {Operations Research}\ }\textbf {\bibinfo {volume} {36}},\ \bibinfo {pages} {493–513} (\bibinfo {year} {1988})}\BibitemShut {NoStop}%
\bibitem [{\citenamefont {Chatziafratis}\ \emph {et~al.}(2021)\citenamefont {Chatziafratis}, \citenamefont {Mahdian},\ and\ \citenamefont {Ahmadian}}]{chatziafratis2021}%
  \BibitemOpen
  \bibfield  {author} {\bibinfo {author} {\bibfnamefont {V.}~\bibnamefont {Chatziafratis}}, \bibinfo {author} {\bibfnamefont {M.}~\bibnamefont {Mahdian}},\ and\ \bibinfo {author} {\bibfnamefont {S.}~\bibnamefont {Ahmadian}},\ }\bibfield  {title} {\bibinfo {title} {Maximizing agreements for ranking, clustering and hierarchical clustering via {MAX-CUT}},\ }\href {https://arxiv.org/abs/2102.11548} {\bibfield  {journal} {\bibinfo  {journal} {arXiv preprint arXiv:2102.11548}\ } (\bibinfo {year} {2021})}\BibitemShut {NoStop}%
\bibitem [{\citenamefont {Wurtz}\ and\ \citenamefont {Love}(2021)}]{wurtz2021maxcut}%
  \BibitemOpen
  \bibfield  {author} {\bibinfo {author} {\bibfnamefont {J.}~\bibnamefont {Wurtz}}\ and\ \bibinfo {author} {\bibfnamefont {P.}~\bibnamefont {Love}},\ }\bibfield  {title} {\bibinfo {title} {{MaxCut} quantum approximate optimization algorithm performance guarantees for $p > 1$},\ }\href {https://doi.org/10.1103/PhysRevA.103.042612} {\bibfield  {journal} {\bibinfo  {journal} {Physical Review A}\ }\textbf {\bibinfo {volume} {103}},\ \bibinfo {pages} {042612} (\bibinfo {year} {2021})}\BibitemShut {NoStop}%
\bibitem [{\citenamefont {Wurtz}\ and\ \citenamefont {Lykov}(2021)}]{wurtz2021fixed}%
  \BibitemOpen
  \bibfield  {author} {\bibinfo {author} {\bibfnamefont {J.}~\bibnamefont {Wurtz}}\ and\ \bibinfo {author} {\bibfnamefont {D.}~\bibnamefont {Lykov}},\ }\bibfield  {title} {\bibinfo {title} {Fixed-angle conjectures for the quantum approximate optimization algorithm on regular {MaxCut} graphs},\ }\href {https://doi.org/10.1103/physreva.104.052419} {\bibfield  {journal} {\bibinfo  {journal} {Physical Review A}\ }\textbf {\bibinfo {volume} {104}},\ \bibinfo {pages} {052419} (\bibinfo {year} {2021})}\BibitemShut {NoStop}%
\bibitem [{\citenamefont {Dupont}\ \emph {et~al.}(2025{\natexlab{a}})\citenamefont {Dupont}, \citenamefont {Sundar}, \citenamefont {Evert}, \citenamefont {Neira}, \citenamefont {Peng}, \citenamefont {Jeffrey},\ and\ \citenamefont {Hodson}}]{dupont2025benchmarking}%
  \BibitemOpen
  \bibfield  {author} {\bibinfo {author} {\bibfnamefont {M.}~\bibnamefont {Dupont}}, \bibinfo {author} {\bibfnamefont {B.}~\bibnamefont {Sundar}}, \bibinfo {author} {\bibfnamefont {B.}~\bibnamefont {Evert}}, \emph {et~al.},\ }\bibfield  {title} {\bibinfo {title} {Benchmarking quantum optimization for the maximum-cut problem on a superconducting quantum computer},\ }\href {https://doi.org/10.1103/PhysRevApplied.23.014045} {\bibfield  {journal} {\bibinfo  {journal} {Physical Review Applied}\ }\textbf {\bibinfo {volume} {23}},\ \bibinfo {pages} {014045} (\bibinfo {year} {2025}{\natexlab{a}})}\BibitemShut {NoStop}%
\bibitem [{\citenamefont {Håstad}(2001)}]{Hstad2001}%
  \BibitemOpen
  \bibfield  {author} {\bibinfo {author} {\bibfnamefont {J.}~\bibnamefont {Håstad}},\ }\bibfield  {title} {\bibinfo {title} {Some optimal inapproximability results},\ }\href {https://doi.org/10.1145/502090.502098} {\bibfield  {journal} {\bibinfo  {journal} {Journal of the ACM}\ }\textbf {\bibinfo {volume} {48}},\ \bibinfo {pages} {798–859} (\bibinfo {year} {2001})}\BibitemShut {NoStop}%
\bibitem [{\citenamefont {Berman}\ and\ \citenamefont {Karpinski}(1999)}]{Berman1999}%
  \BibitemOpen
  \bibfield  {author} {\bibinfo {author} {\bibfnamefont {P.}~\bibnamefont {Berman}}\ and\ \bibinfo {author} {\bibfnamefont {M.}~\bibnamefont {Karpinski}},\ }\bibinfo {title} {On some tighter inapproximability results (extended abstract)},\ in\ \href {https://doi.org/10.1007/3-540-48523-6_17} {\emph {\bibinfo {booktitle} {Automata, Languages and Programming}}}\ (\bibinfo  {publisher} {Springer Berlin Heidelberg},\ \bibinfo {year} {1999})\ p.\ \bibinfo {pages} {200–209}\BibitemShut {NoStop}%
\bibitem [{\citenamefont {Harangi}(2025{\natexlab{a}})}]{harangi2025rsbboundsmaximumcut}%
  \BibitemOpen
  \bibfield  {author} {\bibinfo {author} {\bibfnamefont {V.}~\bibnamefont {Harangi}},\ }\bibfield  {title} {\bibinfo {title} {{RSB} bounds on the maximum cut},\ }\href {https://arxiv.org/abs/2506.21296} {\bibfield  {journal} {\bibinfo  {journal} {arXiv preprint arXiv:2506.21296}\ } (\bibinfo {year} {2025}{\natexlab{a}})}\BibitemShut {NoStop}%
\bibitem [{\citenamefont {Khot}\ \emph {et~al.}(2007)\citenamefont {Khot}, \citenamefont {Kindler}, \citenamefont {Mossel},\ and\ \citenamefont {O’Donnell}}]{Khot2007}%
  \BibitemOpen
  \bibfield  {author} {\bibinfo {author} {\bibfnamefont {S.}~\bibnamefont {Khot}}, \bibinfo {author} {\bibfnamefont {G.}~\bibnamefont {Kindler}}, \bibinfo {author} {\bibfnamefont {E.}~\bibnamefont {Mossel}},\ and\ \bibinfo {author} {\bibfnamefont {R.}~\bibnamefont {O’Donnell}},\ }\bibfield  {title} {\bibinfo {title} {Optimal inapproximability results for {MAX‐CUT} and other 2‐variable {CSPs}?},\ }\href {https://doi.org/10.1137/s0097539705447372} {\bibfield  {journal} {\bibinfo  {journal} {SIAM Journal on Computing}\ }\textbf {\bibinfo {volume} {37}},\ \bibinfo {pages} {319–357} (\bibinfo {year} {2007})}\BibitemShut {NoStop}%
\bibitem [{\citenamefont {Wolkowicz}\ \emph {et~al.}(2012)\citenamefont {Wolkowicz}, \citenamefont {Saigal},\ and\ \citenamefont {Vandenberghe}}]{wolkowicz2012handbook}%
  \BibitemOpen
  \bibfield  {author} {\bibinfo {author} {\bibfnamefont {H.}~\bibnamefont {Wolkowicz}}, \bibinfo {author} {\bibfnamefont {R.}~\bibnamefont {Saigal}},\ and\ \bibinfo {author} {\bibfnamefont {L.}~\bibnamefont {Vandenberghe}},\ }\href {https://doi.org/10.1007/978-1-4615-4381-7} {\emph {\bibinfo {title} {Handbook of Semidefinite Programming: Theory, Algorithms, and Applications}}}\ (\bibinfo  {publisher} {Springer Science \& Business Media},\ \bibinfo {year} {2012})\BibitemShut {NoStop}%
\bibitem [{\citenamefont {Burer}\ \emph {et~al.}(2002)\citenamefont {Burer}, \citenamefont {Monteiro},\ and\ \citenamefont {Zhang}}]{burer2002rank}%
  \BibitemOpen
  \bibfield  {author} {\bibinfo {author} {\bibfnamefont {S.}~\bibnamefont {Burer}}, \bibinfo {author} {\bibfnamefont {R.~D.}\ \bibnamefont {Monteiro}},\ and\ \bibinfo {author} {\bibfnamefont {Y.}~\bibnamefont {Zhang}},\ }\bibfield  {title} {\bibinfo {title} {Rank-two relaxation heuristics for max-cut and other binary quadratic programs},\ }\href {https://doi.org/10.1137/S1052623400382467} {\bibfield  {journal} {\bibinfo  {journal} {SIAM Journal on Optimization}\ }\textbf {\bibinfo {volume} {12}},\ \bibinfo {pages} {503} (\bibinfo {year} {2002})}\BibitemShut {NoStop}%
\bibitem [{\citenamefont {Dupont}\ and\ \citenamefont {Sundar}(2024)}]{dupont2024extending}%
  \BibitemOpen
  \bibfield  {author} {\bibinfo {author} {\bibfnamefont {M.}~\bibnamefont {Dupont}}\ and\ \bibinfo {author} {\bibfnamefont {B.}~\bibnamefont {Sundar}},\ }\bibfield  {title} {\bibinfo {title} {Extending relax-and-round combinatorial optimization solvers with quantum correlations},\ }\href {https://doi.org/10.1103/PhysRevA.109.012429} {\bibfield  {journal} {\bibinfo  {journal} {Physical Review A}\ }\textbf {\bibinfo {volume} {109}},\ \bibinfo {pages} {012429} (\bibinfo {year} {2024})}\BibitemShut {NoStop}%
\bibitem [{\citenamefont {Dupont}\ \emph {et~al.}(2025{\natexlab{b}})\citenamefont {Dupont}, \citenamefont {Oberoi},\ and\ \citenamefont {Sundar}}]{dupont2025optimization}%
  \BibitemOpen
  \bibfield  {author} {\bibinfo {author} {\bibfnamefont {M.}~\bibnamefont {Dupont}}, \bibinfo {author} {\bibfnamefont {T.}~\bibnamefont {Oberoi}},\ and\ \bibinfo {author} {\bibfnamefont {B.}~\bibnamefont {Sundar}},\ }\bibfield  {title} {\bibinfo {title} {Optimization via quantum preconditioning},\ }\href {https://doi.org/10.1103/9prw-684p} {\bibfield  {journal} {\bibinfo  {journal} {Physical Review Applied}\ }\textbf {\bibinfo {volume} {24}},\ \bibinfo {pages} {044013} (\bibinfo {year} {2025}{\natexlab{b}})}\BibitemShut {NoStop}%
\bibitem [{\citenamefont {Fuller}\ \emph {et~al.}(2024)\citenamefont {Fuller}, \citenamefont {Hadfield}, \citenamefont {Glick}, \citenamefont {Imamichi}, \citenamefont {Itoko}, \citenamefont {Thompson}, \citenamefont {Jiao}, \citenamefont {Kagele}, \citenamefont {Blom-Schieber}, \citenamefont {Raymond} \emph {et~al.}}]{fuller2024approximate}%
  \BibitemOpen
  \bibfield  {author} {\bibinfo {author} {\bibfnamefont {B.}~\bibnamefont {Fuller}}, \bibinfo {author} {\bibfnamefont {C.}~\bibnamefont {Hadfield}}, \bibinfo {author} {\bibfnamefont {J.~R.}\ \bibnamefont {Glick}}, \emph {et~al.},\ }\bibfield  {title} {\bibinfo {title} {Approximate solutions of combinatorial problems via quantum relaxations},\ }\href {https://doi.org/10.1109/TQE.2024.3421294} {\bibfield  {journal} {\bibinfo  {journal} {IEEE Transactions on Quantum Engineering}\ }\textbf {\bibinfo {volume} {5}},\ \bibinfo {pages} {1} (\bibinfo {year} {2024})}\BibitemShut {NoStop}%
\bibitem [{\citenamefont {He}\ \emph {et~al.}(2025{\natexlab{b}})\citenamefont {He}, \citenamefont {Raymond}, \citenamefont {Shaydulin},\ and\ \citenamefont {Pistoia}}]{he2025non}%
  \BibitemOpen
  \bibfield  {author} {\bibinfo {author} {\bibfnamefont {Z.}~\bibnamefont {He}}, \bibinfo {author} {\bibfnamefont {R.}~\bibnamefont {Raymond}}, \bibinfo {author} {\bibfnamefont {R.}~\bibnamefont {Shaydulin}},\ and\ \bibinfo {author} {\bibfnamefont {M.}~\bibnamefont {Pistoia}},\ }\bibfield  {title} {\bibinfo {title} {Non-variational quantum random access optimization with alternating operator ansatz},\ }\href {https://doi.org/10.1038/s41598-025-13543-w} {\bibfield  {journal} {\bibinfo  {journal} {Scientific Reports}\ }\textbf {\bibinfo {volume} {15}},\ \bibinfo {pages} {29191} (\bibinfo {year} {2025}{\natexlab{b}})}\BibitemShut {NoStop}%
\bibitem [{\citenamefont {{\v{C}}epait{\.e}}\ \emph {et~al.}(2025)\citenamefont {{\v{C}}epait{\.e}}, \citenamefont {Vaishnav}, \citenamefont {Zhou},\ and\ \citenamefont {Montanaro}}]{vcepaite2025quantum}%
  \BibitemOpen
  \bibfield  {author} {\bibinfo {author} {\bibfnamefont {I.}~\bibnamefont {{\v{C}}epait{\.e}}}, \bibinfo {author} {\bibfnamefont {N.}~\bibnamefont {Vaishnav}}, \bibinfo {author} {\bibfnamefont {L.}~\bibnamefont {Zhou}},\ and\ \bibinfo {author} {\bibfnamefont {A.}~\bibnamefont {Montanaro}},\ }\bibfield  {title} {\bibinfo {title} {Quantum-enhanced optimization by warm starts},\ }\href {https://arxiv.org/abs/2508.16309} {\bibfield  {journal} {\bibinfo  {journal} {arXiv preprint arXiv:2508.16309}\ } (\bibinfo {year} {2025})}\BibitemShut {NoStop}%
\bibitem [{Hel()}]{Helios}%
  \BibitemOpen
  \href@noop {} {\bibinfo {title} {Quantinuum {Helios}}},\ \bibinfo {howpublished} {\url{https://www.quantinuum.com/}},\ \bibinfo {note} {{Nov.} 05 - {Nov.} 15, 2025}\BibitemShut {NoStop}%
\bibitem [{\citenamefont {Goto}\ \emph {et~al.}(2019)\citenamefont {Goto}, \citenamefont {Tatsumura},\ and\ \citenamefont {Dixon}}]{Goto2019}%
  \BibitemOpen
  \bibfield  {author} {\bibinfo {author} {\bibfnamefont {H.}~\bibnamefont {Goto}}, \bibinfo {author} {\bibfnamefont {K.}~\bibnamefont {Tatsumura}},\ and\ \bibinfo {author} {\bibfnamefont {A.~R.}\ \bibnamefont {Dixon}},\ }\bibfield  {title} {\bibinfo {title} {Combinatorial optimization by simulating adiabatic bifurcations in nonlinear {Hamiltonian} systems},\ }\bibfield  {journal} {\bibinfo  {journal} {Science Advances}\ }\textbf {\bibinfo {volume} {5}},\ \href {https://doi.org/10.1126/sciadv.aav2372} {10.1126/sciadv.aav2372} (\bibinfo {year} {2019})\BibitemShut {NoStop}%
\bibitem [{\citenamefont {Ageron}\ \emph {et~al.}(2025)\citenamefont {Ageron}, \citenamefont {Bouquet},\ and\ \citenamefont {Pugliese}}]{Ageron_Simulated_Bifurcation_SB_2023}%
  \BibitemOpen
  \bibfield  {author} {\bibinfo {author} {\bibfnamefont {R.}~\bibnamefont {Ageron}}, \bibinfo {author} {\bibfnamefont {T.}~\bibnamefont {Bouquet}},\ and\ \bibinfo {author} {\bibfnamefont {L.}~\bibnamefont {Pugliese}},\ }\href {https://github.com/bqth29/simulated-bifurcation-algorithm} {\bibinfo {title} {{Simulated Bifurcation (SB) algorithm for Python}}} (\bibinfo {year} {2025})\BibitemShut {NoStop}%
\bibitem [{\citenamefont {Dunning}\ \emph {et~al.}(2018)\citenamefont {Dunning}, \citenamefont {Gupta},\ and\ \citenamefont {Silberholz}}]{Dunning2018}%
  \BibitemOpen
  \bibfield  {author} {\bibinfo {author} {\bibfnamefont {I.}~\bibnamefont {Dunning}}, \bibinfo {author} {\bibfnamefont {S.}~\bibnamefont {Gupta}},\ and\ \bibinfo {author} {\bibfnamefont {J.}~\bibnamefont {Silberholz}},\ }\bibfield  {title} {\bibinfo {title} {What works best when? {A} systematic evaluation of heuristics for {Max-Cut} and {QUBO}},\ }\href {https://doi.org/10.1287/ijoc.2017.0798} {\bibfield  {journal} {\bibinfo  {journal} {INFORMS Journal on Computing}\ }\textbf {\bibinfo {volume} {30}},\ \bibinfo {pages} {608–624} (\bibinfo {year} {2018})}\BibitemShut {NoStop}%
\bibitem [{\citenamefont {Harangi}(2025{\natexlab{b}})}]{harangi2025rsb}%
  \BibitemOpen
  \bibfield  {author} {\bibinfo {author} {\bibfnamefont {V.}~\bibnamefont {Harangi}},\ }\bibfield  {title} {\bibinfo {title} {{RSB} bounds on the maximum cut},\ }\href {https://arxiv.org/abs/2506.21296} {\bibfield  {journal} {\bibinfo  {journal} {arXiv preprint arXiv:2506.21296}\ } (\bibinfo {year} {2025}{\natexlab{b}})}\BibitemShut {NoStop}%
\bibitem [{\citenamefont {Gidney}\ \emph {et~al.}(2024)\citenamefont {Gidney}, \citenamefont {Shutty},\ and\ \citenamefont {Jones}}]{gidney2024magic}%
  \BibitemOpen
  \bibfield  {author} {\bibinfo {author} {\bibfnamefont {C.}~\bibnamefont {Gidney}}, \bibinfo {author} {\bibfnamefont {N.}~\bibnamefont {Shutty}},\ and\ \bibinfo {author} {\bibfnamefont {C.}~\bibnamefont {Jones}},\ }\bibfield  {title} {\bibinfo {title} {Magic state cultivation: growing {T} states as cheap as {CNOT} gates},\ }\href {https://arxiv.org/abs/2409.17595} {\bibfield  {journal} {\bibinfo  {journal} {arXiv preprint arXiv:2409.17595}\ } (\bibinfo {year} {2024})}\BibitemShut {NoStop}%
\bibitem [{\citenamefont {Mohseni}\ \emph {et~al.}(2024)\citenamefont {Mohseni}, \citenamefont {Scherer}, \citenamefont {Johnson}, \citenamefont {Wertheim}, \citenamefont {Otten}, \citenamefont {Aadit}, \citenamefont {Alexeev}, \citenamefont {Bresniker}, \citenamefont {Camsari}, \citenamefont {Chapman} \emph {et~al.}}]{mohseni2024build}%
  \BibitemOpen
  \bibfield  {author} {\bibinfo {author} {\bibfnamefont {M.}~\bibnamefont {Mohseni}}, \bibinfo {author} {\bibfnamefont {A.}~\bibnamefont {Scherer}}, \bibinfo {author} {\bibfnamefont {K.~G.}\ \bibnamefont {Johnson}}, \emph {et~al.},\ }\bibfield  {title} {\bibinfo {title} {How to build a quantum supercomputer: Scaling from hundreds to millions of qubits},\ }\href {https://arxiv.org/abs/2411.10406} {\bibfield  {journal} {\bibinfo  {journal} {arXiv preprint arXiv:2411.10406}\ } (\bibinfo {year} {2024})}\BibitemShut {NoStop}%
\bibitem [{\citenamefont {Basso}\ \emph {et~al.}(2022)\citenamefont {Basso}, \citenamefont {Farhi}, \citenamefont {Marwaha}, \citenamefont {Villalonga},\ and\ \citenamefont {Zhou}}]{Basso2022}%
  \BibitemOpen
  \bibfield  {author} {\bibinfo {author} {\bibfnamefont {J.}~\bibnamefont {Basso}}, \bibinfo {author} {\bibfnamefont {E.}~\bibnamefont {Farhi}}, \bibinfo {author} {\bibfnamefont {K.}~\bibnamefont {Marwaha}}, \emph {et~al.},\ }\bibfield  {title} {\bibinfo {title} {{The Quantum Approximate Optimization Algorithm at High Depth for MaxCut on Large-Girth Regular Graphs and the Sherrington-Kirkpatrick Model}},\ }in\ \href {https://doi.org/10.4230/LIPIcs.TQC.2022.7} {\emph {\bibinfo {booktitle} {17th Conference on the Theory of Quantum Computation, Communication and Cryptography (TQC 2022)}}},\ \bibinfo {series} {Leibniz International Proceedings in Informatics (LIPIcs)}, Vol.\ \bibinfo {volume} {232},\ \bibinfo {editor} {edited by\ \bibinfo {editor} {\bibfnamefont {F.}~\bibnamefont {Le~Gall}}\ and\ \bibinfo {editor} {\bibfnamefont {T.}~\bibnamefont {Morimae}}}\ (\bibinfo  {publisher} {Schloss Dagstuhl -- Leibniz-Zentrum f{\"u}r Informatik},\ \bibinfo {address} {Dagstuhl, Germany},\ \bibinfo {year} {2022})\ pp.\
  \bibinfo {pages} {7:1--7:21}\BibitemShut {NoStop}%
\bibitem [{\citenamefont {Farhi}\ \emph {et~al.}(2025)\citenamefont {Farhi}, \citenamefont {Gutmann}, \citenamefont {Ranard},\ and\ \citenamefont {Villalonga}}]{farhi2025}%
  \BibitemOpen
  \bibfield  {author} {\bibinfo {author} {\bibfnamefont {E.}~\bibnamefont {Farhi}}, \bibinfo {author} {\bibfnamefont {S.}~\bibnamefont {Gutmann}}, \bibinfo {author} {\bibfnamefont {D.}~\bibnamefont {Ranard}},\ and\ \bibinfo {author} {\bibfnamefont {B.}~\bibnamefont {Villalonga}},\ }\bibfield  {title} {\bibinfo {title} {Lower bounding the {MaxCut} of high girth 3-regular graphs using the {QAOA}},\ }\href {https://arxiv.org/abs/2503.12789} {\bibfield  {journal} {\bibinfo  {journal} {arXiv preprint arXiv:2503.12789}\ } (\bibinfo {year} {2025})}\BibitemShut {NoStop}%
\bibitem [{\citenamefont {Apte}\ \emph {et~al.}(2026)\citenamefont {Apte}, \citenamefont {Boulebnane}, \citenamefont {Jin}, \citenamefont {Omanakuttan}, \citenamefont {Perlin},\ and\ \citenamefont {Shaydulin}}]{apte2026}%
  \BibitemOpen
  \bibfield  {author} {\bibinfo {author} {\bibfnamefont {A.}~\bibnamefont {Apte}}, \bibinfo {author} {\bibfnamefont {S.}~\bibnamefont {Boulebnane}}, \bibinfo {author} {\bibfnamefont {Y.}~\bibnamefont {Jin}}, \emph {et~al.},\ }\bibfield  {title} {\bibinfo {title} {Quantum approximate optimization of integer graph problems and surpassing semidefinite programming for {Max-k-Cut}},\ }\href {https://arxiv.org/abs/2602.05956} {\bibfield  {journal} {\bibinfo  {journal} {arXiv preprint arXiv:2602.05956}\ } (\bibinfo {year} {2026})}\BibitemShut {NoStop}%
\bibitem [{\citenamefont {Hao}\ \emph {et~al.}(2025)\citenamefont {Hao}, \citenamefont {He}, \citenamefont {Shaydulin}, \citenamefont {Larson},\ and\ \citenamefont {Pistoia}}]{hao2024end}%
  \BibitemOpen
  \bibfield  {author} {\bibinfo {author} {\bibfnamefont {T.}~\bibnamefont {Hao}}, \bibinfo {author} {\bibfnamefont {Z.}~\bibnamefont {He}}, \bibinfo {author} {\bibfnamefont {R.}~\bibnamefont {Shaydulin}}, \emph {et~al.},\ }\bibfield  {title} {\bibinfo {title} {End-to-end protocol for high-quality quantum approximate optimization algorithm parameters with few shots},\ }\href {https://doi.org/10.1103/24gg-7p8z} {\bibfield  {journal} {\bibinfo  {journal} {Physical Review Research}\ }\textbf {\bibinfo {volume} {7}},\ \bibinfo {pages} {033179} (\bibinfo {year} {2025})}\BibitemShut {NoStop}%
\bibitem [{\citenamefont {He}\ \emph {et~al.}(2024)\citenamefont {He}, \citenamefont {Shaydulin}, \citenamefont {Herman}, \citenamefont {Li}, \citenamefont {Raymond}, \citenamefont {Sureshbabu},\ and\ \citenamefont {Pistoia}}]{ICCAD_qaoapara}%
  \BibitemOpen
  \bibfield  {author} {\bibinfo {author} {\bibfnamefont {Z.}~\bibnamefont {He}}, \bibinfo {author} {\bibfnamefont {R.}~\bibnamefont {Shaydulin}}, \bibinfo {author} {\bibfnamefont {D.}~\bibnamefont {Herman}}, \emph {et~al.},\ }\bibfield  {title} {\bibinfo {title} {Parameter setting heuristics make the quantum approximate optimization algorithm suitable for the early fault-tolerant era},\ }in\ \href {https://doi.org/10.1145/3676536.3697128} {\emph {\bibinfo {booktitle} {Proceedings of the 43rd IEEE/ACM International Conference on Computer-Aided Design}}}\ (\bibinfo {year} {2024})\ pp.\ \bibinfo {pages} {1--7}\BibitemShut {NoStop}%
\bibitem [{\citenamefont {Sciorilli}\ \emph {et~al.}(2025)\citenamefont {Sciorilli}, \citenamefont {Borges}, \citenamefont {Patti}, \citenamefont {García-Martín}, \citenamefont {Camilo}, \citenamefont {Anandkumar},\ and\ \citenamefont {Aolita}}]{Sciorilli2025}%
  \BibitemOpen
  \bibfield  {author} {\bibinfo {author} {\bibfnamefont {M.}~\bibnamefont {Sciorilli}}, \bibinfo {author} {\bibfnamefont {L.}~\bibnamefont {Borges}}, \bibinfo {author} {\bibfnamefont {T.~L.}\ \bibnamefont {Patti}}, \emph {et~al.},\ }\bibfield  {title} {\bibinfo {title} {Towards large-scale quantum optimization solvers with few qubits},\ }\href {https://doi.org/10.1038/s41467-024-55346-z} {\bibfield  {journal} {\bibinfo  {journal} {Nature Communications}\ }\textbf {\bibinfo {volume} {16}},\ \bibinfo {pages} {824} (\bibinfo {year} {2025})}\BibitemShut {NoStop}%
\bibitem [{\citenamefont {Ushijima-Mwesigwa}\ \emph {et~al.}(2021)\citenamefont {Ushijima-Mwesigwa}, \citenamefont {Shaydulin}, \citenamefont {Negre}, \citenamefont {Mniszewski}, \citenamefont {Alexeev},\ and\ \citenamefont {Safro}}]{UshijimaMwesigwa2021}%
  \BibitemOpen
  \bibfield  {author} {\bibinfo {author} {\bibfnamefont {H.}~\bibnamefont {Ushijima-Mwesigwa}}, \bibinfo {author} {\bibfnamefont {R.}~\bibnamefont {Shaydulin}}, \bibinfo {author} {\bibfnamefont {C.~F.~A.}\ \bibnamefont {Negre}}, \emph {et~al.},\ }\bibfield  {title} {\bibinfo {title} {Multilevel combinatorial optimization across quantum architectures},\ }\href {https://doi.org/10.1145/3425607} {\bibfield  {journal} {\bibinfo  {journal} {ACM Transactions on Quantum Computing}\ }\textbf {\bibinfo {volume} {2}},\ \bibinfo {pages} {1–29} (\bibinfo {year} {2021})}\BibitemShut {NoStop}%
\bibitem [{\citenamefont {Schuetz}\ \emph {et~al.}(2025)\citenamefont {Schuetz}, \citenamefont {Yalovetzky}, \citenamefont {Andrist}, \citenamefont {Salton}, \citenamefont {Sun}, \citenamefont {Raymond}, \citenamefont {Chakrabarti}, \citenamefont {Acharya}, \citenamefont {Shaydulin}, \citenamefont {Pistoia},\ and\ \citenamefont {Katzgraber}}]{2503.12551}%
  \BibitemOpen
  \bibfield  {author} {\bibinfo {author} {\bibfnamefont {M.~J.~A.}\ \bibnamefont {Schuetz}}, \bibinfo {author} {\bibfnamefont {R.}~\bibnamefont {Yalovetzky}}, \bibinfo {author} {\bibfnamefont {R.~S.}\ \bibnamefont {Andrist}}, \emph {et~al.},\ }\bibfield  {title} {\bibinfo {title} {{qReduMIS}: A quantum-informed reduction algorithm for the maximum independent set problem},\ }\href {https://arxiv.org/abs/2503.12551} {\bibfield  {journal} {\bibinfo  {journal} {arXiv preprint arXiv:2503.12551}\ } (\bibinfo {year} {2025})}\BibitemShut {NoStop}%
\bibitem [{\citenamefont {Kotil}\ \emph {et~al.}(2025)\citenamefont {Kotil}, \citenamefont {Pelofske}, \citenamefont {Riedm\"{u}ller}, \citenamefont {Egger}, \citenamefont {Eidenbenz}, \citenamefont {Koch},\ and\ \citenamefont {Woerner}}]{Kotil2025}%
  \BibitemOpen
  \bibfield  {author} {\bibinfo {author} {\bibfnamefont {A.}~\bibnamefont {Kotil}}, \bibinfo {author} {\bibfnamefont {E.}~\bibnamefont {Pelofske}}, \bibinfo {author} {\bibfnamefont {S.}~\bibnamefont {Riedm\"{u}ller}}, \emph {et~al.},\ }\bibfield  {title} {\bibinfo {title} {Quantum approximate multi-objective optimization},\ }\href {https://doi.org/10.1038/s43588-025-00873-y} {\bibfield  {journal} {\bibinfo  {journal} {Nature Computational Science}\ }\textbf {\bibinfo {volume} {5}},\ \bibinfo {pages} {1168–1177} (\bibinfo {year} {2025})}\BibitemShut {NoStop}%
\bibitem [{\citenamefont {Brandao}\ \emph {et~al.}(2018)\citenamefont {Brandao}, \citenamefont {Broughton}, \citenamefont {Farhi}, \citenamefont {Gutmann},\ and\ \citenamefont {Neven}}]{1812.04170}%
  \BibitemOpen
  \bibfield  {author} {\bibinfo {author} {\bibfnamefont {F.~G. S.~L.}\ \bibnamefont {Brandao}}, \bibinfo {author} {\bibfnamefont {M.}~\bibnamefont {Broughton}}, \bibinfo {author} {\bibfnamefont {E.}~\bibnamefont {Farhi}}, \emph {et~al.},\ }\bibfield  {title} {\bibinfo {title} {For fixed control parameters the quantum approximate optimization algorithm's objective function value concentrates for typical instances},\ }\href@noop {} {\bibfield  {journal} {\bibinfo  {journal} {arXiv:1812.04170}\ } (\bibinfo {year} {2018})}\BibitemShut {NoStop}%
\bibitem [{\citenamefont {Boulebnane}\ \emph {et~al.}(2025)\citenamefont {Boulebnane}, \citenamefont {Sud}, \citenamefont {Shaydulin},\ and\ \citenamefont {Pistoia}}]{boulebnane2025quantum}%
  \BibitemOpen
  \bibfield  {author} {\bibinfo {author} {\bibfnamefont {S.}~\bibnamefont {Boulebnane}}, \bibinfo {author} {\bibfnamefont {J.}~\bibnamefont {Sud}}, \bibinfo {author} {\bibfnamefont {R.}~\bibnamefont {Shaydulin}},\ and\ \bibinfo {author} {\bibfnamefont {M.}~\bibnamefont {Pistoia}},\ }\bibfield  {title} {\bibinfo {title} {Quantum approximate optimization algorithm in finite size and large depth and equivalence to quantum annealing},\ }\href {https://arxiv.org/abs/2503.09563} {\bibfield  {journal} {\bibinfo  {journal} {arXiv preprint arXiv:2503.09563}\ } (\bibinfo {year} {2025})}\BibitemShut {NoStop}%
\bibitem [{\citenamefont {Kingma}\ and\ \citenamefont {Ba}(2015)}]{adam}%
  \BibitemOpen
  \bibfield  {author} {\bibinfo {author} {\bibfnamefont {D.~P.}\ \bibnamefont {Kingma}}\ and\ \bibinfo {author} {\bibfnamefont {J.}~\bibnamefont {Ba}},\ }\bibfield  {title} {\bibinfo {title} {{Adam}: A method for stochastic optimization},\ }\href {https://arxiv.org/abs/1412.6980} {\bibfield  {journal} {\bibinfo  {journal} {International Conference on Learning Representations}\ } (\bibinfo {year} {2015})}\BibitemShut {NoStop}%
\bibitem [{\citenamefont {Brouwer}\ and\ \citenamefont {Haemers}(2012)}]{Brouwer2012}%
  \BibitemOpen
  \bibfield  {author} {\bibinfo {author} {\bibfnamefont {A.~E.}\ \bibnamefont {Brouwer}}\ and\ \bibinfo {author} {\bibfnamefont {W.~H.}\ \bibnamefont {Haemers}},\ }\href {https://doi.org/10.1007/978-1-4614-1939-6} {\emph {\bibinfo {title} {Spectra of Graphs}}}\ (\bibinfo  {publisher} {Springer New York},\ \bibinfo {year} {2012})\BibitemShut {NoStop}%
\bibitem [{\citenamefont {Ross}\ and\ \citenamefont {Selinger}(2016)}]{ross2016optimal}%
  \BibitemOpen
  \bibfield  {author} {\bibinfo {author} {\bibfnamefont {N.~J.}\ \bibnamefont {Ross}}\ and\ \bibinfo {author} {\bibfnamefont {P.}~\bibnamefont {Selinger}},\ }\bibfield  {title} {\bibinfo {title} {Optimal ancilla-free {Clifford+T} approximation of z-rotations},\ }\href {https://doi.org/10.26421/QIC16.11-12-1} {\bibfield  {journal} {\bibinfo  {journal} {Quantum Information and Computation}\ }\textbf {\bibinfo {volume} {16}},\ \bibinfo {pages} {901–953} (\bibinfo {year} {2016})}\BibitemShut {NoStop}%
\bibitem [{\citenamefont {Bocharov}\ \emph {et~al.}(2015)\citenamefont {Bocharov}, \citenamefont {Roetteler},\ and\ \citenamefont {Svore}}]{Bocharov_efficient_synthesis_2015}%
  \BibitemOpen
  \bibfield  {author} {\bibinfo {author} {\bibfnamefont {A.}~\bibnamefont {Bocharov}}, \bibinfo {author} {\bibfnamefont {M.}~\bibnamefont {Roetteler}},\ and\ \bibinfo {author} {\bibfnamefont {K.~M.}\ \bibnamefont {Svore}},\ }\bibfield  {title} {\bibinfo {title} {Efficient synthesis of universal repeat-until-success quantum circuits},\ }\href {https://doi.org/10.1103/PhysRevLett.114.080502} {\bibfield  {journal} {\bibinfo  {journal} {Physical Review Letters}\ }\textbf {\bibinfo {volume} {114}},\ \bibinfo {pages} {080502} (\bibinfo {year} {2015})}\BibitemShut {NoStop}%
\bibitem [{\citenamefont {Babbush}\ \emph {et~al.}(2021)\citenamefont {Babbush}, \citenamefont {McClean}, \citenamefont {Newman}, \citenamefont {Gidney}, \citenamefont {Boixo},\ and\ \citenamefont {Neven}}]{focus_beyond_quadratic}%
  \BibitemOpen
  \bibfield  {author} {\bibinfo {author} {\bibfnamefont {R.}~\bibnamefont {Babbush}}, \bibinfo {author} {\bibfnamefont {J.~R.}\ \bibnamefont {McClean}}, \bibinfo {author} {\bibfnamefont {M.}~\bibnamefont {Newman}}, \emph {et~al.},\ }\bibfield  {title} {\bibinfo {title} {Focus beyond quadratic speedups for error-corrected quantum advantage},\ }\href {https://doi.org/10.1103/PRXQuantum.2.010103} {\bibfield  {journal} {\bibinfo  {journal} {PRX Quantum}\ }\textbf {\bibinfo {volume} {2}},\ \bibinfo {pages} {010103} (\bibinfo {year} {2021})}\BibitemShut {NoStop}%
\bibitem [{\citenamefont {Beverland}\ \emph {et~al.}(2022)\citenamefont {Beverland}, \citenamefont {Kliuchnikov},\ and\ \citenamefont {Schoute}}]{PRXQuantum.3.020342}%
  \BibitemOpen
  \bibfield  {author} {\bibinfo {author} {\bibfnamefont {M.}~\bibnamefont {Beverland}}, \bibinfo {author} {\bibfnamefont {V.}~\bibnamefont {Kliuchnikov}},\ and\ \bibinfo {author} {\bibfnamefont {E.}~\bibnamefont {Schoute}},\ }\bibfield  {title} {\bibinfo {title} {Surface code compilation via edge-disjoint paths},\ }\href {https://doi.org/10.1103/PRXQuantum.3.020342} {\bibfield  {journal} {\bibinfo  {journal} {PRX Quantum}\ }\textbf {\bibinfo {volume} {3}},\ \bibinfo {pages} {020342} (\bibinfo {year} {2022})}\BibitemShut {NoStop}%
\bibitem [{\citenamefont {Markov}\ and\ \citenamefont {Shi}(2008)}]{doi:10.1137/050644756}%
  \BibitemOpen
  \bibfield  {author} {\bibinfo {author} {\bibfnamefont {I.~L.}\ \bibnamefont {Markov}}\ and\ \bibinfo {author} {\bibfnamefont {Y.}~\bibnamefont {Shi}},\ }\bibfield  {title} {\bibinfo {title} {Simulating quantum computation by contracting tensor networks},\ }\href {https://doi.org/10.1137/050644756} {\bibfield  {journal} {\bibinfo  {journal} {SIAM Journal on Computing}\ }\textbf {\bibinfo {volume} {38}},\ \bibinfo {pages} {963} (\bibinfo {year} {2008})}\BibitemShut {NoStop}%
\bibitem [{\citenamefont {Gray}(2018)}]{gray2018quimb}%
  \BibitemOpen
  \bibfield  {author} {\bibinfo {author} {\bibfnamefont {J.}~\bibnamefont {Gray}},\ }\bibfield  {title} {\bibinfo {title} {{quimb}: A {Python} library for quantum information and many-body calculations},\ }\href {https://doi.org/10.21105/joss.00819} {\bibfield  {journal} {\bibinfo  {journal} {Journal of Open Source Software}\ }\textbf {\bibinfo {volume} {3}},\ \bibinfo {pages} {819} (\bibinfo {year} {2018})}\BibitemShut {NoStop}%
\bibitem [{\citenamefont {Allen}\ \emph {et~al.}(2025)\citenamefont {Allen}, \citenamefont {Anchell}, \citenamefont {Anisimov}, \citenamefont {Applencourt}, \citenamefont {Bagusetty}, \citenamefont {Balakrishnan}, \citenamefont {Balin}, \citenamefont {Bekele}, \citenamefont {Bertoni}, \citenamefont {Blackworth} \emph {et~al.}}]{allen2025}%
  \BibitemOpen
  \bibfield  {author} {\bibinfo {author} {\bibfnamefont {B.~S.}\ \bibnamefont {Allen}}, \bibinfo {author} {\bibfnamefont {J.}~\bibnamefont {Anchell}}, \bibinfo {author} {\bibfnamefont {V.}~\bibnamefont {Anisimov}}, \emph {et~al.},\ }\bibfield  {title} {\bibinfo {title} {{Aurora}: Architecting {Argonne}'s first exascale supercomputer for accelerated scientific discovery},\ }\href {https://arxiv.org/abs/2509.08207} {\bibfield  {journal} {\bibinfo  {journal} {arXiv preprint arXiv:2509.08207}\ } (\bibinfo {year} {2025})}\BibitemShut {NoStop}%
\bibitem [{aur(2024)}]{auroraoverview}%
  \BibitemOpen
  \href@noop {} {\bibinfo {title} {Overview of aurora}},\ \bibinfo {howpublished} {\url{https://www.alcf.anl.gov/sites/default/files/2024-11/Overview-of-Aurora-Oct-2024.pdf}} (\bibinfo {year} {2024})\BibitemShut {NoStop}%
\bibitem [{\citenamefont {Wang}\ \emph {et~al.}(2025)\citenamefont {Wang}, \citenamefont {Chen}, \citenamefont {Sun},\ and\ \citenamefont {Zhang}}]{wang2025unifiedcomplexityalgorithmaccountconstantround}%
  \BibitemOpen
  \bibfield  {author} {\bibinfo {author} {\bibfnamefont {J.}~\bibnamefont {Wang}}, \bibinfo {author} {\bibfnamefont {S.}~\bibnamefont {Chen}}, \bibinfo {author} {\bibfnamefont {X.}~\bibnamefont {Sun}},\ and\ \bibinfo {author} {\bibfnamefont {J.}~\bibnamefont {Zhang}},\ }\bibfield  {title} {\bibinfo {title} {A unified complexity-algorithm account of constant-round {QAOA} expectation computation},\ }\href {https://arxiv.org/abs/2511.20212} {\bibfield  {journal} {\bibinfo  {journal} {arXiv preprint arXiv:2511.20212}\ } (\bibinfo {year} {2025})}\BibitemShut {NoStop}%
\bibitem [{\citenamefont {Verstraete}\ and\ \citenamefont {Cirac}(2006)}]{Verstraete_2006}%
  \BibitemOpen
  \bibfield  {author} {\bibinfo {author} {\bibfnamefont {F.}~\bibnamefont {Verstraete}}\ and\ \bibinfo {author} {\bibfnamefont {J.~I.}\ \bibnamefont {Cirac}},\ }\bibfield  {title} {\bibinfo {title} {Matrix product states represent ground states faithfully},\ }\bibfield  {journal} {\bibinfo  {journal} {Physical Review B}\ }\textbf {\bibinfo {volume} {73}},\ \href {https://doi.org/10.1103/physrevb.73.094423} {10.1103/physrevb.73.094423} (\bibinfo {year} {2006})\BibitemShut {NoStop}%
\bibitem [{\citenamefont {Fishman}\ \emph {et~al.}(2022)\citenamefont {Fishman}, \citenamefont {White},\ and\ \citenamefont {Stoudenmire}}]{ITensor}%
  \BibitemOpen
  \bibfield  {author} {\bibinfo {author} {\bibfnamefont {M.}~\bibnamefont {Fishman}}, \bibinfo {author} {\bibfnamefont {S.~R.}\ \bibnamefont {White}},\ and\ \bibinfo {author} {\bibfnamefont {E.~M.}\ \bibnamefont {Stoudenmire}},\ }\bibfield  {title} {\bibinfo {title} {{The {ITensor} Software Library for Tensor Network Calculations}},\ }\href {https://doi.org/10.21468/SciPostPhysCodeb.4} {\bibfield  {journal} {\bibinfo  {journal} {SciPost Physics Codebases}\ ,\ \bibinfo {pages} {4}} (\bibinfo {year} {2022})}\BibitemShut {NoStop}%
\bibitem [{\citenamefont {Farhi}\ and\ \citenamefont {Harrow}(2019)}]{farhi2019quantumsupremacyquantumapproximate}%
  \BibitemOpen
  \bibfield  {author} {\bibinfo {author} {\bibfnamefont {E.}~\bibnamefont {Farhi}}\ and\ \bibinfo {author} {\bibfnamefont {A.~W.}\ \bibnamefont {Harrow}},\ }\bibfield  {title} {\bibinfo {title} {Quantum supremacy through the quantum approximate optimization algorithm},\ }\href {https://arxiv.org/abs/1602.07674} {\bibfield  {journal} {\bibinfo  {journal} {arXiv preprint arXiv:1602.07674}\ } (\bibinfo {year} {2019})}\BibitemShut {NoStop}%
\bibitem [{\citenamefont {Karloff}(1996)}]{karloff1996good}%
  \BibitemOpen
  \bibfield  {author} {\bibinfo {author} {\bibfnamefont {H.}~\bibnamefont {Karloff}},\ }\bibfield  {title} {\bibinfo {title} {How good is the {Goemans-Williamson} {MAX CUT} algorithm?},\ }in\ \href {https://doi.org/10.1145/237814.237990} {\emph {\bibinfo {booktitle} {Proceedings of the Twenty-Eighth Annual ACM Symposium on Theory of Computing}}},\ \bibinfo {series and number} {STOC '96}\ (\bibinfo  {publisher} {Association for Computing Machinery},\ \bibinfo {address} {New York, NY, USA},\ \bibinfo {year} {1996})\ p.\ \bibinfo {pages} {427–434}\BibitemShut {NoStop}%
\bibitem [{\citenamefont {Alon}\ \emph {et~al.}(2001)\citenamefont {Alon}, \citenamefont {Sudakov},\ and\ \citenamefont {Zwick}}]{alon2001constructing}%
  \BibitemOpen
  \bibfield  {author} {\bibinfo {author} {\bibfnamefont {N.}~\bibnamefont {Alon}}, \bibinfo {author} {\bibfnamefont {B.}~\bibnamefont {Sudakov}},\ and\ \bibinfo {author} {\bibfnamefont {U.}~\bibnamefont {Zwick}},\ }\bibfield  {title} {\bibinfo {title} {Constructing worst case instances for semidefinite programming based approximation algorithms},\ }\href {https://doi.org/10.1137/S0895480100379713} {\bibfield  {journal} {\bibinfo  {journal} {SIAM Journal on Discrete Mathematics}\ }\textbf {\bibinfo {volume} {15}},\ \bibinfo {pages} {58} (\bibinfo {year} {2001})}\BibitemShut {NoStop}%
\end{thebibliography}%


\begin{thebibliography}{13}%
\makeatletter
\providecommand \@ifxundefined [1]{%
 \@ifx{#1\undefined}
}%
\providecommand \@ifnum [1]{%
 \ifnum #1\expandafter \@firstoftwo
 \else \expandafter \@secondoftwo
 \fi
}%
\providecommand \@ifx [1]{%
 \ifx #1\expandafter \@firstoftwo
 \else \expandafter \@secondoftwo
 \fi
}%
\providecommand \natexlab [1]{#1}%
\providecommand \enquote  [1]{``#1''}%
\providecommand \bibnamefont  [1]{#1}%
\providecommand \bibfnamefont [1]{#1}%
\providecommand \citenamefont [1]{#1}%
\providecommand \href@noop [0]{\@secondoftwo}%
\providecommand \href [0]{\begingroup \@sanitize@url \@href}%
\providecommand \@href[1]{\@@startlink{#1}\@@href}%
\providecommand \@@href[1]{\endgroup#1\@@endlink}%
\providecommand \@sanitize@url [0]{\catcode `\\12\catcode `\$12\catcode `\&12\catcode `\#12\catcode `\^12\catcode `\_12\catcode `\%12\relax}%
\providecommand \@@startlink[1]{}%
\providecommand \@@endlink[0]{}%
\providecommand \url  [0]{\begingroup\@sanitize@url \@url }%
\providecommand \@url [1]{\endgroup\@href {#1}{\urlprefix }}%
\providecommand \urlprefix  [0]{URL }%
\providecommand \Eprint [0]{\href }%
\providecommand \doibase [0]{https://doi.org/}%
\providecommand \selectlanguage [0]{\@gobble}%
\providecommand \bibinfo  [0]{\@secondoftwo}%
\providecommand \bibfield  [0]{\@secondoftwo}%
\providecommand \translation [1]{[#1]}%
\providecommand \BibitemOpen [0]{}%
\providecommand \bibitemStop [0]{}%
\providecommand \bibitemNoStop [0]{.\EOS\space}%
\providecommand \EOS [0]{\spacefactor3000\relax}%
\providecommand \BibitemShut  [1]{\csname bibitem#1\endcsname}%
\let\auto@bib@innerbib\@empty
\bibitem [{\citenamefont {Ransford}\ \emph {et~al.}(2025)\citenamefont {Ransford}, \citenamefont {Allman}, \citenamefont {Arkinstall}, \citenamefont {Campora~III}, \citenamefont {Cooper}, \citenamefont {Delaney}, \citenamefont {Dreiling}, \citenamefont {Estey}, \citenamefont {Figgatt}, \citenamefont {Hall} \emph {et~al.}}]{ransford2025}%
  \BibitemOpen
  \bibfield  {author} {\bibinfo {author} {\bibfnamefont {A.}~\bibnamefont {Ransford}}, \bibinfo {author} {\bibfnamefont {M.~S.}\ \bibnamefont {Allman}}, \bibinfo {author} {\bibfnamefont {J.}~\bibnamefont {Arkinstall}}, \emph {et~al.},\ }\bibfield  {title} {\bibinfo {title} {{Helios}: A 98-qubit trapped-ion quantum computer},\ }\href {https://arxiv.org/abs/2511.05465} {\bibfield  {journal} {\bibinfo  {journal} {arXiv preprint arXiv:2511.05465}\ } (\bibinfo {year} {2025})}\BibitemShut {NoStop}%
\bibitem [{\citenamefont {Devoret}\ \emph {et~al.}(2004)\citenamefont {Devoret}, \citenamefont {Wallraff},\ and\ \citenamefont {Martinis}}]{devoret2004superconductingqubitsshortreview}%
  \BibitemOpen
  \bibfield  {author} {\bibinfo {author} {\bibfnamefont {M.~H.}\ \bibnamefont {Devoret}}, \bibinfo {author} {\bibfnamefont {A.}~\bibnamefont {Wallraff}},\ and\ \bibinfo {author} {\bibfnamefont {J.~M.}\ \bibnamefont {Martinis}},\ }\bibfield  {title} {\bibinfo {title} {Superconducting qubits: A short review},\ }\href {https://arxiv.org/abs/cond-mat/0411174} {\bibfield  {journal} {\bibinfo  {journal} {arXiv preprint arXiv:cond-mat/0411174}\ } (\bibinfo {year} {2004})}\BibitemShut {NoStop}%
\bibitem [{\citenamefont {Tate}\ \emph {et~al.}(2023)\citenamefont {Tate}, \citenamefont {Moondra}, \citenamefont {Gard}, \citenamefont {Mohler},\ and\ \citenamefont {Gupta}}]{tate2023warm}%
  \BibitemOpen
  \bibfield  {author} {\bibinfo {author} {\bibfnamefont {R.}~\bibnamefont {Tate}}, \bibinfo {author} {\bibfnamefont {J.}~\bibnamefont {Moondra}}, \bibinfo {author} {\bibfnamefont {B.}~\bibnamefont {Gard}}, \emph {et~al.},\ }\bibfield  {title} {\bibinfo {title} {Warm-started {QAOA} with custom mixers provably converges and computationally beats {Goemans-Williamson}'s max-cut at low circuit depths},\ }\href {https://doi.org/10.22331/q-2023-09-26-1121} {\bibfield  {journal} {\bibinfo  {journal} {Quantum}\ }\textbf {\bibinfo {volume} {7}},\ \bibinfo {pages} {1121} (\bibinfo {year} {2023})}\BibitemShut {NoStop}%
\bibitem [{\citenamefont {Bhattacharyya}\ \emph {et~al.}(2025)\citenamefont {Bhattacharyya}, \citenamefont {Capriotti},\ and\ \citenamefont {Tate}}]{bhattacharyya2025solving}%
  \BibitemOpen
  \bibfield  {author} {\bibinfo {author} {\bibfnamefont {B.}~\bibnamefont {Bhattacharyya}}, \bibinfo {author} {\bibfnamefont {M.}~\bibnamefont {Capriotti}},\ and\ \bibinfo {author} {\bibfnamefont {R.}~\bibnamefont {Tate}},\ }\bibfield  {title} {\bibinfo {title} {Solving general {QUBOs} with warm-start {QAOA} via a reduction to {Max-Cut}},\ }\href {https://arxiv.org/abs/2504.06253} {\bibfield  {journal} {\bibinfo  {journal} {arXiv preprint arXiv:2504.06253}\ } (\bibinfo {year} {2025})}\BibitemShut {NoStop}%
\bibitem [{\citenamefont {Augustino}\ \emph {et~al.}(2024)\citenamefont {Augustino}, \citenamefont {Cain}, \citenamefont {Farhi}, \citenamefont {Gupta}, \citenamefont {Gutmann}, \citenamefont {Ranard}, \citenamefont {Tang},\ and\ \citenamefont {Van~Kirk}}]{augustino2024strategies}%
  \BibitemOpen
  \bibfield  {author} {\bibinfo {author} {\bibfnamefont {B.}~\bibnamefont {Augustino}}, \bibinfo {author} {\bibfnamefont {M.}~\bibnamefont {Cain}}, \bibinfo {author} {\bibfnamefont {E.}~\bibnamefont {Farhi}}, \emph {et~al.},\ }\bibfield  {title} {\bibinfo {title} {Strategies for running the {QAOA} at hundreds of qubits},\ }\href {https://arxiv.org/abs/2410.03015} {\bibfield  {journal} {\bibinfo  {journal} {arXiv preprint arXiv:2410.03015}\ } (\bibinfo {year} {2024})}\BibitemShut {NoStop}%
\bibitem [{\citenamefont {Yu}\ \emph {et~al.}(2025)\citenamefont {Yu}, \citenamefont {Wang}, \citenamefont {Shannon},\ and\ \citenamefont {Joynt}}]{yu2025warm}%
  \BibitemOpen
  \bibfield  {author} {\bibinfo {author} {\bibfnamefont {Y.}~\bibnamefont {Yu}}, \bibinfo {author} {\bibfnamefont {X.-B.}\ \bibnamefont {Wang}}, \bibinfo {author} {\bibfnamefont {N.}~\bibnamefont {Shannon}},\ and\ \bibinfo {author} {\bibfnamefont {R.}~\bibnamefont {Joynt}},\ }\bibfield  {title} {\bibinfo {title} {Warm-start adaptive-bias quantum approximate optimization algorithm},\ }\href {https://doi.org/10.1103/nt3w-j4mj} {\bibfield  {journal} {\bibinfo  {journal} {Physical Review A}\ }\textbf {\bibinfo {volume} {112}},\ \bibinfo {pages} {012422} (\bibinfo {year} {2025})}\BibitemShut {NoStop}%
\bibitem [{\citenamefont {Egger}\ \emph {et~al.}(2021)\citenamefont {Egger}, \citenamefont {Mare{\v{c}}ek},\ and\ \citenamefont {Woerner}}]{egger2021warm}%
  \BibitemOpen
  \bibfield  {author} {\bibinfo {author} {\bibfnamefont {D.~J.}\ \bibnamefont {Egger}}, \bibinfo {author} {\bibfnamefont {J.}~\bibnamefont {Mare{\v{c}}ek}},\ and\ \bibinfo {author} {\bibfnamefont {S.}~\bibnamefont {Woerner}},\ }\bibfield  {title} {\bibinfo {title} {Warm-starting quantum optimization},\ }\href {https://doi.org/10.22331/q-2021-06-17-479} {\bibfield  {journal} {\bibinfo  {journal} {Quantum}\ }\textbf {\bibinfo {volume} {5}},\ \bibinfo {pages} {479} (\bibinfo {year} {2021})}\BibitemShut {NoStop}%
\bibitem [{\citenamefont {Omanakuttan}\ \emph {et~al.}(2025)\citenamefont {Omanakuttan}, \citenamefont {He}, \citenamefont {Zhang}, \citenamefont {Hao}, \citenamefont {Babakhani}, \citenamefont {Boulebnane}, \citenamefont {Chakrabarti}, \citenamefont {Herman}, \citenamefont {Sullivan}, \citenamefont {Perlin} \emph {et~al.}}]{omanakuttan2025threshold}%
  \BibitemOpen
  \bibfield  {author} {\bibinfo {author} {\bibfnamefont {S.}~\bibnamefont {Omanakuttan}}, \bibinfo {author} {\bibfnamefont {Z.}~\bibnamefont {He}}, \bibinfo {author} {\bibfnamefont {Z.}~\bibnamefont {Zhang}}, \emph {et~al.},\ }\bibfield  {title} {\bibinfo {title} {Threshold for fault-tolerant quantum advantage with the quantum approximate optimization algorithm},\ }\href {https://arxiv.org/abs/2504.01897} {\bibfield  {journal} {\bibinfo  {journal} {arXiv preprint arXiv:2504.01897}\ } (\bibinfo {year} {2025})}\BibitemShut {NoStop}%
\bibitem [{\citenamefont {Gidney}\ \emph {et~al.}(2024)\citenamefont {Gidney}, \citenamefont {Shutty},\ and\ \citenamefont {Jones}}]{gidney2024magic}%
  \BibitemOpen
  \bibfield  {author} {\bibinfo {author} {\bibfnamefont {C.}~\bibnamefont {Gidney}}, \bibinfo {author} {\bibfnamefont {N.}~\bibnamefont {Shutty}},\ and\ \bibinfo {author} {\bibfnamefont {C.}~\bibnamefont {Jones}},\ }\bibfield  {title} {\bibinfo {title} {Magic state cultivation: growing {T} states as cheap as {CNOT} gates},\ }\href {https://arxiv.org/abs/2409.17595} {\bibfield  {journal} {\bibinfo  {journal} {arXiv preprint arXiv:2409.17595}\ } (\bibinfo {year} {2024})}\BibitemShut {NoStop}%
\bibitem [{\citenamefont {Dunning}\ \emph {et~al.}(2018)\citenamefont {Dunning}, \citenamefont {Gupta},\ and\ \citenamefont {Silberholz}}]{Dunning2018}%
  \BibitemOpen
  \bibfield  {author} {\bibinfo {author} {\bibfnamefont {I.}~\bibnamefont {Dunning}}, \bibinfo {author} {\bibfnamefont {S.}~\bibnamefont {Gupta}},\ and\ \bibinfo {author} {\bibfnamefont {J.}~\bibnamefont {Silberholz}},\ }\bibfield  {title} {\bibinfo {title} {What works best when? {A} systematic evaluation of heuristics for {Max-Cut} and {QUBO}},\ }\href {https://doi.org/10.1287/ijoc.2017.0798} {\bibfield  {journal} {\bibinfo  {journal} {INFORMS Journal on Computing}\ }\textbf {\bibinfo {volume} {30}},\ \bibinfo {pages} {608–624} (\bibinfo {year} {2018})}\BibitemShut {NoStop}%
\bibitem [{\citenamefont {Dembo}\ \emph {et~al.}(2017)\citenamefont {Dembo}, \citenamefont {Montanari},\ and\ \citenamefont {Sen}}]{Dembo2017}%
  \BibitemOpen
  \bibfield  {author} {\bibinfo {author} {\bibfnamefont {A.}~\bibnamefont {Dembo}}, \bibinfo {author} {\bibfnamefont {A.}~\bibnamefont {Montanari}},\ and\ \bibinfo {author} {\bibfnamefont {S.}~\bibnamefont {Sen}},\ }\bibfield  {title} {\bibinfo {title} {Extremal cuts of sparse random graphs},\ }\href {https://doi.org/10.1214/15-aop1084} {\bibfield  {journal} {\bibinfo  {journal} {The Annals of Probability}\ }\textbf {\bibinfo {volume} {45}},\ \bibinfo {pages} {1190} (\bibinfo {year} {2017})}\BibitemShut {NoStop}%
\bibitem [{\citenamefont {Harangi}(2025)}]{harangi2025rsbboundsmaximumcut}%
  \BibitemOpen
  \bibfield  {author} {\bibinfo {author} {\bibfnamefont {V.}~\bibnamefont {Harangi}},\ }\bibfield  {title} {\bibinfo {title} {{RSB} bounds on the maximum cut},\ }\href {https://arxiv.org/abs/2506.21296} {\bibfield  {journal} {\bibinfo  {journal} {arXiv preprint arXiv:2506.21296}\ } (\bibinfo {year} {2025})}\BibitemShut {NoStop}%
\bibitem [{\citenamefont {Burer}\ \emph {et~al.}(2002)\citenamefont {Burer}, \citenamefont {Monteiro},\ and\ \citenamefont {Zhang}}]{burer2002rank}%
  \BibitemOpen
  \bibfield  {author} {\bibinfo {author} {\bibfnamefont {S.}~\bibnamefont {Burer}}, \bibinfo {author} {\bibfnamefont {R.~D.}\ \bibnamefont {Monteiro}},\ and\ \bibinfo {author} {\bibfnamefont {Y.}~\bibnamefont {Zhang}},\ }\bibfield  {title} {\bibinfo {title} {Rank-two relaxation heuristics for max-cut and other binary quadratic programs},\ }\href {https://doi.org/10.1137/S1052623400382467} {\bibfield  {journal} {\bibinfo  {journal} {SIAM Journal on Optimization}\ }\textbf {\bibinfo {volume} {12}},\ \bibinfo {pages} {503} (\bibinfo {year} {2002})}\BibitemShut {NoStop}%
\end{thebibliography}%


\begin{thebibliography}{0}%
\makeatletter
\providecommand \@ifxundefined [1]{%
 \@ifx{#1\undefined}
}%
\providecommand \@ifnum [1]{%
 \ifnum #1\expandafter \@firstoftwo
 \else \expandafter \@secondoftwo
 \fi
}%
\providecommand \@ifx [1]{%
 \ifx #1\expandafter \@firstoftwo
 \else \expandafter \@secondoftwo
 \fi
}%
\providecommand \natexlab [1]{#1}%
\providecommand \enquote  [1]{``#1''}%
\providecommand \bibnamefont  [1]{#1}%
\providecommand \bibfnamefont [1]{#1}%
\providecommand \citenamefont [1]{#1}%
\providecommand \href@noop [0]{\@secondoftwo}%
\providecommand \href [0]{\begingroup \@sanitize@url \@href}%
\providecommand \@href[1]{\@@startlink{#1}\@@href}%
\providecommand \@@href[1]{\endgroup#1\@@endlink}%
\providecommand \@sanitize@url [0]{\catcode `\\12\catcode `\$12\catcode `\&12\catcode `\#12\catcode `\^12\catcode `\_12\catcode `\%12\relax}%
\providecommand \@@startlink[1]{}%
\providecommand \@@endlink[0]{}%
\providecommand \url  [0]{\begingroup\@sanitize@url \@url }%
\providecommand \@url [1]{\endgroup\@href {#1}{\urlprefix }}%
\providecommand \urlprefix  [0]{URL }%
\providecommand \Eprint [0]{\href }%
\providecommand \doibase [0]{https://doi.org/}%
\providecommand \selectlanguage [0]{\@gobble}%
\providecommand \bibinfo  [0]{\@secondoftwo}%
\providecommand \bibfield  [0]{\@secondoftwo}%
\providecommand \translation [1]{[#1]}%
\providecommand \BibitemOpen [0]{}%
\providecommand \bibitemStop [0]{}%
\providecommand \bibitemNoStop [0]{.\EOS\space}%
\providecommand \EOS [0]{\spacefactor3000\relax}%
\providecommand \BibitemShut  [1]{\csname bibitem#1\endcsname}%
\let\auto@bib@innerbib\@empty
\end{thebibliography}%
\end{document}